\documentclass[12pt]{article}
\usepackage[utf8]{inputenc}

\usepackage[expansion=false]{microtype}
\usepackage{graphicx}
\usepackage{caption}
\usepackage{subfigure}
\usepackage{booktabs}
\usepackage{appendix}
\usepackage[linesnumbered,ruled]{algorithm2e}
\usepackage{xcolor}
\usepackage{cite}
\usepackage[round, sort]{natbib}

\usepackage{algorithmic,eqparbox,array}

\usepackage[hidelinks,colorlinks,citecolor={blue},linkcolor={blue}]{hyperref}
\usepackage[margin = 1in]{geometry}

\usepackage{setspace}
\usepackage{amsmath}
\usepackage{amssymb}
\usepackage{mathtools}
\usepackage{amsthm}

\newtheorem{theorem}{Theorem}[section]
\newtheorem{proposition}[theorem]{Proposition}
\newtheorem{lemma}[theorem]{Lemma}

\newtheorem{assumption}[theorem]{Assumption}
\newtheorem{remark}[theorem]{Remark}

 \newtheoremstyle{PropositionNum}
{\topsep}{\topsep}    
{\itshape}  
{}  
{\bfseries}  
{.} 
{ } 
{\thmname{#1}\thmnote{ \bfseries #3}}
\theoremstyle{PropositionNum}

\usepackage{amstext,mathrsfs}
\usepackage{multirow,comment}
\usepackage{url} 
\usepackage{bbm}
\usepackage{bm}
\usepackage{enumitem}
\usepackage{array}
\usepackage{graphicx}

\usepackage{ifthen}

\usepackage[normalem]{ulem}

\newcommand{\WADD}{{\mathsf{WADD}}}

\newcommand{\ARL}{{\mathsf{ARL}}}

\providecommand{\E}{\mathbb{E}}

\providecommand{\Var}{\mathbb{V}\mathrm{ar}}
\providecommand{\Corr}{\mathbb{C}\mathrm{orr}}

\DeclareMathOperator*{\esssup}{ess\,sup}
\DeclareMathOperator*{\argmax}{arg\,max}

\def\bE{{\mathbb{E}}}

\def\bR{{\mathbb{R}}}

\usepackage[textsize=tiny]{todonotes}
\usepackage{authblk}

\allowdisplaybreaks

\begin{document}
\title{Online Window-Limited Change Detection in Correlation Structure}

\author{Jie Gao$^1$, Liyan Xie$^2$, and Zhaoyuan Li$^1$\footnote{Corresponding author, Email: lizhaoyuan@cuhk.edu.cn.}
}

\affil{
$^1$School of Data Science, The Chinese University of Hong Kong, Shenzhen\\
Guangdong, China, 518172 \\
$^2$Department of Industrial and Systems Engineering, University of Minnesota}

\date{}
\maketitle

\vspace{-1cm}
\begin{abstract}
\noindent {\it Abstract: }
We consider detecting change points in the correlation structure of streaming data with minimum assumptions posed on the underlying data distribution. Detection statistics are constructed for dense and sparse change settings, based on $\ell_1$ and $\ell_{\infty}$ norms of the squared difference of vectorized pre- and post-change correlation matrices, respectively. We also propose a novel threshold determination algorithm based on sign-flip permutations that enhances the efficiency of our procedure, particularly when the data dimension is large compared to the window size. 
Theoretical evaluations of the proposed methods are conducted in terms of average run length in the no-change regime and expected detection delay in the post-change regime. We evaluate the performance of the proposed methods in a range of simulated datasets and demonstrate their effectiveness, with small detection delays that are only slightly larger than those of the exact optimal CUSUM test. Finally, we apply our methods to forecast El Ni{\~n}o events (achieving a state-of-the-art hit rate over 0.86 with near-zero false alarms) and to detect seismic events. These applications illustrate the efficiency and effectiveness of our proposed methodology in detecting fundamental changes with a small delay. 
\end{abstract}


\section{Introduction}

The detection of abrupt changes in the statistical behavior of streaming data is a classic and fundamental problem in signal processing and statistics \citep{siegmund1985sequential,poor-hadj-QCD-book-2008,veeravalli2013quickest,tartakovsky2014sequential}. A change point refers to the time when the underlying data distribution changes, and it may correspond to a triggering event that could have catastrophic consequences if not detected promptly. Therefore, the goal is to detect the change in distribution as quickly as possible, subject to false alarm constraints. Such problems have been extensively studied in the univariate setting, especially for detecting mean shifts \citep{wang2020univariate,chan2013detection,fryzlewicz2014wild,yu2023note,li2015m}.

In the multivariate setting where multiple variables are observed, in addition to changes in univariate characteristics, crucial events are often marked by abrupt correlational changes \citep{cabrieto2018capturing}. Such multivariate settings are common in network data, including social networks \citep{raginsky_OCP, PeelClauset2014,wang2021optimal},  sensor networks \citep{VenuSensor2010}, and cyber-physical systems \citep{cyberphysical15,chen2015quickest,LakhinaCrovellaDiot2004}. The correlation changes are significant in various real-world applications. For example, in economics, increasing associations among diverse financial assets are often associated with financial crises \citep{galeano2017dating,wied2017nonparametric}. In climate science, the El Ni{\~n}o event, the most important phenomenon of contemporary natural climate variability, exemplifies a phenomenon in which interactions between different locations in the Pacific strengthen before and weaken during events \citep{ludescher2013improved}. In seismology, correlations between seismic sensors typically strengthen when a seismic event occurs \citep{xie2019-isit}. 

Change point detection in correlation structures has been mostly studied in conventional low-dimensional offline settings \citep{wied2017nonparametric, cabrieto2018capturing, madrid2023change_nips, killick2013wavelet, dette2019change}.
However, when the dimension is large compared to the (available) sample size, methods designed for the low-dimensional settings often perform poorly or are not well-defined. For example, kernel-based methods are sensitive to kernel function and parameter choices, particularly in moderate to large dimensions, and the method proposed by \citet{wied2017nonparametric} is unstable when the dimension is large relative to the sample size due to the (near) singularity of the variance estimator. For large-dimensional offline setting, \citet{choi2021self} proposed a break test based on the self-normalization method, which avoids variance estimation. \citet{li2023efficient} introduced a sign-flip parallel analysis-based detection method and a two-step approach to locate change points. 

In addition, since the correlation matrix is a particular case of the covariance matrix, some methods designed to detect changes in covariance may be used to detect changes in correlation. \citet{dette2022estimating} proposed a two-stage bootstrap approach to detect and locate change points in the covariance. There is a line of work on detecting changes in covariance matrices\citep{buzun2018change, xie2020sequential, li2023online}, as well as changes in the inverse covariance matrix\citep{ryan2023detecting} in online settings.
However, it is worth mentioning that relying solely on the covariance matrix for analysis may be susceptible to noise interference, making the correlation matrix more reliable for certain applications. For example, the method proposed by  \citet{dette2022estimating} performed poorly in detecting changes in correlation, since it mainly detects changes in univariate variances. This is also evidenced by the real data analysis in \citet{li2023efficient}, which accurately detected changes in the coronary artery disease index in rearranged RNA data using correlation, while the covariance-based methods failed to do so. A similar observation is demonstrated in our second application on seismic event detection.

This paper focuses on online change detection of correlation matrices. We do not restrict a particular parametric form for the data distribution, but only assume the availability of a reference dataset generated from the pre-change distribution in order to estimate the pre-change correlation matrix. The post-change correlation matrix is estimated from the data sequences and compared with the reference one to compute detection statistics. We propose a series of detection statistics based on the $\ell_1$ and $\ell_\infty$ norms of the squared difference between vectorized pre- and post-change correlation matrices, accommodating dense and sparse changes in the correlation structure, respectively. Meanwhile, a combined detection statistic that integrates both norms is also proposed to ensure good detection performance across varying sparsity levels. For additional studies in dense and sparse settings, see \citet{korostelev2008multi, enikeeva2019high,xie2013sequential}.

We mainly consider window-limited detection statistics as window-limited versions are widely used to improve computational efficiency. Such window-limited approaches have been explored for various statistics, including the generalized likelihood ratio \citep{willsky2003generalized,lai-ieeetit-1998} and the CUSUM statistic \citep{xie2023window}.
In addition, the Shewhart chart-type statistic \citep{shewhart1925application,chen2022high}, as a special version of the window-limited statistic, is also used in our real data analysis due to its simplicity and ease of implementation. Unlike the offline setting, there is no fixed sample size in the online setting because the streaming data is constantly coming. We consider the data dimension to be fixed, but it can be very large compared to the window size.

The main contributions of this paper are threefold. First, our methods are the first to detect correlation changes in streaming data without any dimensionality restrictions. 
Second, we characterize the theoretical performance of the proposed methods, specifically in terms of the average run length and the expected detection delay. Last, we apply the proposed method to critical applications of El Ni{\~n}o forecasting and seismic event detection. 
For example, using observational data from 1974 to 2024, our method successfully detected 13 out of 15 El Ni{\~n}o events, achieving a hit rate of more than 0.86 and a false alarm rate close to zero. In addition, our method is capable of early detection of microearthquakes and tremor-like signals in seismic datasets, which are triggered by subtle subsurface changes within the Earth. These results demonstrate the potential of our method to address a variety of practical problems.

The paper is organized as follows. Section \ref{sec:classic} presents our problem setup and preliminaries. Section \ref{sec:online detection} introduces the proposed online detection procedure and two improved methods incorporating the synthetic minority oversampling technique (SMOTE) and knockoff techniques. More importantly, a sign-flip permutation method is proposed to select thresholds used in detection procedures. Section \ref{sec:theoretical} presents the theoretical results of the proposed methods in terms of average run length and expected detection delay. Simulations are presented in Section \ref{sec:simulation}, and two real data applications including El Ni{\~n}o forecasting and seismic detection are discussed in Section \ref{sec:application}. Section \ref{sec:conclusion} concludes with a brief discussion. The Supplementary Material (SM) includes all proofs and some auxiliary  remarks and simulation results.

\paragraph{Notation.} 
We use $\mathbb{P}_\infty$ to denote the probability measure on the sequence of observations when the change never occurs, and $\bE_\infty$ is the corresponding expectation. Similarly, $\mathbb{P}_\nu$ denotes the probability measure when the actual change point is equal to $\nu$ and $\bE_\nu$ is the corresponding expectation; thus $\bE_1$ refers to the expectation under the post-change regime. The functions $f_0$ and $f_1$ represent the pre- and post-change probability density functions of the observations, respectively. For a vector $\boldsymbol{x}=(x_1,\ldots,x_p)$,  $\left\Vert \boldsymbol{x} \right\Vert_1:=\sum_{i=1}^p |x_i|$ denotes its $\ell_1$ norm, and $\left\Vert \boldsymbol{x} \right\Vert_\infty:=\max_{i=1,\ldots,p} |x_i|$ denotes its $\ell_\infty$ norms. For a matrix $\boldsymbol{A}=(a_{ij})_{1\leq i \leq m, 1\leq j\leq n}$, $\left\Vert \boldsymbol{A} \right\Vert_F:=\sqrt{\sum_{i=1}^m\sum_{j=1}^n a_{ij}^2}$ is its Frobenius norm. We denote $\boldsymbol{A}_{k:k'}$ as the $k$-th to $k'$-th columns of the matrix $\boldsymbol{A}$, and let $ncol(\boldsymbol{A})$ denote the number of columns of $\boldsymbol{A}$. We use $O(\cdot)$ and $o(\cdot)$ for the standard big-O and little-o notation. Finally, the symbol $``\circ"$ is the Hadamard product (element-wise).

\section{Problem Setup and Preliminaries} \label{sec:classic}

\subsection{Problem Setup}

Assume an i.i.d. data sequence $\{\boldsymbol{x}_t, t \in \mathbb N\}$, where each $\boldsymbol{x}_t=(
x_{t1}, \ldots, x_{tp})^\top$ has bounded fourth-moment entries and correlation matrix $\boldsymbol{R}$. In general, we do not impose restrictions on the dimension $p$, except in the knockoff enhancement discussed in Section~\ref{sec:SMOTE-knockoff}. We are interested in detecting the change point in the underlying correlation structure. We assume the following data-generating model:
\begin{equation}\label{eq:model}
	\begin{aligned}
		& \  \Corr(\boldsymbol{x}_t)=\boldsymbol{R}_0, \ t=1,2,\ldots,\nu-1 \\
		&\  \Corr(\boldsymbol{x}_t)=\boldsymbol{R}_1, \  t=\nu,\nu+1,\ldots.
	\end{aligned}    
\end{equation}
Here, $\nu$ is the change-point at which the correlation structure of $\boldsymbol{x}_t$ changes. The matrices $\boldsymbol{R}_0=[\rho_0(i,j)]_{i, j=1,\ldots,p}$ and $\boldsymbol{R}_1=[\rho_1(i,j)]_{i, j=1,\ldots,p}$ are the pre- and post-change correlation matrices, respectively. Both $\boldsymbol{R}_0$ and $\boldsymbol{R}_1$ are unknown, and the change point $\nu$ is deterministic and unknown. The difference between $\boldsymbol{R}_0$ and $\boldsymbol{R}_1$ indicates the magnitude and pattern of the change. Generally, the correlation change is considered sparse if the number of differing entries between $\boldsymbol{R}_0$ and $\boldsymbol{R}_1$ is relatively small compared to the total number of entries, otherwise the change is considered dense.

Our goal is to detect the unknown change-point $\nu$ in the data model \eqref{eq:model} as quickly as possible, subject to false alarm constraints. This is usually achieved by designing a stopping time \citep{poor-hadj-QCD-book-2008}, which is a random variable $T$ relating to the data sequence $\{\boldsymbol{x}_t, t \in \mathbb N\}$ such that for each $t$, the event $\{ T=t\}$ belongs to the sigma-algebra generated by $\boldsymbol{x}_1, \ldots, \boldsymbol{x}_t$, i.e., the event $\{ T =t\}$ depends only on the observations up to time $t$.

We begin by introducing the sample covariance and correlation matrices, which serve as fundamental building blocks in the proposed detection methods and the subsequent analysis. Let $\boldsymbol{ \hat{\Sigma}}_{s:t}$ be the sample covariance estimate for a subset of data from time $s$ to $t$, that is, 
\begin{equation}
	\label{eq:def_cov}
	\boldsymbol{ \hat{\Sigma}}_{s:t} = \frac{1}{t-s}\sum_{k=s}^t  (\boldsymbol{x}_k - \boldsymbol{\bar{x}}_{s:t}) ( \boldsymbol{x}_k - \boldsymbol{\bar{x}}_{s:t} )^\top,
\end{equation}
where $\boldsymbol{\bar{x}}_{s:t} = \frac{1}{t-s+1}\sum_{k=s}^t \boldsymbol{x}_k$. Then the corresponding sample correlation matrix is 
\begin{equation}\label{eq:def_corr}
	\boldsymbol{\hat{R}}_{s:t} = \frac{1}{t-s} \sum_{k=s}^t \boldsymbol{y}_k\boldsymbol{y}_k^\top,  
\end{equation}
where $\boldsymbol{y}_k  =\boldsymbol{  D_{s:t} }^{-1/2}(\boldsymbol{x}_k - \boldsymbol{\bar{x}}_{s:t})$ is the standardized vector of $\boldsymbol{x}_k$, for $k=s,\ldots,t$, with $\boldsymbol{D}_{s:t} = diag (\boldsymbol{\hat{\Sigma}}_{s:t})$.


We assume a set of reference (historical) data $\{\boldsymbol{x}_{-H},\ldots,\boldsymbol{x}_0\}$ of size $H+1$, which are known to be generated from the pre-change distribution with the unknown correlation matrix $\boldsymbol{R}_0$. According to equation \eqref{eq:def_corr}, the sample correlation matrix of all reference data can be calculated as:
\begin{equation}\label{eq:def_R0}
	\boldsymbol{\hat{R}}_0 = 
	\boldsymbol{\hat{R}}_{-H:0} = \frac{1}{H}\sum_{k=-H}^{0} \boldsymbol{y}_k \boldsymbol{y}_k^\top.
\end{equation}
The size of the historical data, $H+1$, is assumed to be sufficiently large. 

Given the online sequence $\{\boldsymbol{x}_t, t\in \mathbb N\}$, at each time $t$, we investigate all values of $t'<t$ as potential change-points. For each $t'$, we calculate the sample correlation matrix $\boldsymbol{ \hat{R}}_{t':t}= [\hat{\rho}_{t':t}(i,j)]_{i,j=1,\ldots,p}$ according to \eqref{eq:def_corr} using {potential post-change samples} $\{\boldsymbol{x}_{t'},\ldots,\boldsymbol{x}_t\}$.
Then, we calculate the squared difference between $\boldsymbol{ \hat{R}}_{t':t}$ and $ \boldsymbol{\hat{R}}_0$, i.e.,
\begin{equation}\label{eq:v}
	\boldsymbol{v}_{t',t} =  vecho \left\{ \left[\left( \hat{\rho}_0(i,j) - \hat{\rho}_{t':t}(i,j) \right)^2 \right]_{i,j=1,\ldots,p} \right\}\in \mathbb{R}^{\frac{p(p-1)}{2}},
\end{equation}
where ${vecho}(\cdot)$ indicates the half-vectorized vector by vectorizing only the lower triangular part (excluding the diagonal) of the symmetric matrix. If a change point exists at $t'$, then $\boldsymbol{v}_{t',t}$ is obviously an estimator of $ vecho \big\{ [\left( \rho_0(i,j) - \rho_1(i,j) \right)^2 ]_{i,j=1,\ldots,p} \big\}$, which measures the entry-wise differences
between the population correlation matrices before and after the change.

More specifically, for any $(i,j)$-th entry of the correlation matrix, we present the expectation of $\boldsymbol{v}_{1,t}(i,j)$ when the data sequence $\{\boldsymbol{x}_t, t=1, 2, \ldots\}$ is entirely from the post-change regime, to illustrate its effectiveness for detection. For notational convenience, we write $\rho_0 = \rho_0(i, j)$ and $\rho_1 = \rho_1(i, j)$ in the following Lemma.
\begin{lemma}\label{lem:mean_V_zero_mean}
	Assume $\bE [x_{ki}] = 0$ and $\bE [x_{ki}]^2= 1$ for $i=1,\ldots,p$, $\forall k$. For $\boldsymbol{v}_{1,t}(i,j) = (\hat{\rho}_0(i,j) - \hat{\rho}_{1:t}(i,j))^2$, we have 
	\begin{equation}
		\label{eq:lem1}
		\begin{multlined}
			\bE_1[\boldsymbol{v}_{1,t}(i,j)]  = (  \rho_0 - \rho_1 )^2  + \frac{1}{H+1}(\beta_{20} -\rho_0^2 )+ \frac{1}{t}( \beta_{21} -\rho_1^2 ),   
		\end{multlined}
	\end{equation} 
	where $\beta_{20} :=\mathbb{E}_\infty[(x_{ki}x_{kj})^2]$ is the expectation under the pre-change regime, and $\beta_{21} :=\mathbb{E}_1[(x_{ki}x_{kj})^2]$ is the expectation under the post-change regime.
\end{lemma}
We note that in \eqref{eq:lem1}, the first term $(\rho_0 - \rho_1 )^2$ dominates the expectation when $H$ and $t$ are sufficiently large. Therefore, larger values of $\boldsymbol{v}_{t',t}(i,j)$
suggest a significant difference between $\boldsymbol{ \hat{R}}_{t':t}$ and $ \boldsymbol{\hat{R}}_0$, thus indicating that a change may have occurred. In contrast, under the pre-change regime, all elements in $\boldsymbol{v}_{t',t}$ are considerably small.

\begin{remark}[The case with unknown mean]\label{lem:mean_V}
	In the more general case with unknown mean and unit variance $\Var(x_{ki}) =1$ for $i=1,\ldots,p$, $\forall k$, the expectation of $\boldsymbol{v}_{1,t}(i,j) = (\hat{\rho}_0(i,j) - \hat{\rho}_{1:t}(i,j))^2$ is 
	\begin{equation*}  
		\begin{multlined}      \bE_1[\boldsymbol{v}_{1,t}(i,j)]  = \left(\rho_0 - \rho_1 \right)^2 + \frac{1}{H}\beta_{20}   + \frac{1}{t-1}\beta_{21} + \frac{5-t}{t(t-1)} \rho_1^2 + \frac{3H+4 - H^2}{H(H+1)^2} \rho_0^2, 
		\end{multlined}
	\end{equation*}
	thus the first term $(\rho_0 - \rho_1 )^2$ remains dominant. When the variance $\Var(x_{ki})$ is also unknown, the sample variances serve as consistent estimators for $\Var(x_{ki})$ and the expectation of $\boldsymbol{v}_{t',t}(i,j)$ can be approximated similarly as above. 
\end{remark}

\begin{remark}
	Although the data are assumed to be i.i.d. to facilitate theoretical analysis, our proposed methods are not limited to such settings. As demonstrated in Section~\ref{sec:simulation} and Section~\ref{sec:application}, our method demonstrates robust performance in more complex scenarios, including non-Gaussian distributions and serial dependence.
\end{remark}



\subsection{Performance Measures}\label{sec:optimality}

A fundamental objective in sequential change point detection is to optimize the trade-off between false alarm rate and average detection delay. Controlling the false alarm rate is commonly achieved by setting an appropriate threshold in the detection procedure. However, the threshold also affects the average detection delay. A larger threshold incurs fewer false alarms but leads to a larger detection delay and vice versa. 

We introduce the two commonly used performance metrics in sequential detection, the average run length (ARL) and the worst-case average detection delay (WADD). ARL is used to characterize false alarms, and it is defined, for a given stopping time $T$, as:
\begin{equation}\label{eq:ARLDef}
	\ARL(T)=\mathbb{E}_\infty[T].
\end{equation}
ARL can be interpreted as the expected time duration between two consecutive false alarms. Its reciprocal, $1/{\ARL(T)}$, corresponds to the false alarm rate---the frequency at which false alarms occur in the pre-change regime. 
We typically focus on the test procedures that satisfy a pre-specified lower bound $\gamma$ on the ARL, i.e., consider the set of tests:
\begin{equation*}\label{eq:DalphaDef}
	\mathcal D_{\gamma}=\{T: \ARL(T) \geq \gamma\}.
\end{equation*}

Finding a uniformly powerful test that minimizes delay over all possible values of the change-point $\nu$, subject to an ARL constraint, is generally intractable. Usually, it is more tractable to pose the problem in the so-called minimax setting. 
We adopt the WADD metric, defined as the supremum of the average detection delay conditioned on the worst possible realizations \citep{Lorden1971}. More specifically,
\begin{equation}\label{eq:WADDdef}
	\WADD(T) = \underset{\nu \geq 1}{\operatorname{\sup}}\ \esssup \ \bE_{\nu}\left[(T-\nu+1)^+| \boldsymbol{x}_1, \dots, \boldsymbol{x}_{\nu-1}\right], 
\end{equation}
where $\esssup$ is essential supremum, and $(x)^+ = \max\{x, 0\}$.
Under such a definition of WADD, the formulation of interest is:
\begin{equation}\label{Lorden}
	\mbox{minimize } \WADD(T), \ \mbox{ subject to } \ARL(T) \geq \gamma. 
\end{equation}

\begin{remark}[Information-theoretic lower bound]
	The information-theoretic lower bound to $\WADD(T)$, i.e., the minimum value of \eqref{Lorden}, is known to be obtained by the CUSUM procedure  \citep{page-biometrica-1954}, which is a commonly used sequential change detection procedure that enjoys efficient implementation and exact optimality properties \citep{Lorden1971,mous-astat-1986,ritov-astat-1990}. By accumulating the log-likelihood ratios, the CUSUM statistic is constructed as $W_t
	= \max_{1 \leq i \leq t }\sum_{k=i}^t \ell (\boldsymbol{x}_k)$,  
	where $\ell(\boldsymbol{x}_k) = \log ({f_1(\boldsymbol{x}_k)}/{f_{0}(\boldsymbol{x}_k)})$ is the log-likelihood ratio statistic, and the stopping time is thus $T_{\text{c}} = \inf\left\{t\geq 1: W_t    \geq b \right\}$, 
	for a pre-specified threshold $b$ such that $\ARL(T_{\text{c}})= \gamma$. The WADD of the CUSUM test is known to be ${(\log \gamma)}/{D (f_1\| f_0)}\cdot (1+o(1))$, where $D (f_1\| f_0) = \int  f_1 (x) \log ({f_1 (x)}/{f_0(x)}) dx$ is the KL divergence between the post- and pre-change distributions \citep{tartakovsky2014sequential}. 
	In Section \ref{sec:simulation}, we compare the performance of the proposed detection methods with that of the CUSUM procedure to validate their effectiveness. It is important to note that the CUSUM test requires {\it full} knowledge of the pre- and post-change density functions, which is not available in our setting and is rarely available in real applications.
\end{remark}

\begin{remark}[Effect of dimension $p$ on the detection procedure]
	Generally, as $p$ increases, the detection delay may increase due to the diminished signal-to-noise ratio and the increased sensitivity to noise. However, since the sample covariance matrix is an entry-wise consistent estimator of the population covariance, regardless of the dimension, the proposed statistics based on the difference in the segment covariance matrices are also $``$entry-wise" consistent and therefore do not suffer from the curse of dimensionality. In Section~\ref{sec:simulation}, our methods have a shorter detection delay as $p$ increases, indicating that the proposed methods effectively take advantage of the additional information embedded in larger dimensional correlation structures.
\end{remark}

\section{Online Detection Procedure} \label{sec:online detection}

In this section, we propose sum-type and max-type detection statistics, together with their window-limited variants to improve computational efficiency, in Sections \ref{sec:sum} and \ref{sec:max}, respectively. We discuss SMOTE and knockoff enhancements in Section \ref{sec:SMOTE-knockoff}, providing solutions for situations with limited sample sizes. Finally, in Section \ref{sec:threshold}, we present a practical method for determining thresholds that ensures computational efficiency while approximately maintaining the desired ARL condition. 

\subsection{Sum-Type Detection Statistics} \label{sec:sum}

To detect dense changes in the correlation structure, we form a sum-type statistic by aggregating the element‐wise differences between estimated correlation matrices using the \(\ell_1\) norm. Recall from \eqref{eq:v} that $\boldsymbol{v}_{t',t}$ represents the pairwise correlation differences between the pre-change estimate (computed from historical data) and the current estimate (based on data within the time window $[t',t]$). Therefore, the $\ell_1$ norm of $\boldsymbol{v}_{t',t}$ captures the total magnitude of deviation from the historical correlation structure. 

Since the true change-point location $\nu$ is unknown, at each time $t$, we evaluate all potential candidate change-points $t'<t$ and define the detection statistic at time $t$ as the maximum of the scaled $\ell_1$ norms over all such $t'$. The resulting detection statistic at $t$ is thus defined as
\begin{equation}
	S_{t}^{(\text{sum})} = \max_{1 \le t' \le t-1} \frac{(t-t')H }{H+t-t'}  \left\Vert \boldsymbol{v}_{t',t}\right\Vert_1,
\end{equation}
where the scaling $\frac{(t-t')H }{H+t-t'}$ is incorporated to balance the variance of $\boldsymbol{v}_{t',t}$, as the standard deviation of $\boldsymbol{v}_{t',t}$ under the pre-change regime is $O(\frac{1}{t-t'} + \frac{1}{H})$. When $H$ is relatively large, we may simply use $S_t^{(\text{sum})} = \max_{1\le t'\le t-1} (t-t')\Vert \boldsymbol{v}_{t',t}\Vert_1$. From the definition of $S_{t}^{(\text{sum})}$, it can be seen that when there is no change, the statistic is expected to remain relatively small and stable. After a change-point, however, the $\ell_1$ norm is expected to increase due to a shift in correlations, causing the detection statistic to increase steadily---thereby enabling effective and timely change detection.


In practice, searching over all possible time points $1\le t'<t$ becomes increasingly expensive as $t$ grows. To reduce computational cost, we use a window-limited variant of the sum-type statistic, referred to as WL-Sum. This variant restricts the search to a fixed-size sliding window of the most recent $w$ time points, and the resulting detection statistic is modified as follows:
\begin{equation}\label{wlsum}
	S_t^{(\text{WL-sum})}=\max_{t-w\le t'\le t-1} \frac{(t-t') H}{H+t-t'} \left\Vert \boldsymbol{v}_{t',t} \right\Vert_1,
\end{equation}
where $w$ is a pre-specified window size. Then the change detection is performed by the stopping time defined as
\begin{equation}\label{eq:wlsum-test}
	T^{(\text{WL-sum})}:=\inf\{t: S_t^{(\text{WL-sum})} \geq b_1\},    
\end{equation}
where $b_1$ is the pre-specified threshold selected to meet the false alarm rate constraint, as detailed later.

\subsection{Max-Type Detection Statistics} \label{sec:max}

The sum-type detection statistics above may not be efficient in detecting sparse changes, where only a relatively small number of elements change in the correlation structure. Under the sparse setting, the conventional test statistic is the maximum difference between the pre- and post-change correlation matrices. Following the same principle used in the sum-type statistic, we define the max-type test statistic as
\begin{equation}
	S_t^{(\max)} = \underset{1 \le t' \le t-1} {\max}\frac{(t-t')H}{H+t-t'} {\left\Vert \boldsymbol{v}_{t',t} \right\Vert_\infty},
\end{equation}
where the $\ell_\infty$ norm replaces the $\ell_1$ norm used in the sum-type statistic, and it captures the largest entry-wise deviation in correlation estimates.

Similarly, a window-limited variant of the max-type statistic (WL-Max) is employed to mitigate the computational burden for a given window size $w$,
\begin{equation}\label{wlmax}
	S_t^{(\text{WL-max})} = \underset{t-w \le t' \le t-1} {\max}\frac{(t-t')H}{H+t-t'} \left\Vert \boldsymbol{v}_{t',t} \right\Vert_\infty. 
\end{equation}
And the corresponding stopping time is denoted as,
\begin{equation}\label{eq:wlmax-test}
	T^{(\text{WL-max})}:=\inf\{t: S_t^{(\text{WL-max})} \geq b_2\},    
\end{equation}
where $b_2$ is again the pre-specified threshold selected to meet the false alarm rate constraint.

\begin{remark} [The choice of window size] In many applications, domain-specific knowledge may be used to determine the window size. For example, in climate science, window sizes can be determined by the properties of climate systems. However, it is also important to analyze the smallest window size that ensures reliable detection in general settings, since the choice of window size involves a tradeoff between computational efficiency and detection delay. 
	Generally, smaller changes require more data to detect, so the minimal window size tends to increase as the magnitude of the change decreases. The key idea is to select a window size that is large enough to ensure reliable detection, yet as small as possible for efficiency. In Proposition 
	1, we provide a lower bound for selecting the window size. 
\end{remark}

\begin{remark}
	We can further enhance the computational efficiency of window-limited tests in \eqref{eq:wlsum-test} and \eqref{eq:wlmax-test} by using {\it lagged} sliding windows. Instead of evaluating the detection statistic at every time step, we only update it at regularly spaced intervals. This leads to the stopping rule: $T':=\inf\{t: t = w+s\times k, k \in \mathbb{N}, \, S_t \geq b \}$,
	where $S_t$ denotes either the sum- or max-type window-limited detection statistic, and $1\leq s \leq w$ is the lag between consecutive window evaluations. When $s=1$, this reduces to tests in \eqref{eq:wlsum-test} and \eqref{eq:wlmax-test}. Setting $s=w$ results in non-overlapping (i.e., independent) windows.
\end{remark}

\subsection{SMOTE and Knockoff Enhancements}\label{sec:SMOTE-knockoff}

For window-limited-variant statistics, at each time $t$, we search time points $t'$ in $[t-w, t-1]$, and the corresponding $\boldsymbol{ \hat{R}}_{t':t}$ and statistics can be unrobust due to small number of samples, especially when $w$ is small compared to the data dimension $p$. We propose two modified approaches that incorporate the SMOTE and Knockoff techniques, respectively (see Figure 8 in the SM for illustration). 

First, the SMOTE technique is an oversampling method originally designed to deal with imbalanced data in classification problems \citep{chawla2002SMOTE}. Specifically, for each sample $\boldsymbol{x}$ in the minority class, the algorithm randomly identifies one of its closest neighbors $\boldsymbol{x}^*$ based on Euclidean distance and produces a new SMOTE sample as $\boldsymbol{x}_{\text{SMOTE}}=\boldsymbol{x} +u \cdot (\boldsymbol{x}^* – \boldsymbol{x} )$, where $u$ is randomly chosen from a uniform distribution on $[0,1]$. \citet{blagus2013SMOTE} investigated SMOTE's theoretical properties and performance on large-dimensional data. We adapt SMOTE to mitigate the sample size imbalance between recent and historical data. SMOTE samples are used to increase recent samples; see the Algorithm \ref{alg:SMOTE} for details.

\begin{algorithm}[ht!]
	\SetKwData{Left}{left}\SetKwData{This}{this}\SetKwData{Up}{up}
	\SetKwFunction{Union}{Union}\SetKwFunction{FindCompress}{FindCompress}
	\SetKwInOut{Input}{Input}
	\SetKwInOut{Output}{Output}
	\Input{Data dimension $p$, online data $\{ \boldsymbol{x}_t,t=1,2,\ldots\}$, reference data $\{\boldsymbol{x}_{-H},\ldots,\boldsymbol{x}_0\}$, the pre-specified ARL $\gamma$, window size $w$, threshold $b$.}
	\Output{Stopping time $T$.}
	Use the reference data to calculate $\boldsymbol{\widehat{R}}_0$, initialize $S'_1 = 0$; \\
	\While{$S'_t < b$}{
		\eIf{$ t< w+1 $}{
			$\boldsymbol{X} = [\boldsymbol{x}_1, \ldots,\boldsymbol{x}_t] \in \bR^{p \times t}$\;
		}{$\boldsymbol{X} = [\boldsymbol{x}_{t-w}, \ldots,\boldsymbol{x}_t]  \in \bR^{p \times (w+1)}$\;
		}
		$m_0 = ncol(\boldsymbol{X}), \boldsymbol{\Tilde{X}} = [\boldsymbol{X},\boldsymbol{X}_{\text{SMOTE}}], \Tilde{m}_0 = ncol(\boldsymbol{\Tilde{X})}=2m_0$;\\
		\For{$m\leftarrow 1$ \KwTo $m_0 - 1$}{
				$\boldsymbol{X}_s = \boldsymbol{\Tilde{X}}_{m:\Tilde{m}_0};$\\
			Use $\boldsymbol{X}_s$ to calculate the sample correlation matrix $\boldsymbol{\hat{R}}_{m:\Tilde{m}_0}$ and obtain the corresponding test statistic $S_t^{(m)} = g(\boldsymbol{v}_{m,\Tilde{m}_0})$\;
		}
		$S'_t = \max_{m=1,\ldots,m_0-1}\{S_t^{(m)}\}$;
	}
	$\text{Stopping time} \ T = t$.\\   
	\caption{SMOTE-enhanced Detection Procedure}
	\label{alg:SMOTE}
\end{algorithm}

In Algorithom~\ref{alg:SMOTE}, $\boldsymbol{X}_{\text{SMOTE}}$ denotes $m_0$ number of SMOTE variables generated from $\boldsymbol{X}$.
$g(\cdot)$ is a function depending on which type of test statistics is used.

Second, the knockoff method proposed by \citet{barber2015controlling} can generate knockoff variables that imitate the original variables' correlation structure. The detection procedures can be enhanced by incorporating the knockoff method when $p>2w+2$ (due to the constrain of knockoff).
Specifically, for a matrix $\boldsymbol{X} \in \mathbb{R}^{p\times (w+1)}$, let $\boldsymbol{X_{\text{knockoff}}} = \boldsymbol{X}(\boldsymbol{I} - \boldsymbol{\Sigma}^{-1} diag\{ \boldsymbol{z}\} ) + \boldsymbol{U C}$, where $\boldsymbol{\Sigma} = \boldsymbol{X}^\top \boldsymbol{X}$ and satisfies $\boldsymbol{\Sigma}_{j,j} = 1$ for all $j$, $\boldsymbol{z}$ is a non-negative vector of dimensions $p$, $\boldsymbol{U}$ is an orthonormal matrix $p\times (w+1)$ which is orthogonal to the span of the feature of $\boldsymbol{X}$, and $\boldsymbol{C}^\top \boldsymbol{C} = 2 diag\{\boldsymbol{z}\} - diag\{\boldsymbol{z}\} \Sigma^{-1} diag\{\boldsymbol{z}\} $ is a Cholesky decomposition, see \citet{barber2015controlling} for more details to choose $\boldsymbol{z}$. Then $\boldsymbol{X_{\text{knockoff}}}$ has the same correlation structure as $\boldsymbol{X}$. 

The knockoff-enhanced algorithm follows the same procedure as Algorithm~\ref{alg:SMOTE}, except that in line 8, $\boldsymbol{\Tilde{X} }= [\boldsymbol{X},\boldsymbol{X_{\text{knockoff}}}]$, $\boldsymbol{X_{\text{knockoff}}}$ denotes $m_0$ knockoff variables generated from
$\boldsymbol{X}$.
Using the knockoff method, additional independent random variables are added that maintain the same correlation structure as the original variables. This approach improves the estimation of the correlation matrix, particularly when the sample size is small. Regarding detection delay, for a fixed ARL, this technique would allow one to use a smaller window size to detect change points effectively, as shown in Section~\ref{sec:simulation}. 

The effects of these two enhancement methods on correlation estimation and detection effectiveness are illustrated by the simulation results in B.2 and B.3 of the SM. Compared to the sample correlation matrix
calculated from the original samples, knockoff method can improve the accuracy of correlation
estimation; while the SMOTE method does not improve accuracy, it has a smaller bias. Importantly, both the SMOTE and knockoff-enhanced methods
have shorter detection delays than the original WL-Sum procedure.

\subsection{Signflip-Based Threshold Selection}\label{sec:threshold}

Selecting a proper threshold $b$ for stopping time is crucial in the sequential change point detection procedure. In general, there are two types of methods for determining the appropriate threshold. One way is to infer an exact threshold based on the distribution of test statistics under the pre-change regime. The other way is to determine the threshold based on the empirical distribution of numerous simulated test statistics, which is more common in practice as it requires little knowledge about the distribution of test statistics. 

In real applications where the number of pre-change samples is limited or generating samples is time-consuming, the available data size is often small. We propose a new heuristic method for selecting the threshold using a sign-flip permutation method, which works well even for small sample sizes; see the Algorithm \ref{alg:threshold}. The SMOTE and Knockoff techniques should also be applied in the threshold selection procedure if they are used during monitoring.

\begin{algorithm}[ht!]
	\caption{Signflip-based Threshold Selection}\label{alg:threshold}
	\SetKwData{Left}{left}\SetKwData{This}{this}\SetKwData{Up}{up}
	\SetKwFunction{Union}{Union}\SetKwFunction{FindCompress}{FindCompress}
	\SetKwInOut{Input}{Input}\SetKwInOut{Output}{Output}
	\Input{Data dimension $p$, number of signflip trials $q$, pre-change correlation estimator $\boldsymbol{\hat{R}}_0$, a sequence of data $\{ \boldsymbol{x}_1,\ldots,\boldsymbol{x}_M\}$, the pre-specified ARL $\gamma$, window size $w$.   }
	\Output{Threshold $b$.}
	\For{$ l\leftarrow 1$ \KwTo $q$}{
		Randomly signflip entries: form $\boldsymbol{x}_k^{*} = \boldsymbol{ \mathcal{R}}\circ \boldsymbol{x}_k, \forall k=1,\ldots,M$ where\\
		$ \boldsymbol{\mathcal{R}}(i)_{i=1,\ldots,p}\overset{i.i.d.}{\sim} \left\{\begin{array}{cc} +1, & \text{with probability 1/2},\\
			-1, & \text{with probability 1/2}, \end{array} \right.$\\
		that is, $\boldsymbol{\mathcal{R}}\in \mathbb{R}^{p\times 1}$ has independent identically distributed Rademacher entries;\\
		\For{$t\leftarrow 2$ \KwTo $M$}{
			\eIf{$t < w+1$}{
				$\boldsymbol{X} = [\boldsymbol{x}^*_1,\ldots,\boldsymbol{x}^*_t] \in \mathbb{R}^{p \times t}$;
			}{
				$\boldsymbol{X} =[\boldsymbol{x}^*_{t-w},\ldots,\boldsymbol{x}^*_t]\in \mathbb{R}^{p \times (w+1)}$;
			}
			$m_0 = ncol(\boldsymbol{X})$; \\
			\For{$m \leftarrow 1$ \KwTo $m_0-1$}{
				Use $\boldsymbol{X}_{m:m_0}$ to calculate the sample correlation matrix $\boldsymbol{\hat{R}}_{m:m_0}$ and then obtain the statistics $S_{t,m}=g(\boldsymbol{v}_{m,m_0})$ based on that;}
			Calculate test statistic $S_t^{(l)} = \max_{1\le m \le m_0-1}\{S_{t,m}\}$; }
	}
	Set the threshold $b$ using $q$ sequences of detection statistics $\{S_t^{(l)},t=1,\ldots,M\}_{l=1}^q$ such that their ARL equals to $\gamma$. 
\end{algorithm}

The following lemma illustrates why the threshold selected via Algorithm \ref{alg:threshold} is valid. 
\begin{lemma} 
	Given a random vector $\boldsymbol{x}_t=(x_{t1},\ldots,x_{tp})^\top \in \mathbb{R}^{p\times 1}$ and two independent Rademacher vectors $\boldsymbol{\mathcal{R}}_1:=(r_1^{(1)},\ldots,r_p^{(1)})^\top$ and $\boldsymbol{\mathcal{R}}_2:=(r_1^{(2)},\ldots,r_p^{(2)})^\top$ with i.i.d. Rademacher entries, we have $\boldsymbol{x}_t \circ \boldsymbol{\mathcal{R}}_1$ and $\boldsymbol{x}_t \circ \boldsymbol{\mathcal{R}}_2$ are uncorrelated as $Cov(x_{ti}r_i^{(1)} , x_{tj}r_j^{(2)} ) =0$ for $1\le i,j\le p.$
	\label{lem:signflip_indep}
\end{lemma}
The lemma~\ref{lem:signflip_indep} demonstrates that for a given vector $\boldsymbol{x}_t$, 
the data shuffled through various sign-flip steps using different $\boldsymbol{\mathcal{R}}$ exhibit weak dependence, and are independent if the data follow Gaussian distributions. This property ensures that the test statistics computed from $q$ sign-flip trials are weakly dependent or independent, allowing for appropriate threshold selection. The sign-flip permutation in Algorithm \ref{alg:threshold} does not change the distribution of detection statistic $S_t$ under the pre-change; see B.4 of the SM.

\begin{remark}[Support Recovery]
	Once a change is detected, it is often of great interest to identify which pairwise correlations have changed, which is known as \textit{support recovery}. Accurate support recovery can provide valuable insights into the underlying structural shifts in a system and lay the foundation for post-change analysis or domain-specific interpretations, such as clustering of affected variables, network visualization, or physical interpretation in natural science applications. 
	
	Based on the proposed WL-Max detection statistic, support recovery is relatively straightforward: the location of the statistical value directly indicates the most significantly changed correlation pairs. Based on the WL-Sum statistic, which aggregates information across multiple correlation pairs, support can be inferred by ranking the element-wise absolute differences of the estimated correlation matrices before and after the change and identifying a subset of correlation pairs whose differences exceed a certain threshold (e.g., the average), thus highlighting the factors that contribute the most to the detected change.
\end{remark}

\section{Theoretical Analysis}  \label{sec:theoretical}

In this section, we provide the theoretical guarantees for the proposed tests with respect to two key performance measures: the average run length and the expected detection delay. The detailed proofs are deferred to SM.

For simplicity, we let the size of the reference data $H$ equal the chosen window size $w$. Note that we can regard the weight $\frac{(t-t')H}{H+t-t'}$ before the test statistic $S_t$ as a constant for a given $t'$, and we may unify both sum-type statistics for non-sparse settings and max-type statistics for sparse settings since they are, in essence, the combination of element-wise entries. 
We first characterize the temporal correlation of the statistics as follows, which will be utilized later for the ARL approximation.

\begin{lemma}[Temporal correlation of sequential detection statistics] \label{lem:corr}
	Suppose all samples are i.i.d. from pre-change distribution with $\bE [x_{ti}] = 0$ and $\Var[x_{ti}] =1$ for $i=1,\ldots,p$, $t=1,2,\ldots$, then the correlation between the detection statistics $S_{t+s}$ and $S_t$ is
	\[\Corr(S_{t+s},S_t) 
	= 1 - \frac{s}{w} + o\left(\frac{s}{w}\right).\]
	Here $s = o(w)$ is a small time shift from $t$ and $S_t$ can be both sum-type and max-type statistics defined in Section~\ref{sec:online detection}.  
\end{lemma}

Based on this temporal correlation, we have the following Theorem characterizing the approximate ARL of the proposed detection procedures. The main idea is to use a linear approximation for the correlation between detection statistics $S_t$ and $S_{t+s}$. Then, the behavior of the detection procedure can be related to a random field. Using the localization theorem \citep{siegmund2010tail}, we can obtain an asymptotic approximation for ARL of the window-limited procedures when the threshold $b$ is large enough.

\begin{theorem}[ARL Approximation] Assume under the pre-change measure we have $\bE_\infty [x_{ti}^4] < \infty$ for $1\le i\le p$, $t=1,2,\ldots$, and under Assumption 1-4
	detailed in the section C.5 of the SM,
	as threshold $b\rightarrow \infty$, the ARL of the stopping time $T$ can be approximated as
	\begin{equation}  \label{eq:ARL approximation}
		\mathbb{E}_{\infty}[T] 
		= \frac{1}{2} [2\pi /\kappa ]^{1/2} e^{\kappa/2} \bigg/ \int_{\xi_1 }^{\xi_2} y \zeta^2(y) d y (1+o(1)),
	\end{equation}
	where $\xi_1 =\frac{2b}{\sqrt{w \sigma_d^2}}$, $\xi_2 =\frac{2b}{\sigma_d}$, $\kappa = \frac{(b-\mu)^2}{\sigma_d^2}$, $\mu = \bE_\infty [ S_t  ]$, and $\sigma_d^2 = \Var_\infty[S_t]$. $T$ serves as the stopping time for either the sum-type or the max-type statistic.
	$\zeta(\cdot)$ is a special function closely related to the Laplace transform of the overshoot over the boundary of a random walk \citep{siegmund2007statistics}: 
	\[
	\zeta(y) \approx \frac{\frac{2}{y}[\Phi(\frac{y}{2}) -0.5 ] }{ \frac{y}{2}\Phi(\frac{y}{2}) + \phi(\frac{y}{2}) },
	\]
	where $\phi(y)$ and $\Phi(y)$ are the probability density function and the cumulative density function of the standard Gaussian distribution.
	\label{theo:ARL}
\end{theorem}

The bounded fourth-moment assumption is needed to ensure that the estimated correlation coefficients are reliable estimators, i.e., $\hat{\rho}(i,j)$ converge to $\rho(i,j)$ in probability. Additionally, the Assumption 1-4
are required to ensure the applicability of the localization theorem. In practice, all assumptions can be easily satisfied by data from some commonly used distributions, such as Gaussian distributions and exponential families.

The main contribution of Theorem~\ref{theo:ARL} is to provide a theoretical approximation for setting the detection threshold, which eliminates the need for time-consuming Monte Carlo simulation, especially when the desired ARL is large. From Theorem~\ref{theo:ARL}, we can numerically compute the threshold value $b$ by setting the right-hand side of Equation~(\ref{eq:ARL approximation}) to the desired ARL value. 
Table~\ref{tab:ARL} shows the high precision of this approximation result by comparing the threshold obtained from Equation~(\ref{eq:ARL approximation}) with that obtained from a simulation study. For a sequence of i.i.d. standard normal samples, we conduct 1000 sign-flip trials to find the threshold for different ARL values. 
The results of the WL-Sum and WL-Max procedures are presented in Table~\ref{tab:ARL}, indicating that the approximation is reasonably accurate as the relative errors are no greater than approximately 1\% for both procedures. 
We also note that the magnitude of thresholds varies for different test statistics. For example, the WL-Sum statistic involves the summation of $p(p-1)/2$ terms and results in consistently higher thresholds than max-type statistics.

\begin{table}
	\centering
	\caption{Comparison of the threshold $b$ obtained from simulations and the theoretical approximation (\ref{eq:ARL approximation}). $p$=50, $w$=20, $H$=100, sign-flip trials $q$=1000. }
	\label{tab:ARL}
	\begin{tabular}{cccccccc}
		\hline
		& ARL & 5,000 & 10,000 & 20,000 & 30,000 & 40,000 & 50,000 \\ \hline
		\multirow{2}{*}{WL-Sum ($10^3$) } & Simulated & 1.3271 & 1.3379 & 1.3478 & 1.3530 & 1.3589 & 1.3598 \\
		& Theoretical & 1.3388 & 1.3499 & 1.3603 & 1.3659 & 1.3698 & 1.3729 \\ \hline
		\multirow{2}{*}{WL-Max} & Simulated & 17.3070 & 17.9350 & 18.5248 & 18.8585 & 19.0879 & 19.1525 \\
		& Theoretical & 16.7333 & 17.0234 & 17.2975 & 17.4572 & 17.5667 & 17.6469 \\ \hline
	\end{tabular}
\end{table}

When there is a change point $\nu$, the performance of the detection procedure is measured by the detection delay as defined in \eqref{eq:WADDdef}, which can be interpreted as the expected number of post-change samples needed to detect the change. We note that the supremum over all possible change points $\nu$ in the definition \eqref{eq:WADDdef} is typically intractable for a general detection procedure. Therefore, we use a simplified definition of the expected detection delay (EDD) as $\bE_{1}[T]$, i.e., the expected stopping time when the change point equals $1$, this is also what is typically simulated in practice.
We present the following proposition for the detection delay of the proposed stopping times.

\begin{proposition}
	[EDD Approximation]\label{prop:edd} Assume under the post-change measure we have $\bE_1 [x_{ti}^4] < \infty$ for $1\le i\le p$, $t=1,2,\ldots$, then the EDD of WL-Sum and WL-Max test statistics can be approximated as follows. As threshold $b\to\infty$ and for window size $w\ge \frac{b}{\min(\Delta_1, \Delta_2)}$, 
	$$
	\mathbb{E}_1[T^{(\text{sum})}] = \frac{b(1+o(1))}{\Delta_1}, \ \text{and} \quad \mathbb{E}_1[ T^{(\text{max})}] = \frac{b(1+o(1))}{\Delta_2},$$
	where $\Delta_1 := \sum_{1\leq i< j \leq p} (\rho_0(i,j) - \rho_1(i,j))^2$ and $\Delta_2 := \max_{1\leq i< j \leq p} (\rho_0(i,j) - \rho_1(i,j))^2$ are two notions of signal strength, and large $\Delta_2/\Delta_1$ indicates sparse changes in correlation.  
\end{proposition}

A small simulation study is presented in Section B.1 of the SM. 
to show the impact of window sizes. For a fixed small ARL value, a wide range of window sizes generally achieves low detection delays, allowing us to select smaller window sizes to ensure computational efficiency. However, as ARL increases, the detection delay for smaller window sizes can grow quadratically, necessitating the use of larger window sizes to maintain detection performance. In addition, larger window sizes are needed to perform more challenging detection tasks effectively.

\section{Simulations}  \label{sec:simulation}

In this section, we conduct extensive simulations to illustrate the performance of the proposed methods. Simulations are carried out on varying dimensions $p=50, 60, 100, 300$ and window sizes $w=20,40,50,60$. We set historical data size $H=100$, number of signflip trials $q=1000$ and $M=1000$ in Algorithm \ref{alg:threshold} for threshold selection, and all detection delays are averaged at 1000 replications. We consider two types of data-generating distributions: (i) multivariate normal $N(\boldsymbol{0},\boldsymbol{R})$, and (ii) Student's $t$, $t_5$ with degree of freedom 5, with correlation matrix $\boldsymbol{R}$. For each distribution, only the correlation matrix changes after the change-point. We examine three scenarios: Cases 1 and 2 represent non-sparse settings, Case 3 represents a sparse setting, and Case 4 is a general setting. In the following, $r$ denotes the non-zero off-diagonal element of the post-change correlation matrix, and $\lfloor \cdot \rfloor$ is the floor function. 

\begin{itemize}
	
	\item  Case 1: $\boldsymbol{R}_0 = \mathbf{I}_p$; $\rho_1(i,j) = r$ for $1\le i\neq j \le p$.
	
	\item  Case 2: $\boldsymbol{R}_0 = \mathbf{I}_p$; $\rho_1(i,j) = r$ for $1\le i\neq j \le \lfloor p/2 \rfloor$ and 0 otherwise.
	
	\item Case 3: $\rho_0(i,j) = -0.3$, $\rho_1(i,j) = 0.9$ for $1\le i \neq j \le \lfloor p^{0.3} \rfloor$ and 0 otherwise.
	
	\item Case 4: $\rho_0(i,j) = 0.3$ for $1\le i\neq j \le \lfloor p/2 \rfloor$ and 0 otherwise, $\rho_1(i,j) = 0.5$ for $\lfloor p/2 \rfloor +1  \le i \neq j \le p$ and 0 otherwise.
\end{itemize}

The expected detection delay $\bE_1[T]$ is calculated by setting the change point $\nu=1$, that is, all streaming data is drawn from the post-change distribution. This simplifies the simulation process, as computing the exact WADD, defined in (\ref{eq:WADDdef}), which requires considering all possible change points, is often impractical. Importantly, for some detection procedures, such as the CUSUM test, WADD is often achieved when the change occurs at $\nu=1$ \citep{xie2021sequential}. As a result, using $\bE_1[T]$ closely approximates the worst-case scenario, proving a reliable alternative to evaluate detection performances.


Following the definition of sparse in \citet{feng2022max}, we define a term ``dense'' to represent the extent of the change in the correlation structure. Dense refers to the proportion of changed elements within the entire lower triangular of the correlation matrix, excluding the diagonal elements. Therefore, a smaller dense value suggests a more sparse alteration in the correlation structure and vice versa. In Case 1, the dense level is 1. In Case 2, the dense level is $\frac{\lfloor \frac{p}{2} \rfloor (\lfloor \frac{p}{2} \rfloor -1 )}{p(p-1)} \approx \frac{1}{4}.$ In Case 3, the dense level is $\frac{\lfloor p^{0.3} \rfloor (\lfloor p^{0.3} \rfloor -1 ) }{p(p-1) } \approx p^{-1.4}$.  From Case 1 to Case 3, the extent of the change in the correlation structure becomes more and more sparse.

\begin{figure}[ht!]   
	\centering
	\subfigure[]{\includegraphics[width=0.45\textwidth]{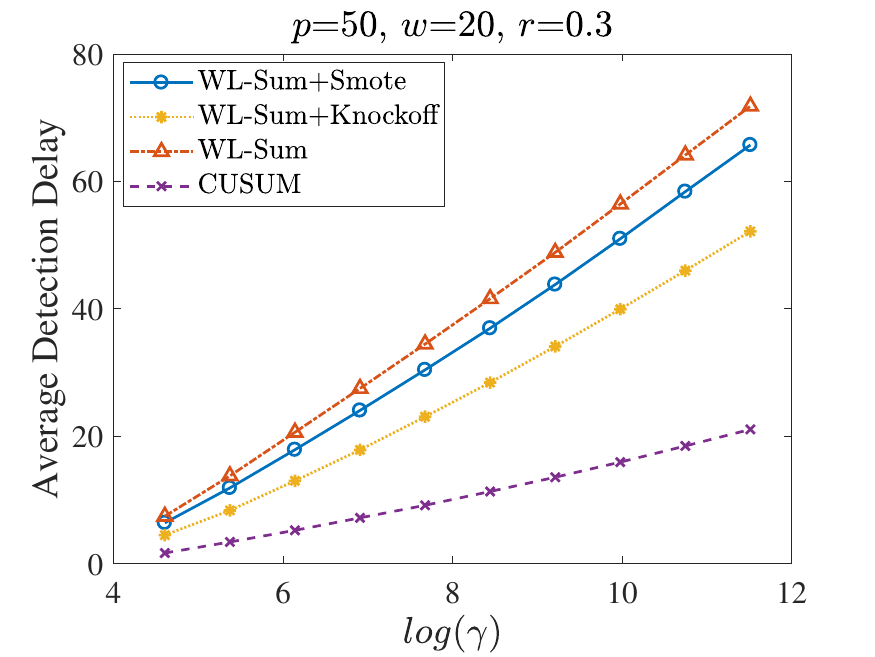}}
	\subfigure[]{\includegraphics[width=0.45\textwidth]{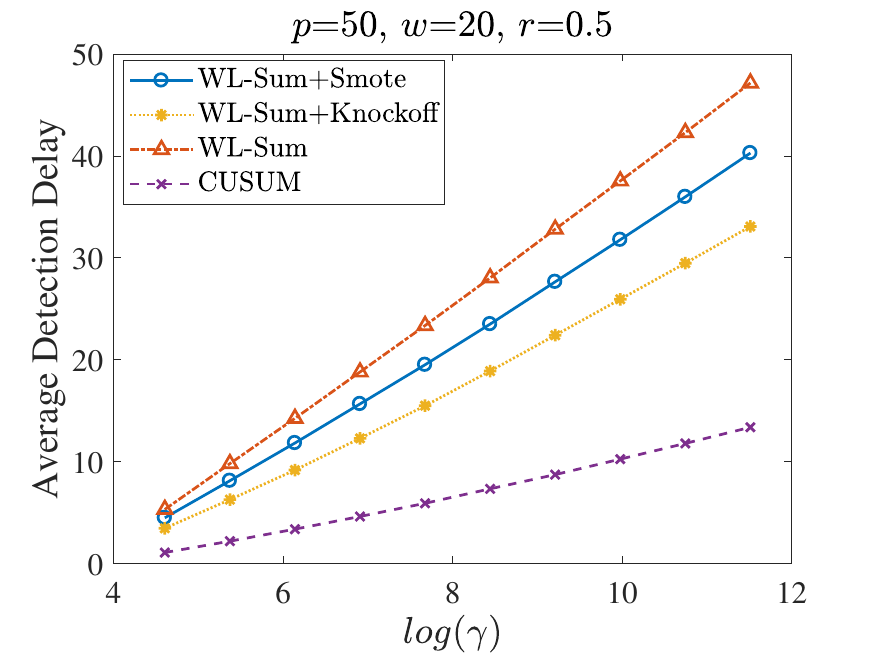}}
	\subfigure[]{\includegraphics[width=0.45\textwidth]{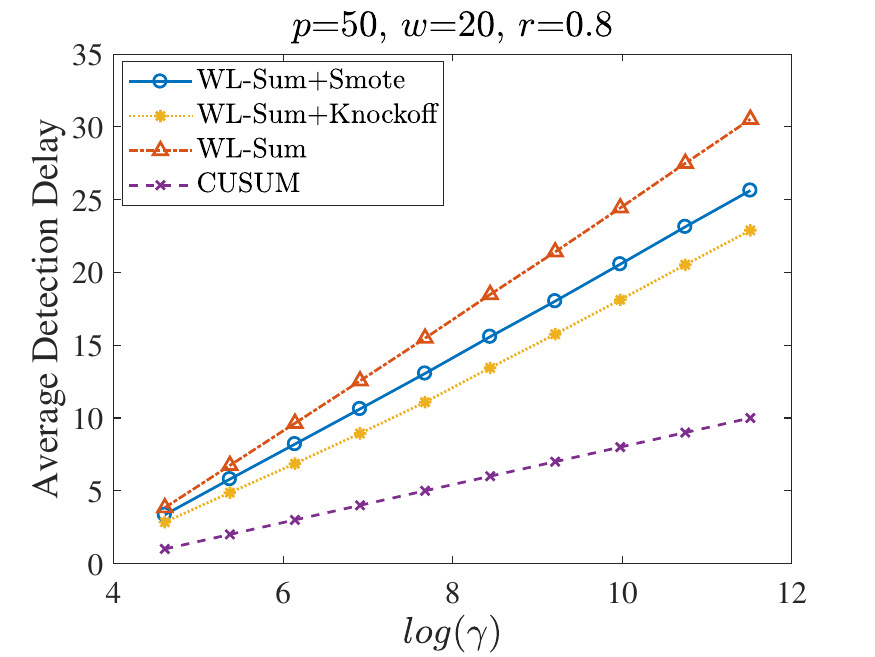}}
	\subfigure[]
	{\includegraphics[width=0.45\textwidth]{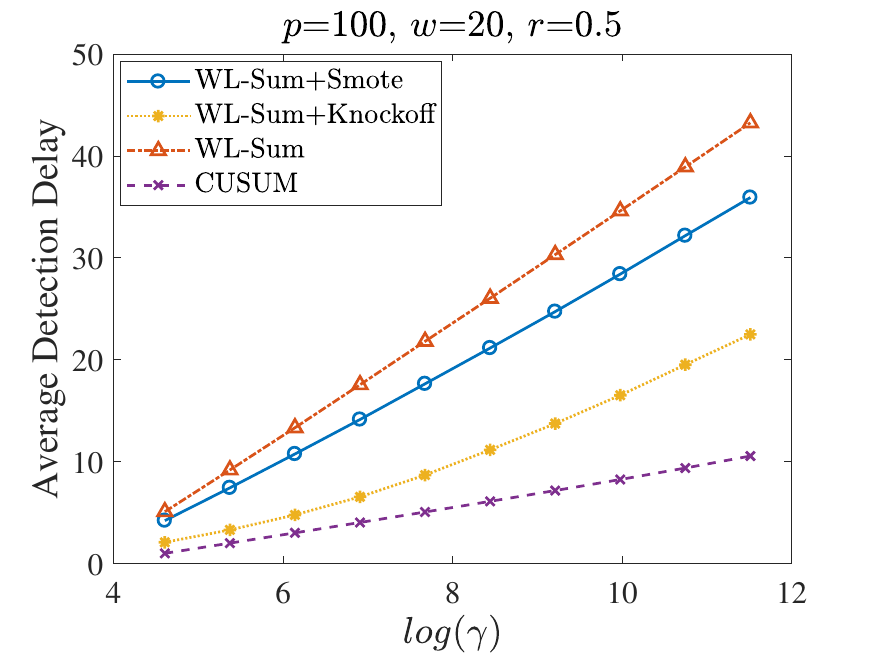}}
	\subfigure[]
	{\includegraphics[width=0.45\textwidth]{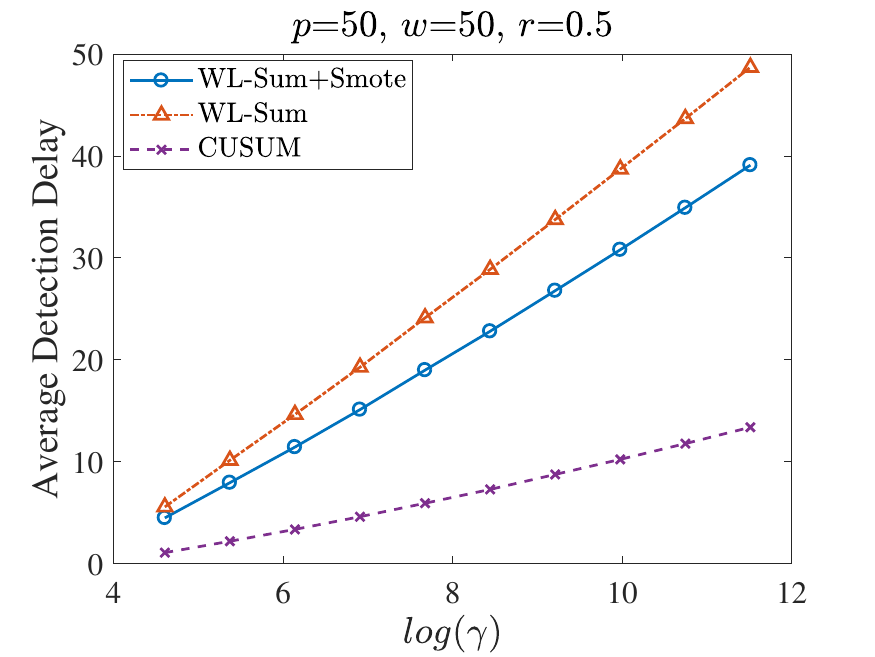}}
	\subfigure[]
	{\includegraphics[width=0.45\textwidth]{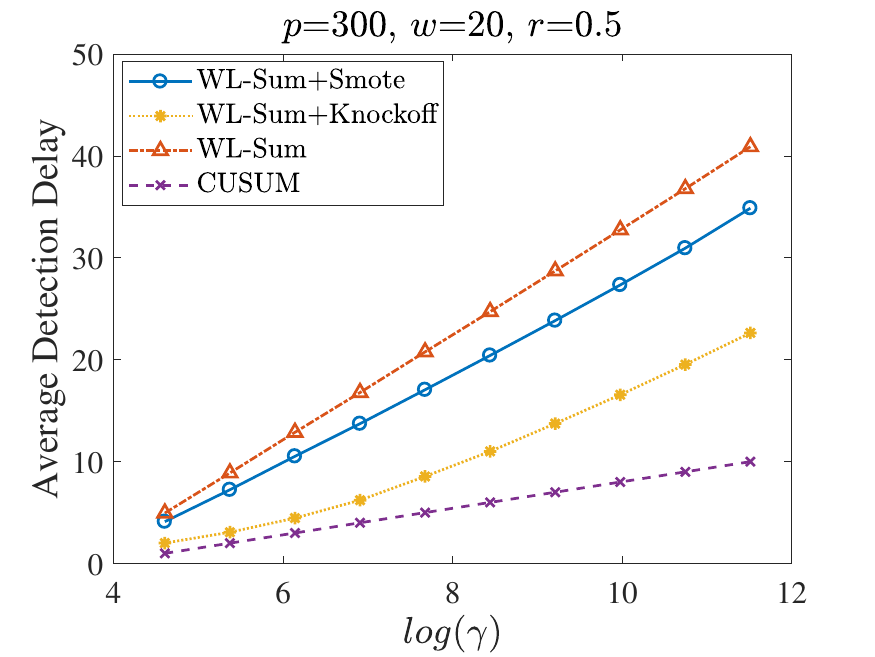}}
	\caption{Comparison of ADDs of WL-Sum, WL-Sum+SMOTE, WL-Sum+Knockoff, and CUSUM under Normal distribution, Case 1 (dense level of change is 1) with varying $r$ values in $\{0.3,0.5,0.8\}$.}
	\label{fig:non-sparse1}
\end{figure}

\begin{figure}[ht!]
	\centering
	\subfigure[]{\includegraphics[width=0.45\textwidth]{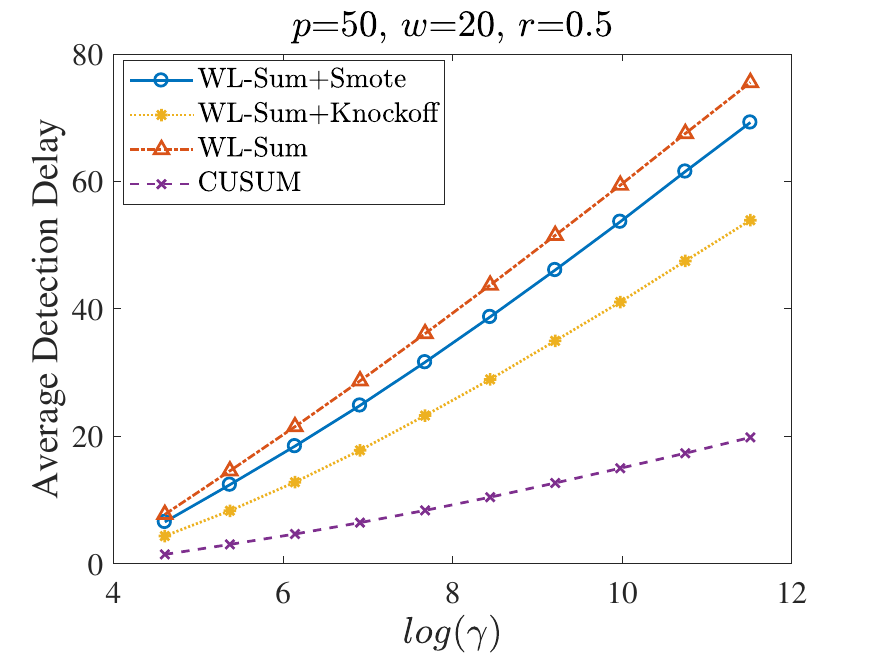}}
	\subfigure[]{\includegraphics[width=0.45\textwidth]{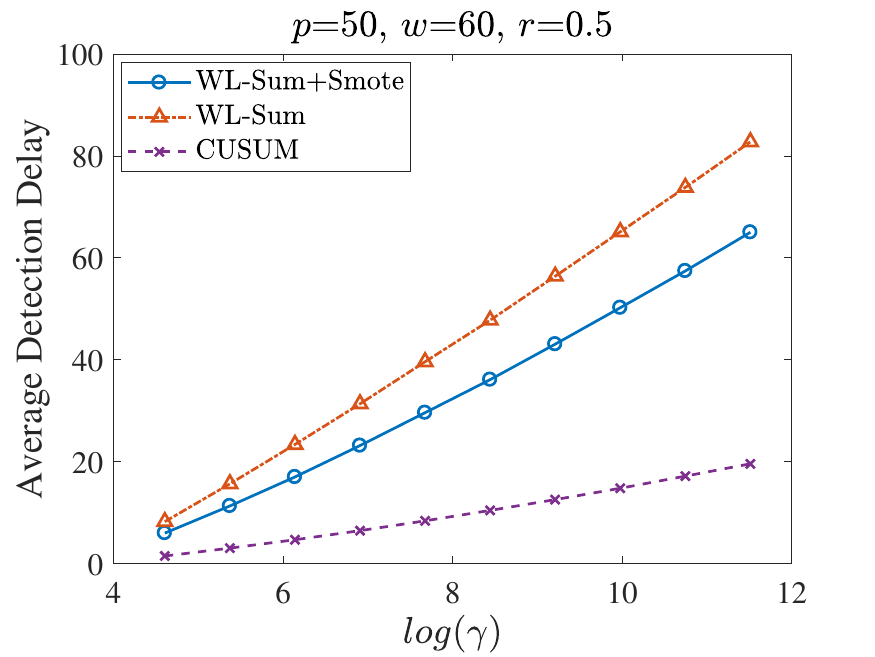}}
	\subfigure[]{\includegraphics[width=0.45\textwidth]{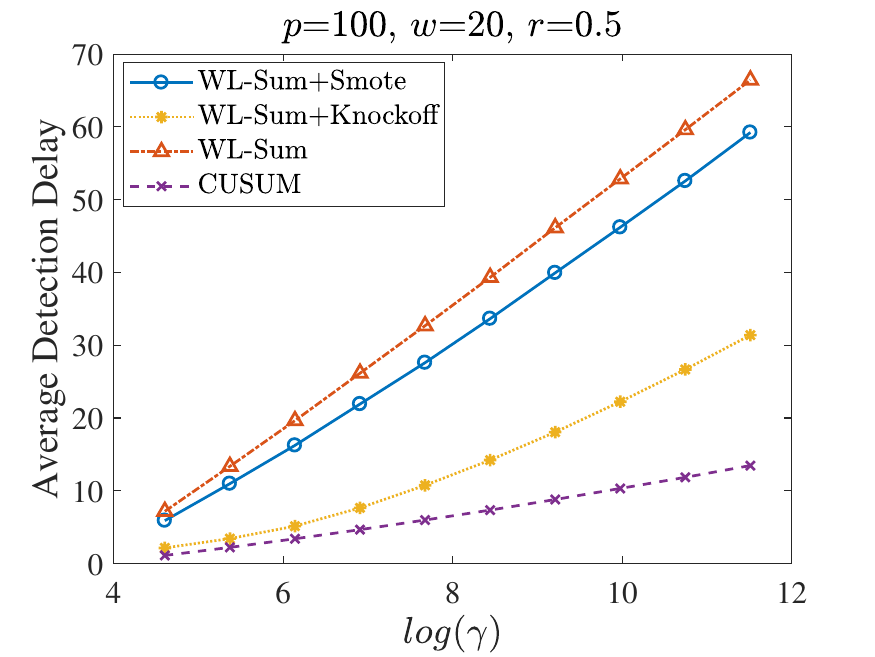}}
	\subfigure[]{\includegraphics[width=0.45\textwidth]{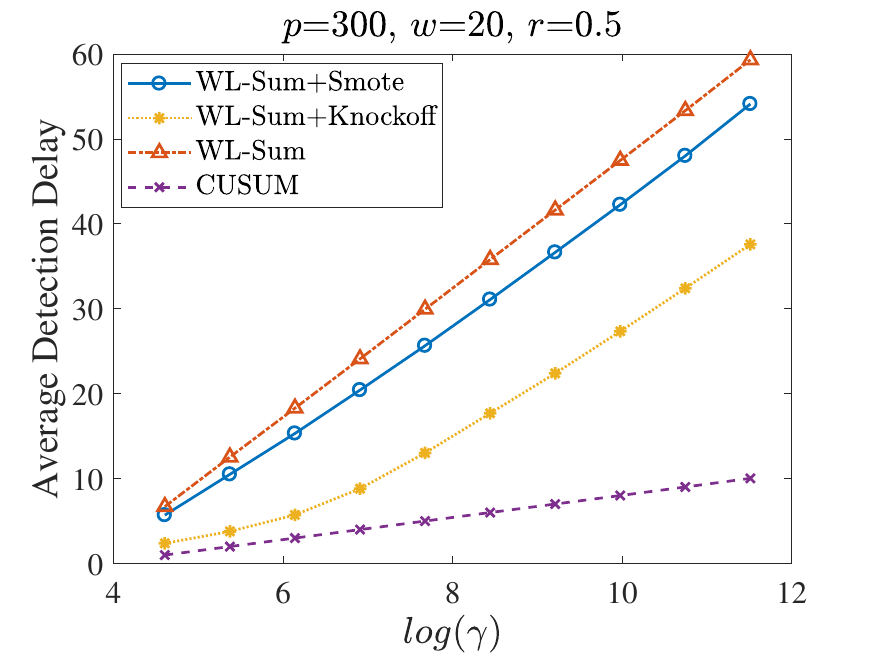}}
	\caption{Comparison of ADDs of WL-Sum, WL-Sum+SMOTE, WL-Sum+Knockoff, and CUSUM under Normal distribution, Case 2 (dense level of change is 0.25) with $r=0.5$.}
	\label{fig:non-sparse2}
\end{figure}

\begin{figure}[htbp]  
	\centering
	\subfigure[]{\includegraphics[width=0.45\textwidth]{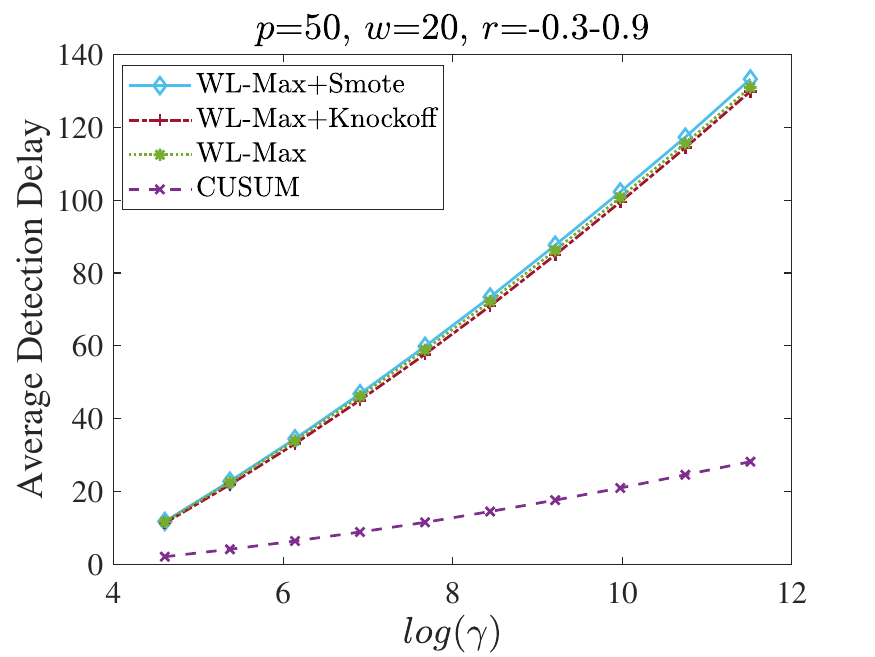}}
	\subfigure[]{\includegraphics[width=0.45\textwidth]{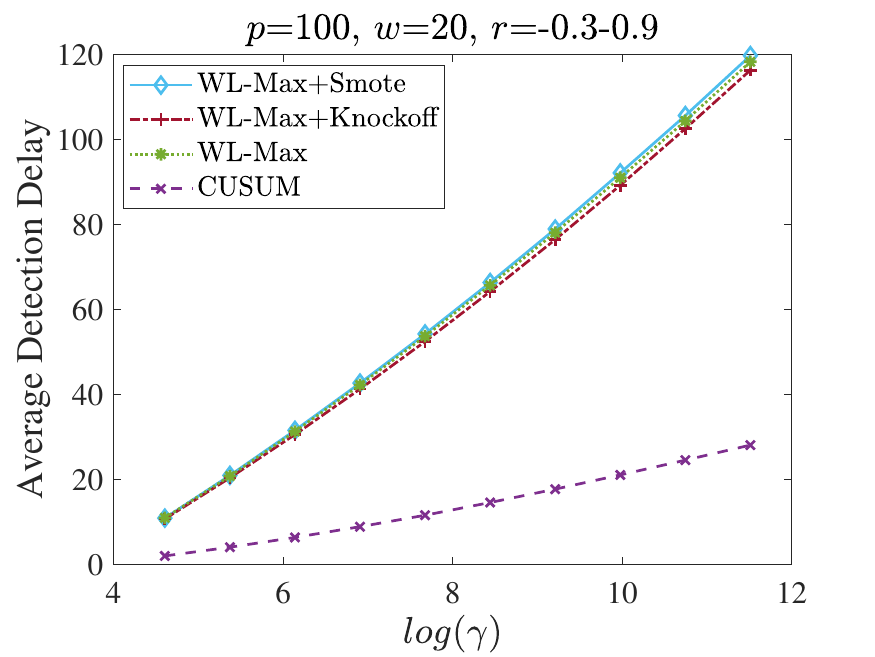}}
	\subfigure[]{\includegraphics[width=0.45\textwidth]{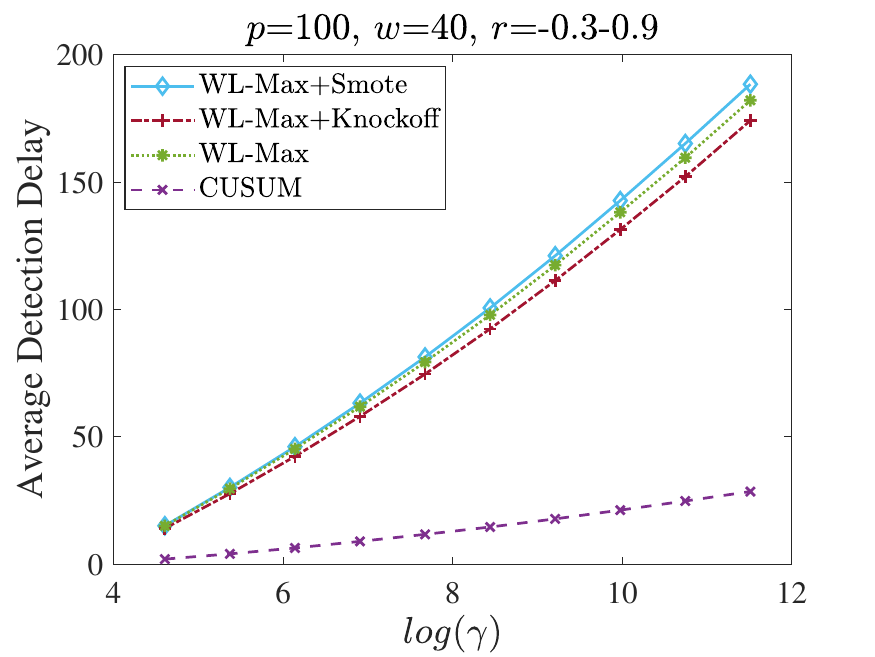}}
	\subfigure[]{\includegraphics[width=0.45\textwidth]{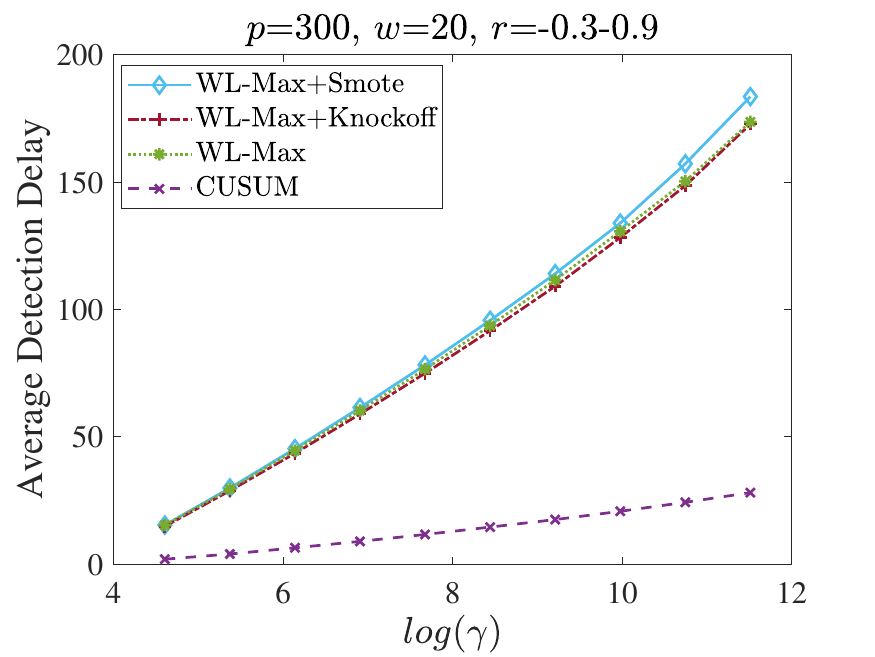}}
	\caption{Comparison of ADDs of WL-Max, WL-Max+SMOTE, WL-Max+Knockoff and CUSUM under Normal distribution, Case 3 (dense level of change is $p^{-1.4}$).}
	\label{fig:sparse}
\end{figure}

Figure~\ref{fig:non-sparse1} shows the average detection delay (ADD) of different methods in Case 1, with ARL ranging from $10^2$ to $10^5$. Figure~\ref{fig:non-sparse1} (a), (b) and (c) are the results when $r$ changes from 0 to 0.3, 0.5 and 0.8, respectively. CUSUM is the optimum detection method that provides the theoretical lower bound of the detection delay for all test procedures \citep{mous-astat-1986}. For a fixed dense level, when the magnitude of the change in the correlation structure increases, ADD decreases accordingly. Moreover, the SMOTE and knockoff enhanced methods can effectively reduce detection delay, and the knockoff-enhanced method performs better. Comparing (b), (d), and (f), as $p$ increases from 50 to 300, the maximum ADD of WL-Sum (WL-Sum+SMOTE, WL-Sum+Knockoff) decreases, indicating that our proposed methods perform better as the dimension grows. By comparing (b) and (e), as the window size $w$ increases from 20 to 50, ADD does not change much for WL-Sum and WL-Sum+SMOTE. Thus, for this setting, the better detection approach is to use a small window size 20 along with knock-off enhancements.

Figure~\ref{fig:non-sparse2} compares the ADDs of different methods in Case 2. Comparing Figure~\ref{fig:non-sparse2} (a) with Figure~\ref{fig:non-sparse1} (b), as the dense level drops from 1 to 0.2449, the ADD of WL-Sum increases from 47.2 to 75.6, indicating a more challenging detection situation when the change becomes scarce. Moreover, (a) and (b) differ in window size, with little impact on ADD, showing that the smaller window size 20 suffices for detecting the change and we do not necessarily need to adopt a larger window in this case. But SMOTE can reduce the detection delay when $w$ is larger. Figure~\ref{fig:non-sparse2} (a), (c), and (d) show that as $p$ increases, the ADD decreases, further demonstrating the effectiveness of our proposed method in dealing with large-dimensional data.

Figure~\ref{fig:sparse} compares the ADDs of WL-Max and its corresponding enhancement methods in Case 3. Across all settings, the max-type detection statistic exhibits shorter detection delays, indicating that it is more suitable for sparse change scenarios. Moreover, the enhancement methods only yield modest improvements to the ADD of WL-Max statistic. Compared with the ADD of the exact CUSUM test, we note that the ADDs of our proposed methods, especially the knockoff-enhanced methods, are comparable to those of exact CUSUM, demonstrating the effectiveness of our method.

The ADD results of WL-Sum and its enhanced versions in Case 4, and results under Student's $t$ distributed data, are presented in B.5 of the SM. In general, our methods show strong and robust performance for the Student $t$ distribution.

As shown above, sum-type statistics perform well for dense correlation changes, while max-type statistics excel for sparse changes. Since the dense level is typically unknown in practice, we can utilize both statistics and calibrate them as follows:
\[
S_t^{c} = \max\left( \frac{1}{b_1}S_t^{(\text{WL-sum})} , \frac{1}{b_2}S_t^{(\text{WL-max})}  \right),
\]
and the stopping time is $T^c= \inf\left\{t: S_t^{c} \ge 1 \right\}$, where $b_1$ and $b_2$ are the estimated thresholds for WL-Sum and WL-Max statistics, respectively, given a fixed $\gamma$. These thresholds are used to calibrate the WL-Sum and WL-Max statistics to the same magnitude. We denote this hybrid method as WL-Sum-Max-Combined, as it operates by running both sum-type and max-type detection procedures in parallel and stopping as soon as either one raises an alarm.

\begin{figure}
	\centering
	\begin{minipage}[t]{0.3\linewidth}
		\centering
		f\includegraphics[width=2.2in, height=2.0in,keepaspectratio]{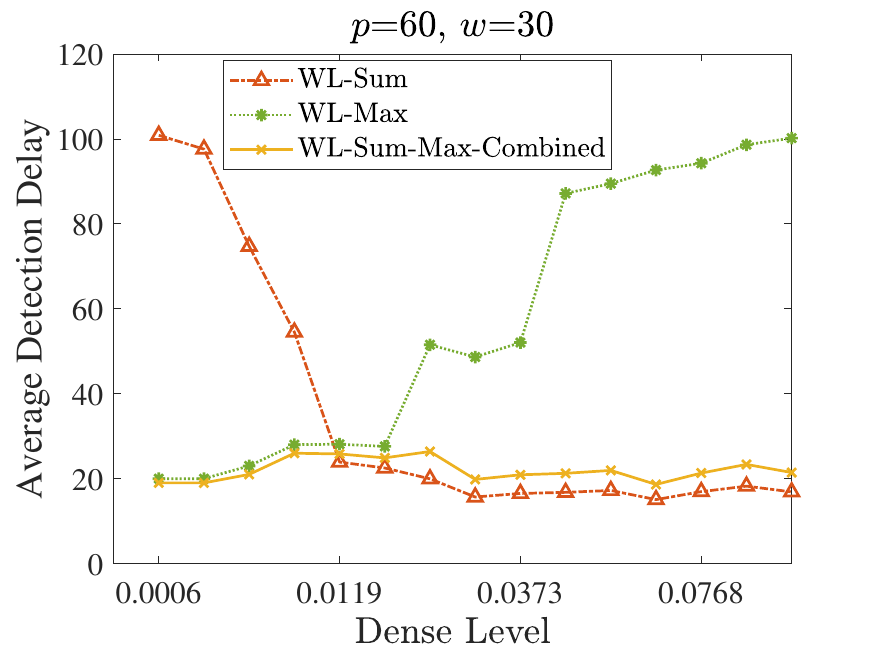}
	\end{minipage}
	\begin{minipage}[t]{0.3\linewidth}
		\centering
		\includegraphics[width=2.2in, height=2.0in,keepaspectratio]{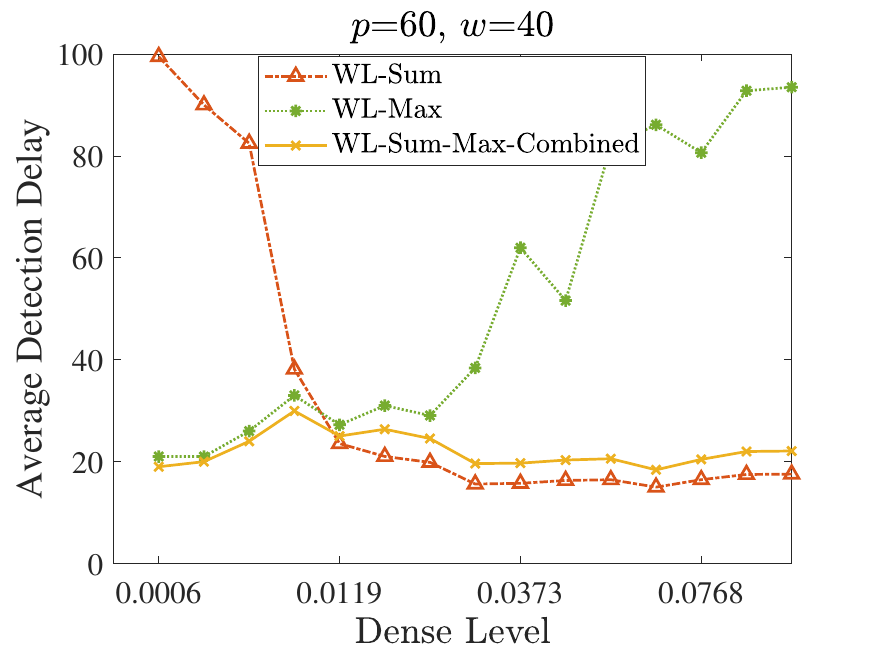}
	\end{minipage}
	\begin{minipage}[t]{0.3\linewidth}
		\centering
		\includegraphics[width=2.2in, height=2.2in,keepaspectratio]{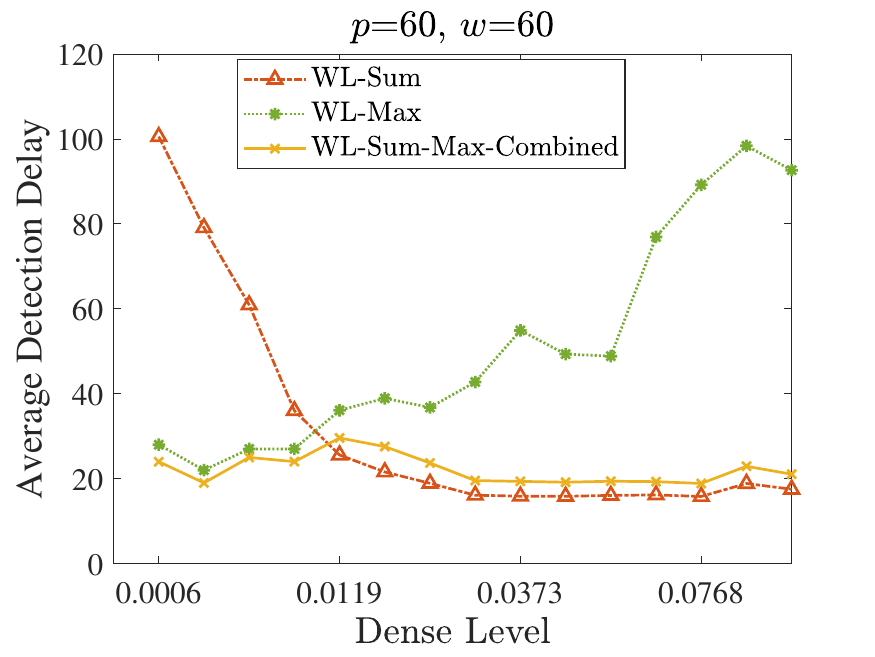}
	\end{minipage}
	\caption{Comparison of ADDs of WL-Sum, WL-Max and WL-Sum-Max-Combined under Normal distribution, with the dense level varying from 0.0006 to 0.1073. $\gamma=10^5$, $p=60$.}
	\label{fig:combine}
\end{figure}

The combined $S_t^{c}$ is compared with the WL-Sum and WL-Max approaches in terms of ADD in Figure~\ref{fig:combine}. We fix ARL to $10^5$ in this setting and let $p=60$, window sizes $w=30,40,60$. For a given integer $n\leq p$, let $\rho_0(i,j)= \rho_1(i,j)=1$ for $i=j$, $\rho_0(i,j) = r_0$ for $1\le i\neq j \le n$ and 0 otherwise, and $\rho_1(i,j) = r_1$ for $1\le i\neq j \le n$ and 0 otherwise. When $n=2$, only the correlation of variable 1 and variable 2 changes, which is extremely sparse. As $n$ increases from 2 to 20, the dense level increases from 0.0006 to 0.1073. The correlation coefficients $r_0$ and $r_1$ are selected to ensure that the largest eigenvalue of $\boldsymbol{R}_1\boldsymbol{R}_0^{-1}$ is between $5$ and $8$.

The WL-Max statistic performs well when detecting very sparse changes, and as the dense level increases, the WL-Sum statistic performs better. This finding aligns with the literature on two-sample testing, indicating that the max-type statistic demonstrates higher power in sparse settings. In contrast, the sum-type statistic exhibits higher power in non-sparse settings. The yellow line represents the performance of the combined method, and it works consistently well under both sparse and non-sparse settings. It has a lower ADD than WL-Max in sparse cases, and nearly the same ADD as WL-Sum in non-sparse cases.

\section{Real Data Analysis}\label{sec:application}
In this section, we present two real-data applications. The first is to detect early warning signals of El Ni{\~n}o events. To our knowledge, this is the first attempt to apply an online change point detection procedure to El Ni{\~n}o prediction, and the results are promising. The second application involves the rapid detection of tremor-like signals in a seismic data. In both applications, we employ a Shewhart-type sum statistic (ST-Sum), a special case of the WL-Sum statistic. 
Specifically, the maximization step on $t-w \le t' \le t-1$ at time $t$ in WL-Sum is eliminated, i.e., $S_t^{(\text{ST-sum})} = \|\boldsymbol{v}_{t-w,t} \|_1$, which significantly reduces computational cost while maintaining effective detection performance.

\subsection{El Ni{\~n}o Event Prediction}

El Ni{\~n}o episodes are part of the El Ni{\~n}o-Southern Oscillation (ENSO), which is the strongest driver of interannual climate variability and can trigger extreme weather events and disasters in various parts of the world. Early warning signals of El Ni{\~n}o events would be instrumental for avoiding some of the worst damages. 

To forecast El Ni{\~n}o events, many state-of-the-art coupled
climate models, as well as a variety of statistical approaches, have been suggested; see the review paper \citep{bunde2024evaluation}. Although these
forecasts are quite successful at shorter lead times (say, 6 months), they have limited anticipation power at longer lead times. In particular,
they generally fail to overcome the so-called $``$spring barrier", that is, in spring, most methods tend to make wrong predictions for El Ni{\~n}o event. Some methods have been constructed to predict El Ni{\~n}o events beyond 9 months in advance \citep{ludescher2013improved}. However, it is almost impossible to forecast these events accurately with more than 12 months in advance. \cite{lenssen2024strong} found that ENSO is predictable at least two years in advance only when forecasts are made during strong EI Ni{\~n}o events, while forecasts initialized during other states do not have predictive skill over one year. Our method can predict El Ni{\~n}o events more than one year most of the time.

\begin{figure}[ht!]
	\centering
	\subfigure[]{\includegraphics[width=0.45\textwidth,height=0.22\textwidth]{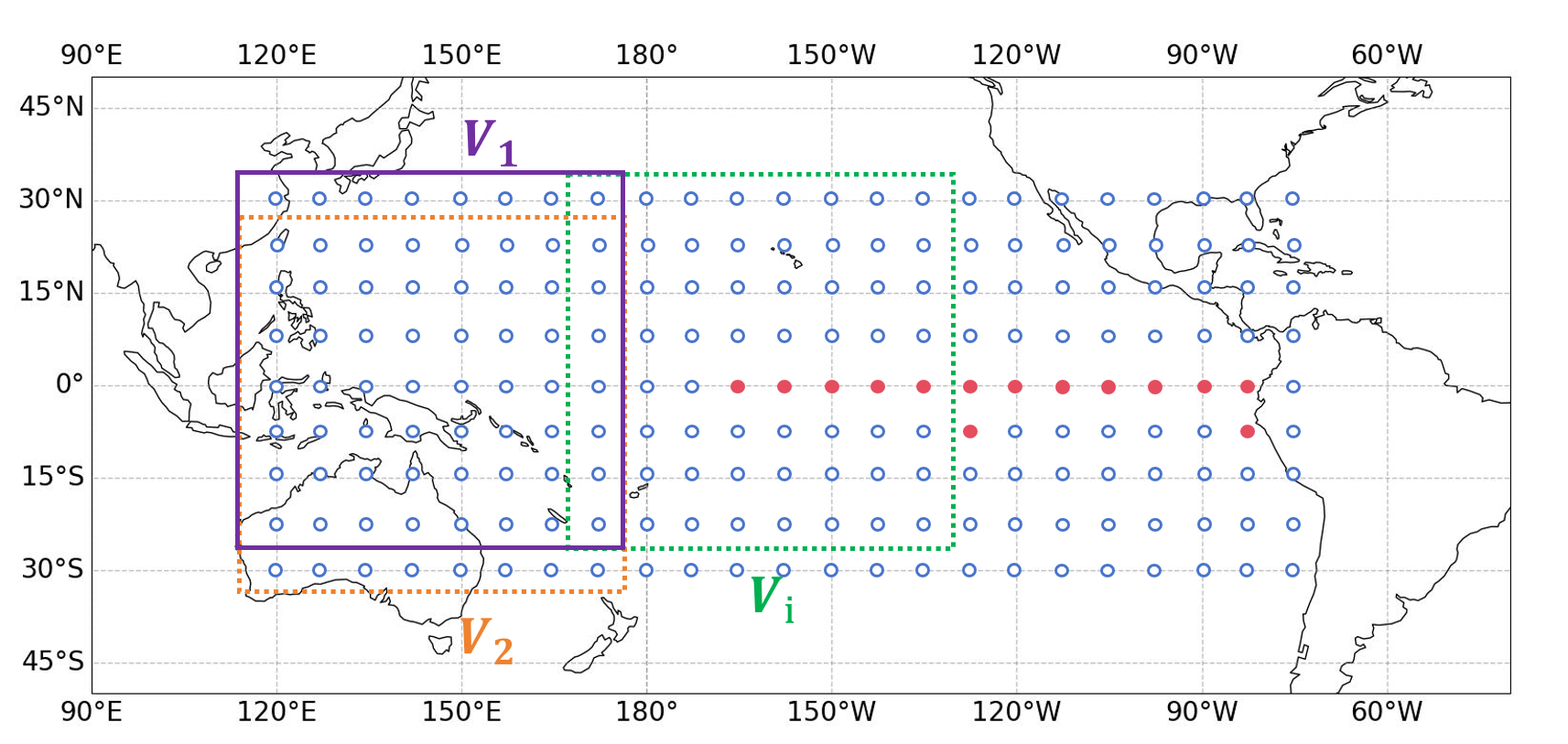}}
	\subfigure[]{\includegraphics[width=0.49\textwidth,height=0.22\textwidth]{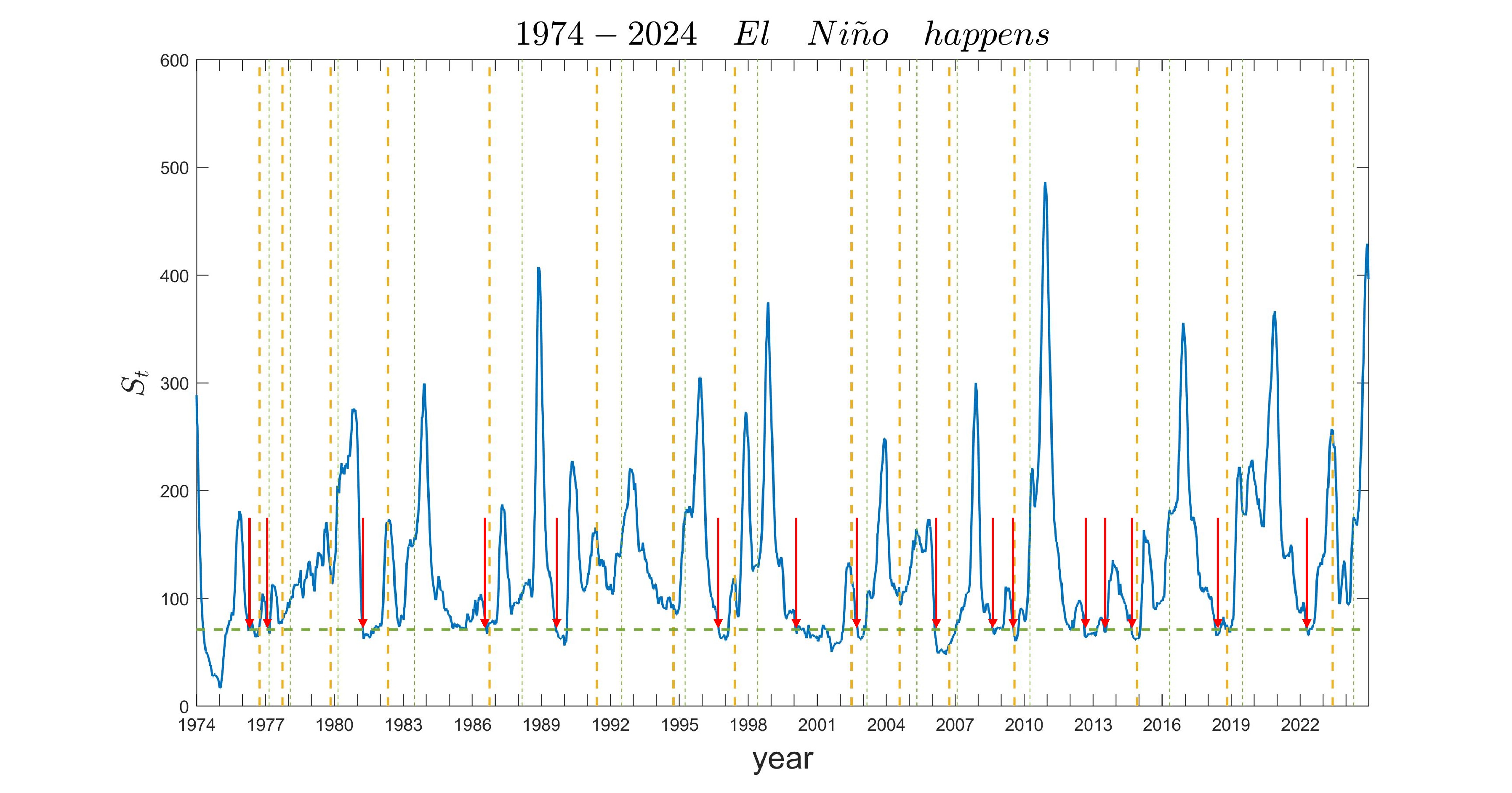}}
	\subfigure[]{\includegraphics[width=0.49\textwidth,height=0.22\textwidth]{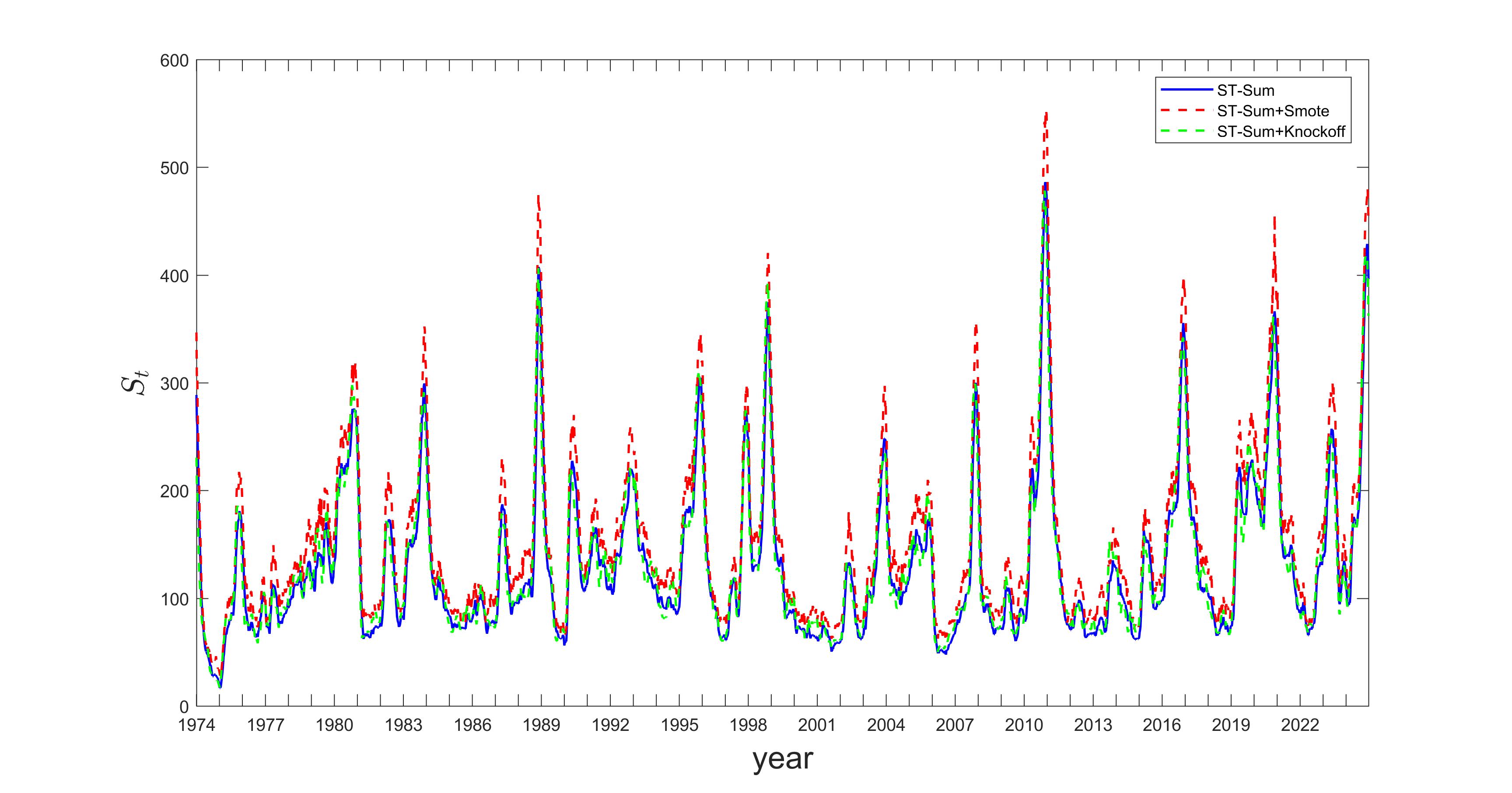}}
	\subfigure[]{\includegraphics[width=0.49\textwidth,height=0.22\textwidth]{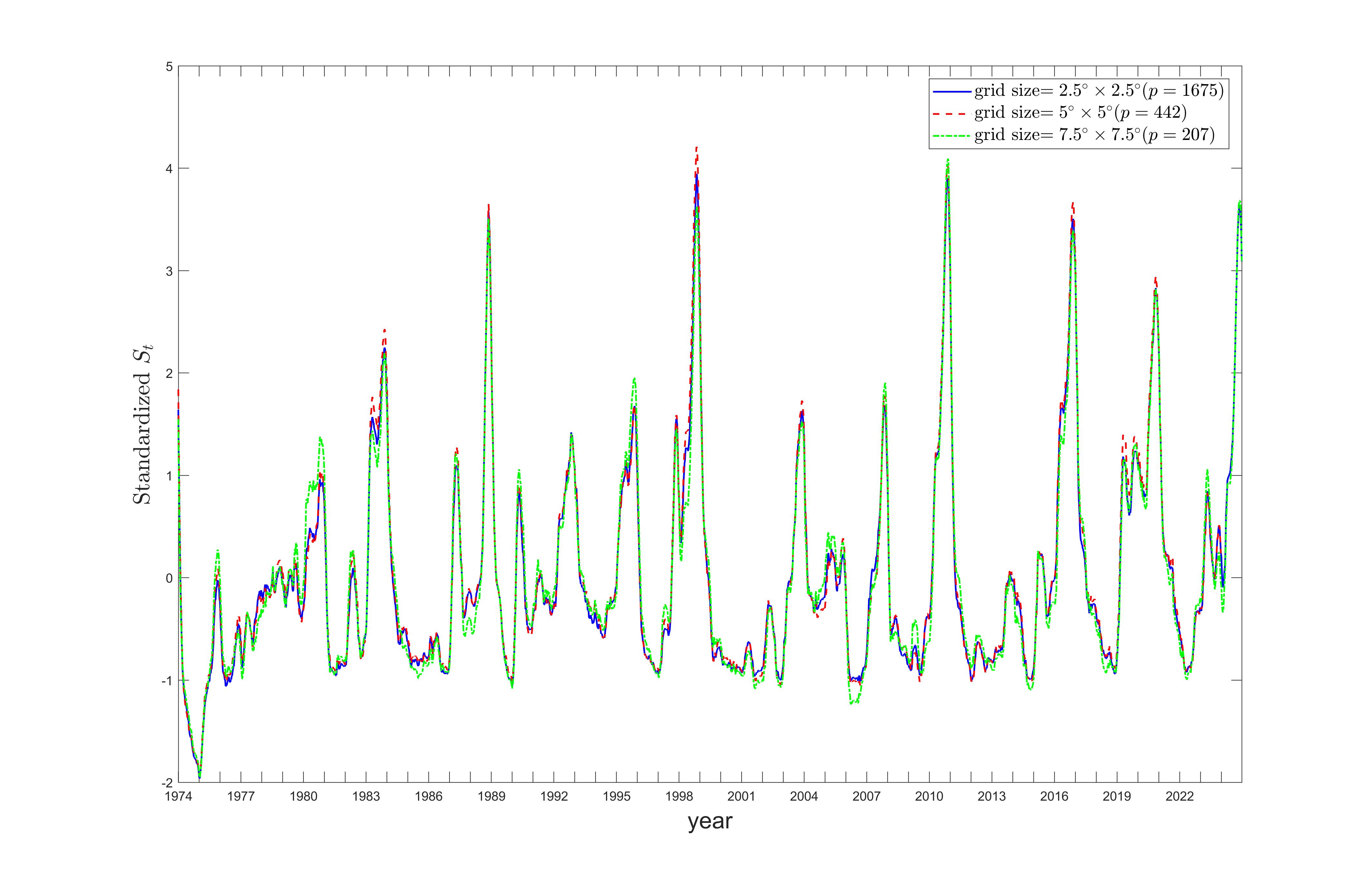}}
	\caption{(a) The region of nodes; (b) The prediction of El Ni{\~n}o events between 1974 and 2024; (c) Comparison of ST-Sum test statistic and its corresponding SMOTE and knockoff enhancement versions; (d) Comparison of ST-Sum test statistic for different grid sizes.}
	\label{fig:climate}
\end{figure}

An El Ni{\~n}o episode is featured by rather irregular warm excursions from the long-term mean state. An El Ni{\~n}o episode starts when the ONI index, the 3-month running mean of the anomaly in sea surface temperature averaged in the NINO3.4 region (5$^o$S-$5^o$N, 170$^o$W-12$0^o$W), is above $0.5^\circ$C for at least five consecutive months. \citet{ludescher2013improved} showed that the strengths of the cross-correlations between the El Ni{\~n}o Basin (red dots in Figure~\ref{fig:climate}(a)) and the surrounding sites (blue dots in Figure~\ref{fig:climate}(a)) tend to strengthen before El Ni{\~n}o episodes and then weaken significantly. Therefore, we concentrate on the changes in the correlations and show that well before an El Ni{\~n}o episode, the correlations tend to increase first and then decrease sharply. We use this robust observation to forecast El Ni{\~n}o development in advance.

We use the 1000hPa daily temperature data from the ERA5 database at time 00:00 UTC on the 7th, 14th, 21st, and 28th days of each month from 1974 to 2024. The spatial domain covers 30$^o$S-30$^o$N and 120$^o$E-75$^o$W, including the ENSO basin (Figure~\ref{fig:climate}(a)). With a grid size of $7.5^\circ \times 7.5^\circ$, there are 207 nodes. The data dimension is $p=207$, and $T=2448$ is the length of the time series. Reference data are chosen from two neutral years, 1971 and 1974, as no El Ni{\~n}o event occurred in these two years. Since ENSO is an interannual phenomenon, which is phase locked in the seasonal cycle, we set the window size $w$ to be one year ($w=48$). We select 8×8 = 64 nodes that are geographically close to each other as a subset; therefore, there are 32 subsets V1, V2, etc., as in panel (a) of Figure~\ref{fig:climate}. The ST-Sum statistic is calculated for each subset and the maximum value over the 32 subsets is used as the final detection statistic.

The statistic is calculated in a time-moving window of one year (blue line in Figure~\ref{fig:climate}(b)). The yellow and green vertical lines mark the beginning and end of 15 El Ni{\~n}o events from 1974 to 2024. The horizontal green line represents the threshold, the maximum value of empirical statistics based on $q=100$ times sign-flip permutations, which is selected by cross-validation from all quantiles (actually, the prediction remains stable with respect to the choice of threshold, that is, similar quantiles of empirical statistics lead to similar results in hit rate and false alarm rate). 
An alarm indicating the arrival of an El Ni{\~n}o event is triggered every time the test statistics cross the threshold from the top. The alarm results in a correct prediction if, in the next two years, El Ni{\~n}o episode sets in; otherwise, it is considered a false alarm. The correct predictions are marked with red arrows. 
From 1974 to 2024, there were 15 El Ni{\~n}o events and 36 event-free years. 
Our method successfully predicts 13 El Ni{\~n}o events, resulting in a hit rate of $13/15=0.867$ and a false alarm rate of $0$, as shown in Table~\ref{tab:hit_rate}. And the average lead time of these signals is 13.95 months. The total computation time is approximately 2 hours on a computer equipped with an Intel Xeon Gold 6248R CPU, 384 GB of RAM, and Windows 10 operating system, without GPU acceleration.

To investigate the effects of two enhancement methods in real data, we compute the ST-Sum statistic with SMOTE and knockoff, respectively, which are shown in Figure~\ref{fig:climate}(c). The statistics with two enhancement methods keep patterns similar to the original statistic, so that they have the same prediction results in terms of hit rate and false alarm rate. But they increase the lead time to 14.86 (knockoff) and 14.15 (SMOTE) months.

We also test 1675 nodes with a grid size of $2.5^\circ \times 2.5^\circ$ and 442 nodes with a grid size of $5^\circ \times 5^\circ$, respectively. And the same predictions are obtained as shown in Figure~\ref{fig:climate}(d). Therefore, our method is robust to different grid sizes, that is, the dimension $p$.

In general, the El Ni{\~n}o events and La Ni{\~n}a events appear in return; that is, an El Ni{\~n}o event is often followed by a La Ni{\~n}a event and vice versa. However, this is not always the case. Sometimes, several El Ni{\~n}o events occur consecutively without an intervening of La Ni{\~n}a event, and similarly, La Ni{\~n}a events can sometimes appear in succession. Of our 13 predictions, 3 prediction signals, in 1977, 1981 and 2002, appear during or at the end of El Ni{\~n}o event, indicating that La Ni{\~n}a will not follow, but another El Ni{\~n}o event will continue to occur. Furthermore, for the strong El Ni{\~n}o event from winter 2014 to summer 2016 with the highest strength ever recorded (peak of the NINO3.4 index), our method generates the first signal in winter 2012, then another signal in summer 2013 and a third signal before the start of this event. Together, these three signals not only forecast the existence of El Ni{\~n}o event in the following two years, but also indicate an extreme event.

\citet{ludescher2013improved} forecasted El Ni{\~n}o events with a hit rate of 0.7 and a false alarm rate of 0.1 in the training data (1951-1980), and a hit rate of 0.667 and a false alarm rate of 0.046 in the testing data (1981-2011) when the threshold is set at 2.82. For the same period of test data, our method shows superior performance with two more correct predictions, increasing the hit rate by 0.222 as shown in Table~\ref{tab:hit_rate} (and Figure 9 in the SM).
In addition, their method failed to detect the three most recent El Ni{\~n}o events, dating back to 2011. Their method contains two steps: an optimum prediction algorithm is first learned from the training data and then applied to the testing data. So, their method can produce different results based on different thresholds. In contrast, our method is easily implemented without optimizing in a training dataset, the threshold is determined by the proposed algorithm, and theoretical guarantees are provided.

\begin{table}
	\centering
	\footnotesize
	\caption{Comparison of our method and \citet{ludescher2013improved}'s method in El Ni{\~n}o forecasting. Dashes (---) indicate that results for Ludescher's method are not available for that period, which was not covered in their original study.}
	\label{tab:hit_rate}
	\begin{tabular}{ccc|cc}
		\hline
		Period & \multicolumn{2}{c|}{1974--2024} & \multicolumn{2}{c}{1981--2011} \\
		\cmidrule(lr){2-3} \cmidrule(lr){4-5}
		& Our method  & Ludescher's & Our method  & Ludescher's \\
		\midrule
		Hit Rate & 0.867(13/15) & --- & 0.889(8/9) & 0.667(6/9) \\
		False Alarm Rate & 0(0/36) & --- & 0(0/22) & 0.046(1/22) \\ \hline
	\end{tabular}
\end{table}



\subsection{Seismic Event Detection}
\label{seismic}

\begin{figure}
	\centering
	\subfigure[]{\includegraphics[width=0.45\textwidth]{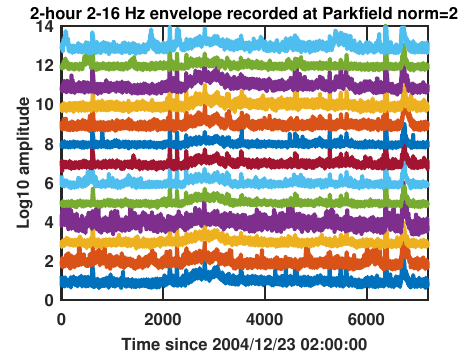}}
	\subfigure[]{\includegraphics[width=0.45\textwidth]{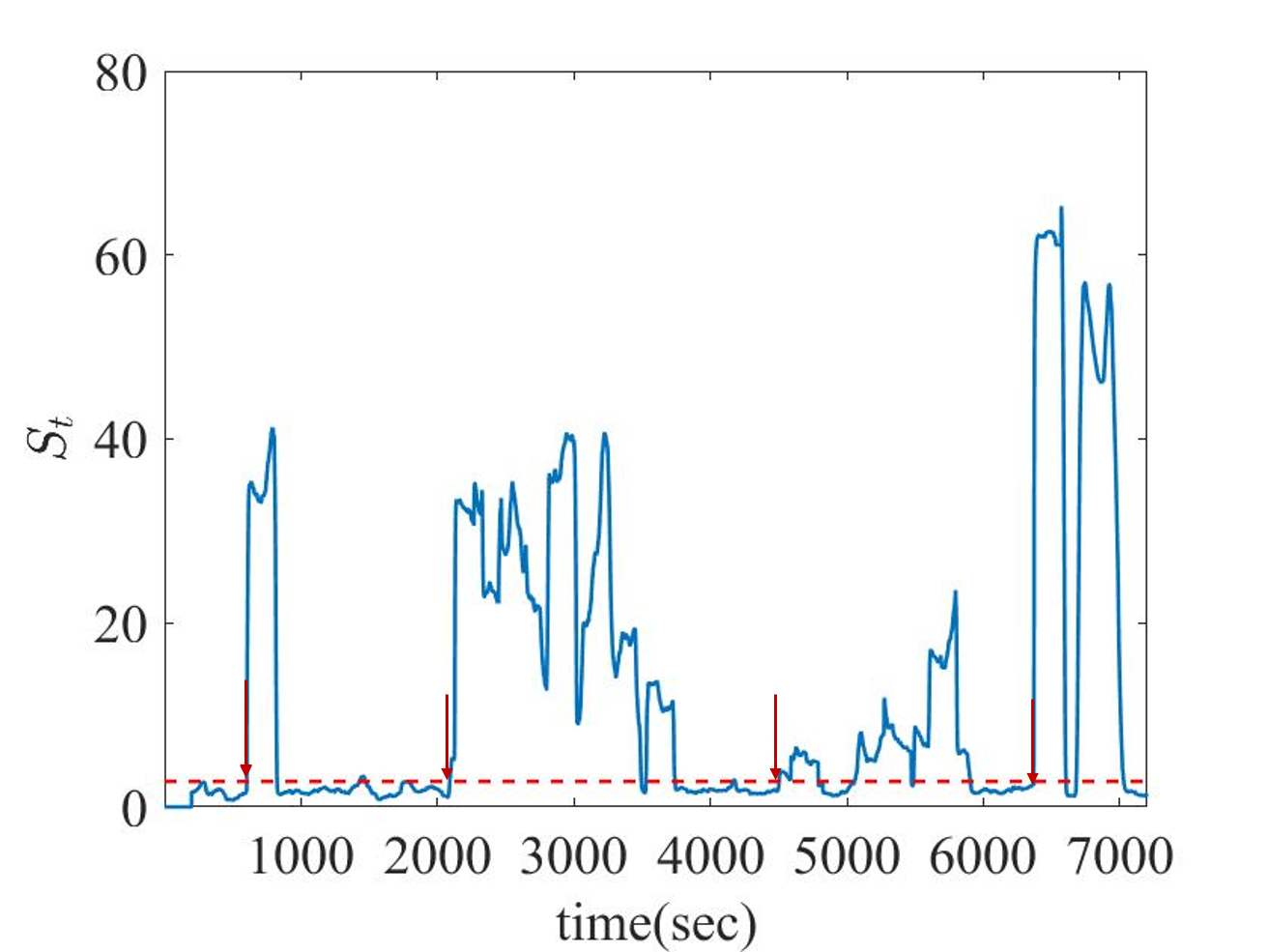}}
	\caption{(a) Raw data; (b) ST-Sum test statistic.}
	\label{fig:seismic data}
\end{figure}

Our proposed methods are applied to a real seismic dataset collected at Parkfield, California, from 2 to 4 a.m. on Dec 23rd, 2004. The raw data contains records at 13 seismic sensors that simultaneously record a continuous data stream with a frequency of 250HZ, see Figure~\ref{fig:seismic data}(a).  
The goal is to detect micro-earthquakes and tremor-like signals as soon as possible, which are weak signals caused by minor subsurface changes in the Earth. The tremor signal is propagated to each sensor from the source, and the affected sensors observe a similar waveform corrupted by noise. These tremor signals are helpful for geophysical study and predicting potential earthquakes.

To improve computational efficiency, we first perform a down-sampling procedure to reduce the data length to $T=7198$. The first 590 observations do not contain tremor signals and are thus selected as reference data. We use a sliding window of length 200s, i.e., at each time point, the previous 200 observations are used to estimate the correlation matrix denoted as $\boldsymbol{\hat{R}}_t$. The trajectory of ST-Sum test statistic is shown in Figure~\ref{fig:seismic data}(b). We use 50 sign-flip trials to get the detection threshold. In practice, we slightly inflate the threshold chosen by the algorithm to ensure a zero false alarm rate in the reference data. 

\begin{figure}[ht!]
	\centering
	\subfigure[]{\includegraphics[width=0.45\textwidth]{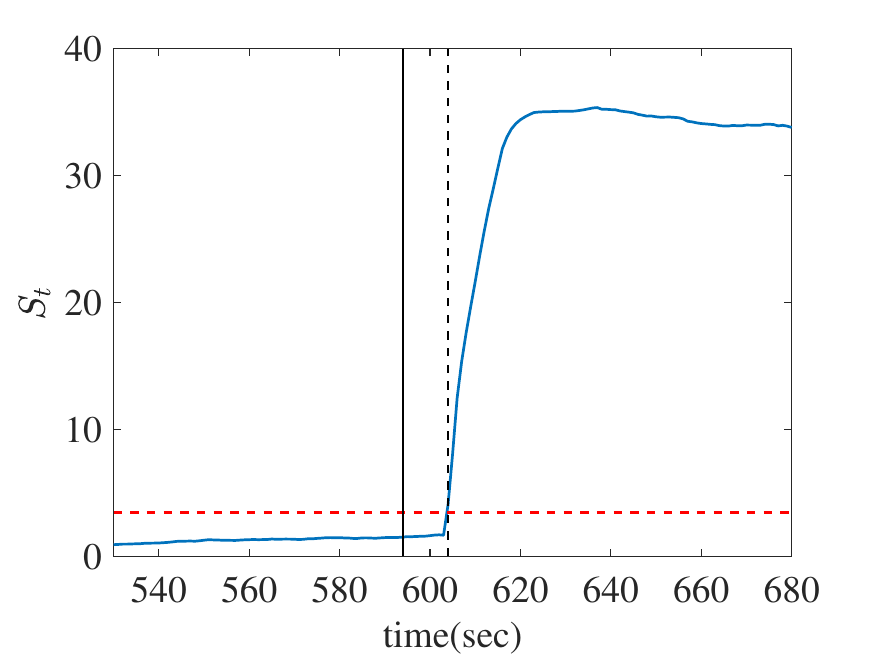}}
	\subfigure[]{\includegraphics[width=0.45\textwidth]{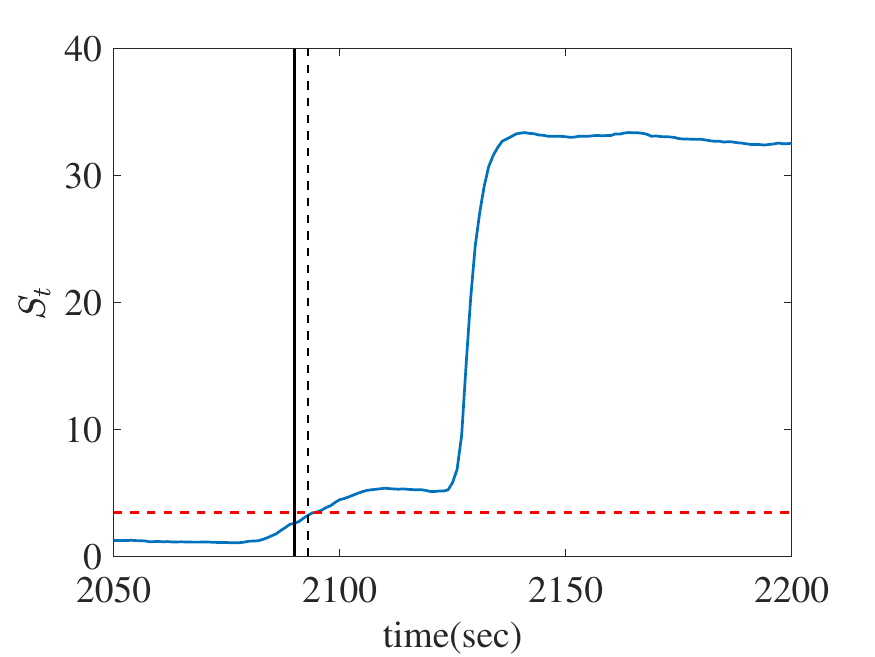}}
	\subfigure[]{\includegraphics[width=0.45\textwidth]{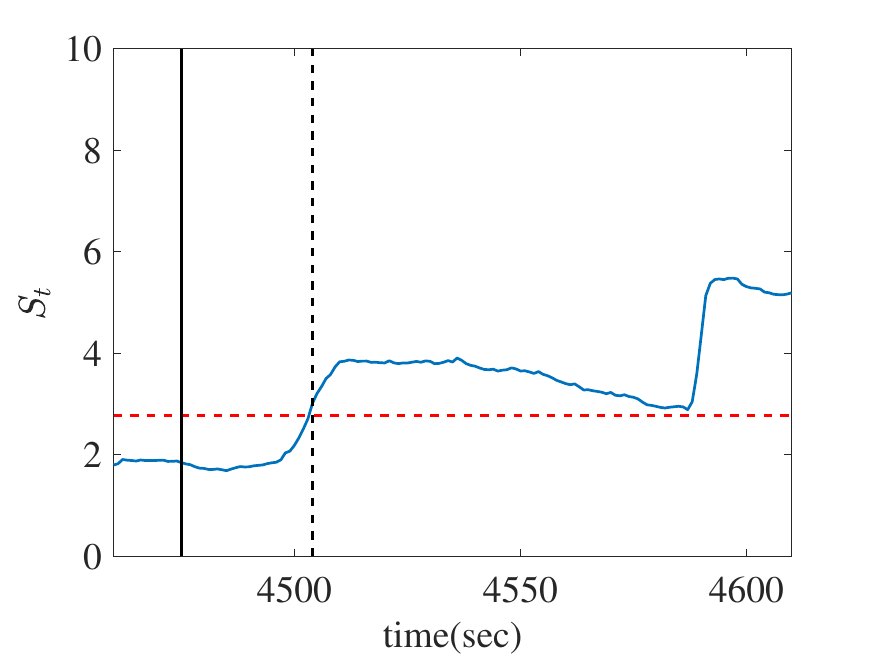}}
	\subfigure[]{\includegraphics[width=0.45\textwidth]{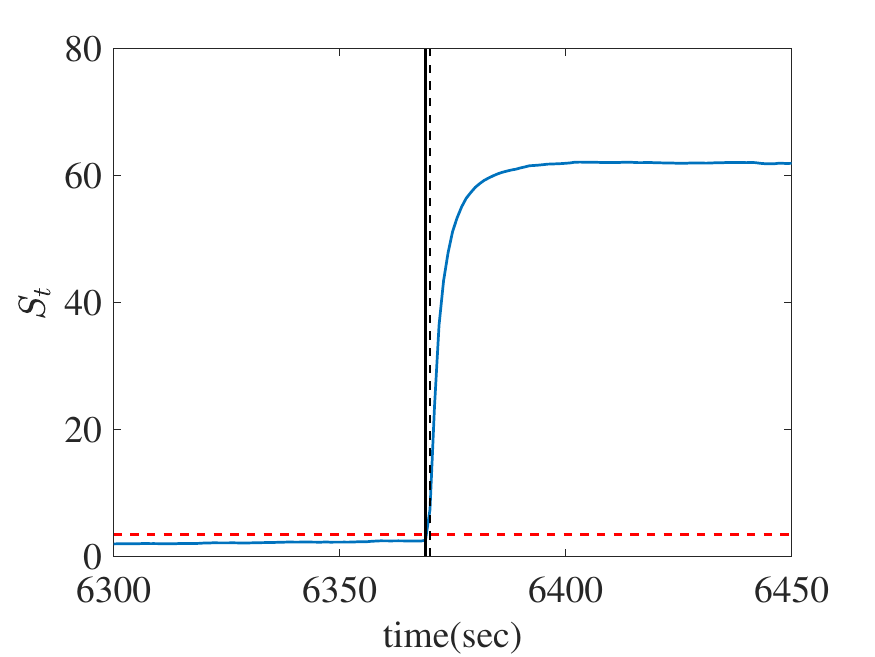}}
	\caption{Four change points detected in seismic data. Each plot covers a time range of 150 seconds. The vertical solid black line denotes a real event, and the vertical dotted black line denotes the location of the detected change point.}
	\label{fig:4changepoints}
\end{figure}

In Figure~\ref{fig:4changepoints}, the vertical solid line denotes the true event time and the dashed line represents the detected change point. Our proposed method finds three main events at 604, 2093, and 6370 seconds as shown in Figure~\ref{fig:4changepoints}(a), (b), (d), respectively. They match well with the true event catalog, 594, 2090, and 6369 seconds, obtained from the Northern California Earthquake Data Center. Using the same dataset, \citet{xie2020sequential} detected three main events at 615, 2127, and 6371 seconds, corresponding to three actual events at 594, 2124, and 6369 seconds. So, both studies detect the same first and third events. The one that happened at 2090s has a magnitude of 1.66 while the one at 2124s has a magnitude of 1.1, so that they are very close to each other. Compared to \citet{xie2020sequential}, our method has a shorter detection delay in the first event and a comparable delay for the last two events. Our method also detects the tremor-like signals at 4504s, which corresponds to an actual event at 4475 seconds with a magnitude of 1.58, with a time delay of 29 seconds, see Figure~\ref{fig:4changepoints}(c), while the subspace method in \citet{xie2020sequential} fails to detect this event. It should be noted that the change in Figure~\ref{fig:4changepoints}(c) is challenging to detect, as the data variation between 2000 and 4000 seconds is very large.

\section{Discussions and Conclusion}
\label{sec:conclusion}

In this paper, we propose an online change point detection procedure for correlation structures. Both sum- and max-type statistics are proposed for non-sparse and sparse settings, respectively. These two types of statistics can also be combined in practice, providing more flexibility and efficiency for detection tasks. Simulation studies illustrate the performance of the combined method when the change in correlation structure varies from sparse to dense regimes. In addition, we propose to combine SMOTE and Knockoff techniques to increase sample efficiency, and the enhanced detection procedures have shown smaller detection delays in most simulation scenarios. Furthermore, an efficient sign-flip algorithm is proposed to select the threshold for our detection statistics. Theoretical approximations are also provided for ARL and EDD of detection procedures. More importantly, the proposed detection procedure is applied to forecast El Ni{\~n}o events. Our method can predict these events with one or two years in advance and achieves a state-of-the-art hit rate of exceeding 0.86 with a false alarm rate close to 0. There is immense societal benefit from these high-accuracy multi-year forecasts, as many human systems make decisions on this timescale. 

Given the superior performance in forecasting El Ni{\~n}o events, we believe that our methods can be applied to predict other important climate phenomena, especially those involving varying relationships between different regions of the Earth. The application of our proposed algorithm to forecasting La Ni{\~n}a events will be reported in our future work.

There are several possible extensions of our methods. A direct extension is to detect change points in the covariance structure of streaming large-dimensional data, including general covariance structure changes and special covariance structure changes. In addition, while we constructed the detection statistics based on \(\ell_1\) and \(\ell_\infty\) norms of the difference between two vectorized correlation matrices, other norms can be used to adapt to different circumstances.

\section*{Supplementary Material}

Proofs and additional numerical results are relegated to the online Supplementary Material.

\bibliographystyle{plainnat}
\bibliography{ref}

\newpage
\appendix

\section*{Supplementary Material}

\section{More figures}

\begin{figure}[ht]
	\centering    \includegraphics[width=0.8\textwidth,height=0.3\textwidth]{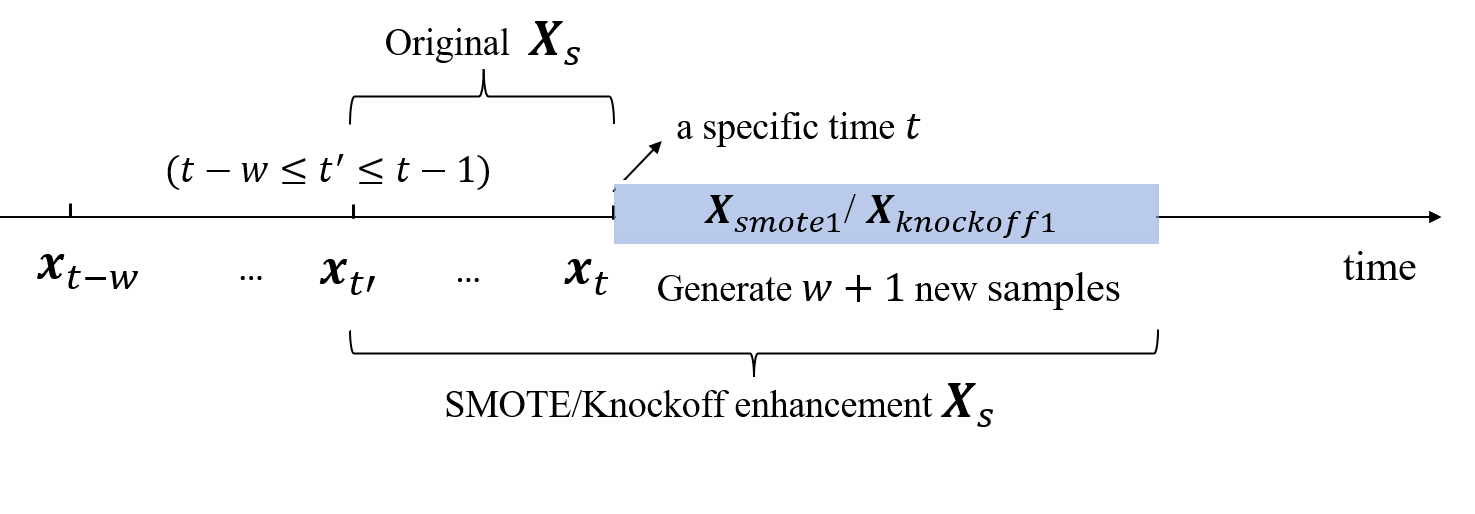}
	\caption{Illustration of SMOTE and Knockoff enhancement techniques.}
	\label{fig:SMOTE illustration}
\end{figure}

\begin{figure}
	\centering
	\includegraphics[width=0.95\textwidth]{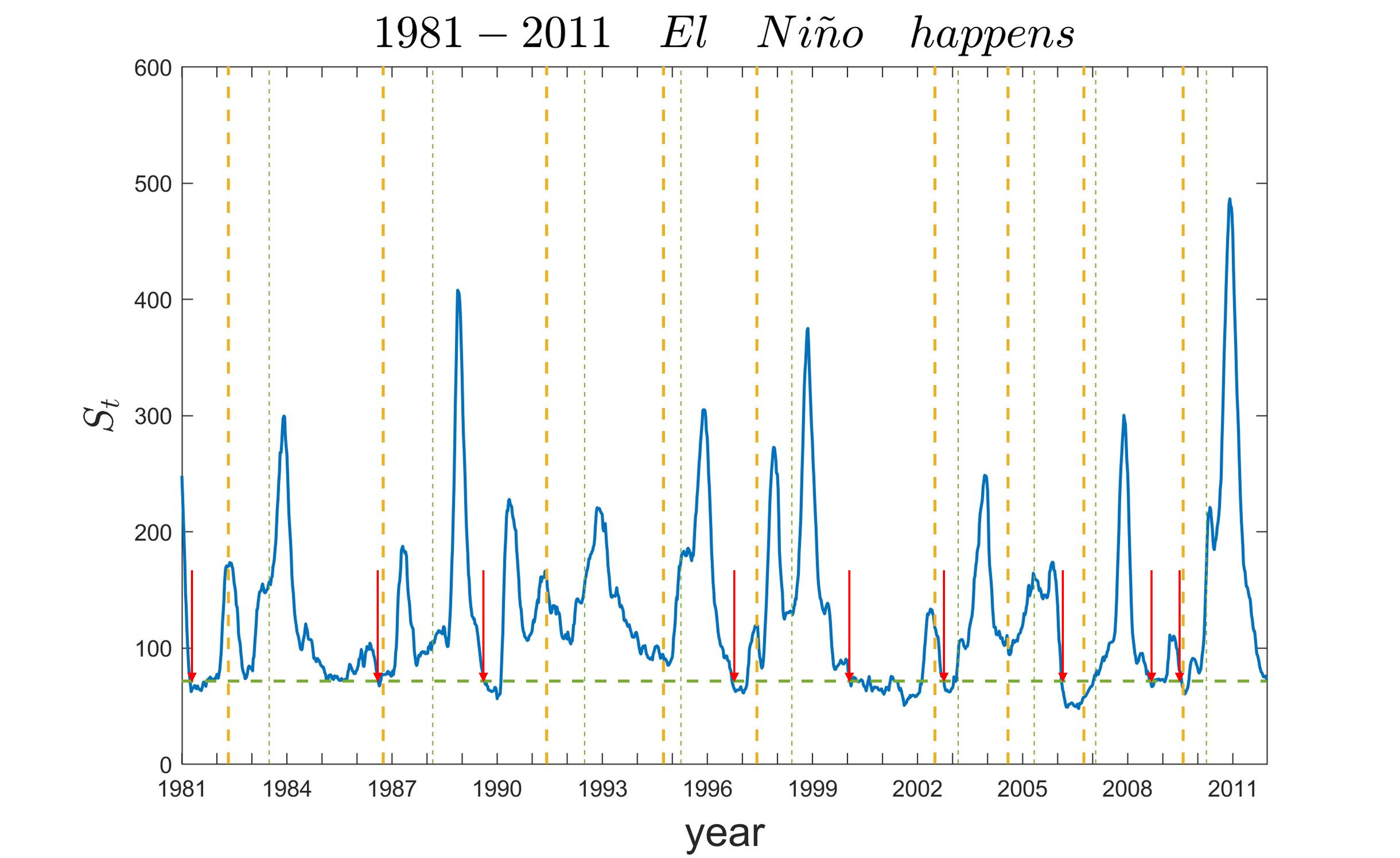}
	\caption{The prediction of El Ni{\~n}o events between 1981 and 2011.}
	\label{fig:climate_1981_2011}
\end{figure}


\section{Additional Numerical Results}
\label{appendix:plot}

\subsection{Auxiliary simulation 1}
\label{appendix:simu2}
\begin{figure}
	\centering
	\begin{minipage}[t]{0.3\linewidth}
		\centering
		\includegraphics[width=2in, height=1.8in,keepaspectratio]{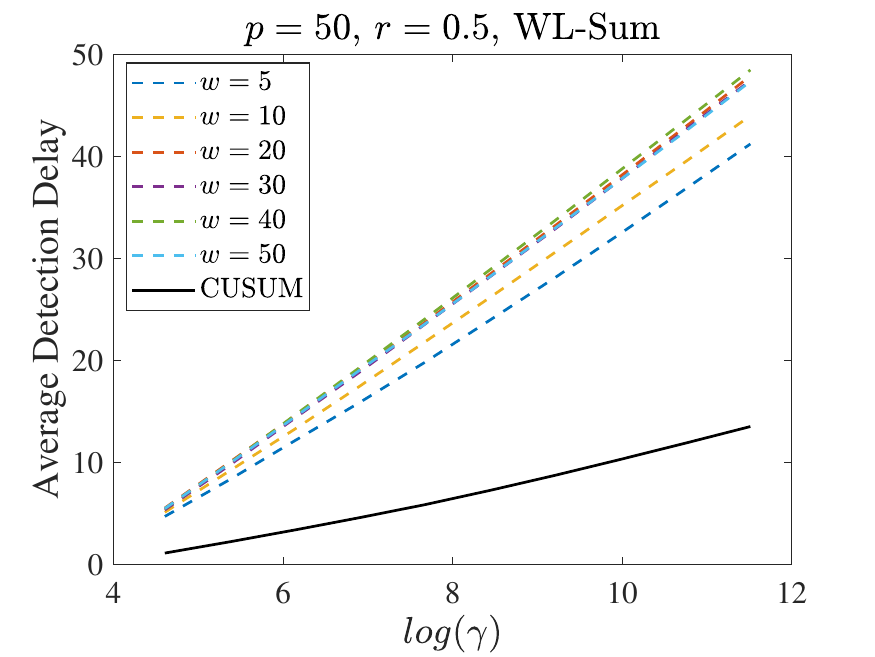}
		\label{fig:w_change_a}
	\end{minipage}
	\begin{minipage}[t]{0.3\linewidth}
		\centering
		\includegraphics[width=2in, height=1.8in,keepaspectratio]{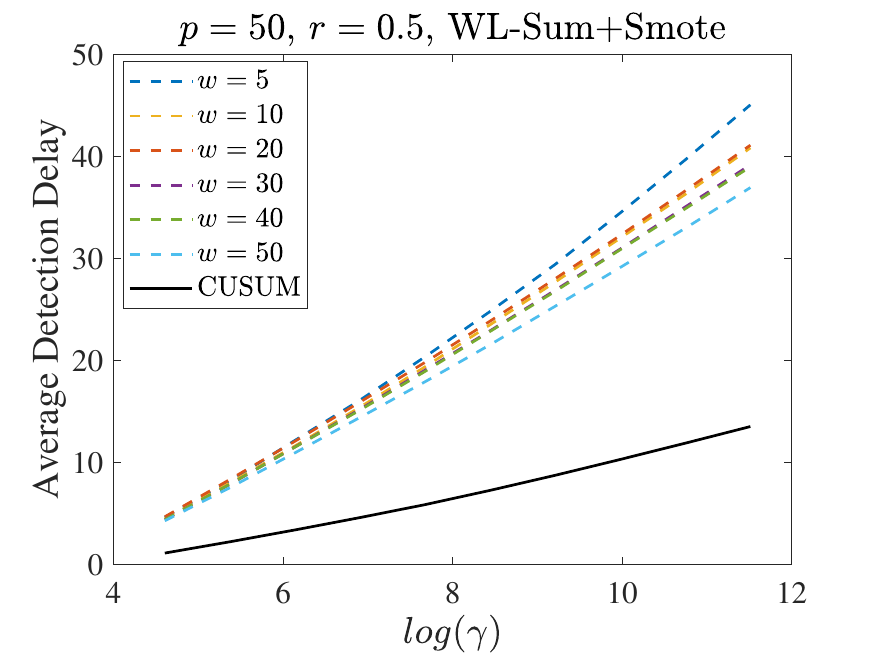}
		\label{fig:w_change_b}
	\end{minipage}
	\begin{minipage}[t]{0.3\linewidth}
		\centering
		\includegraphics[width=2in, height=1.8in,keepaspectratio]{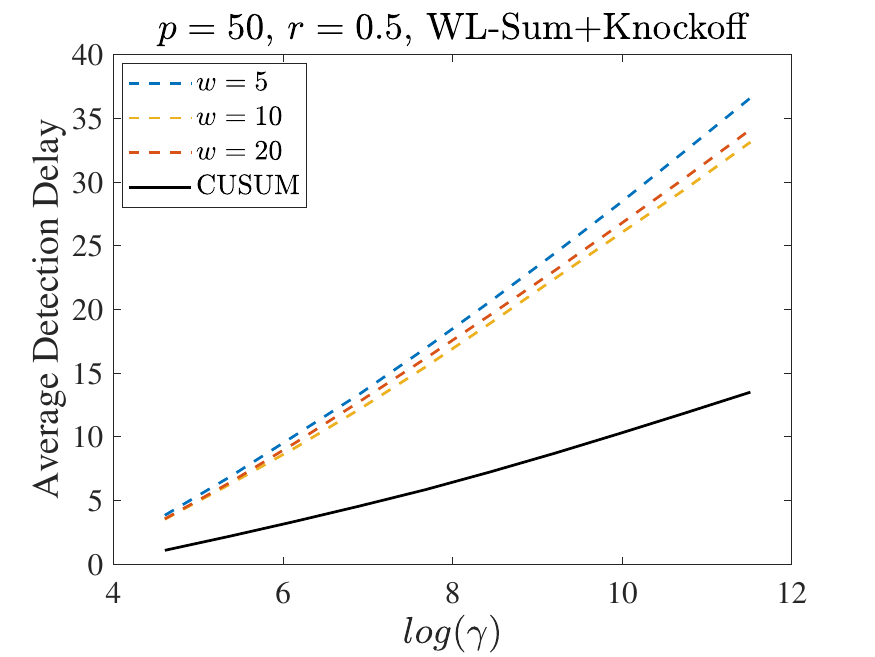}
		\label{fig:w_change_c}
	\end{minipage}
	
	\begin{minipage}[t]{0.3\linewidth}
		\centering
		\includegraphics[width=2in, height=1.8in,keepaspectratio]{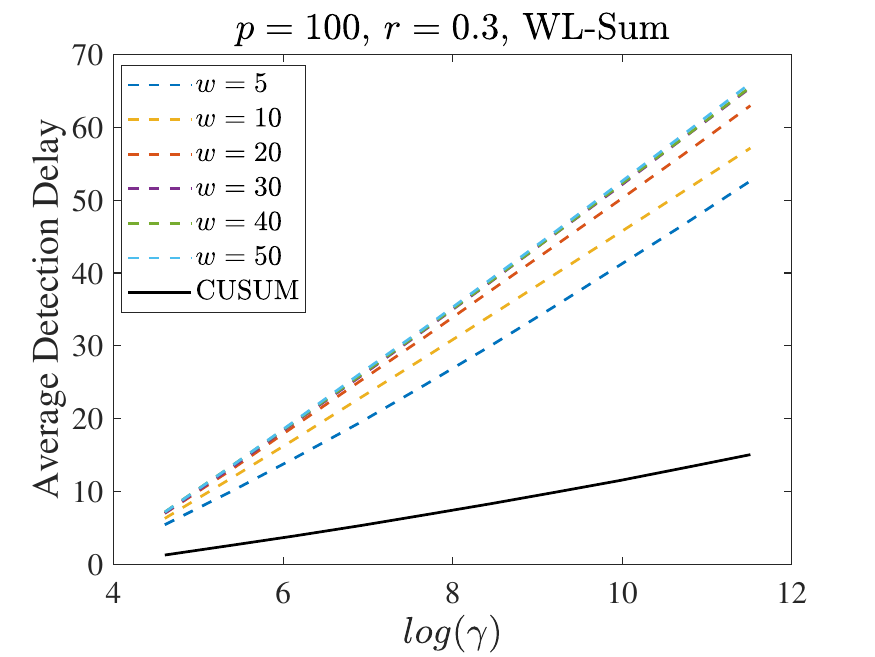}
		\label{fig:w_change_d}
	\end{minipage}
	\begin{minipage}[t]{0.3\linewidth}
		\centering
		\includegraphics[width=2in, height=1.8in,keepaspectratio]{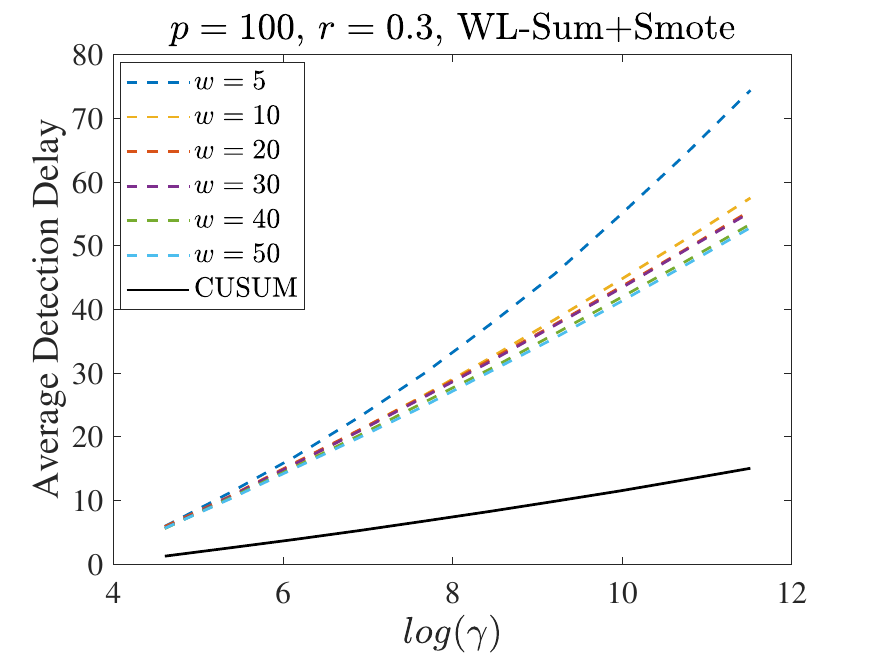}
		\label{fig:w_change_e}
	\end{minipage}
	\begin{minipage}[t]{0.3\linewidth}
		\centering
		\includegraphics[width=2in, height=1.8in,keepaspectratio]{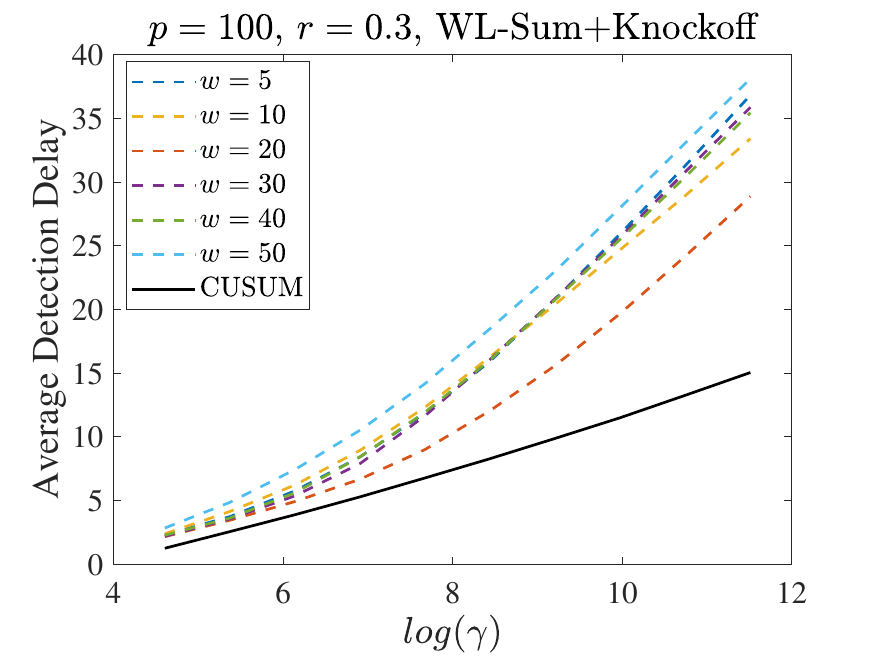}
		\label{fig:w_change_f}
	\end{minipage}

	\begin{minipage}[t]{0.3\linewidth}
		\centering
		\includegraphics[width=2in, height=1.8in,keepaspectratio]{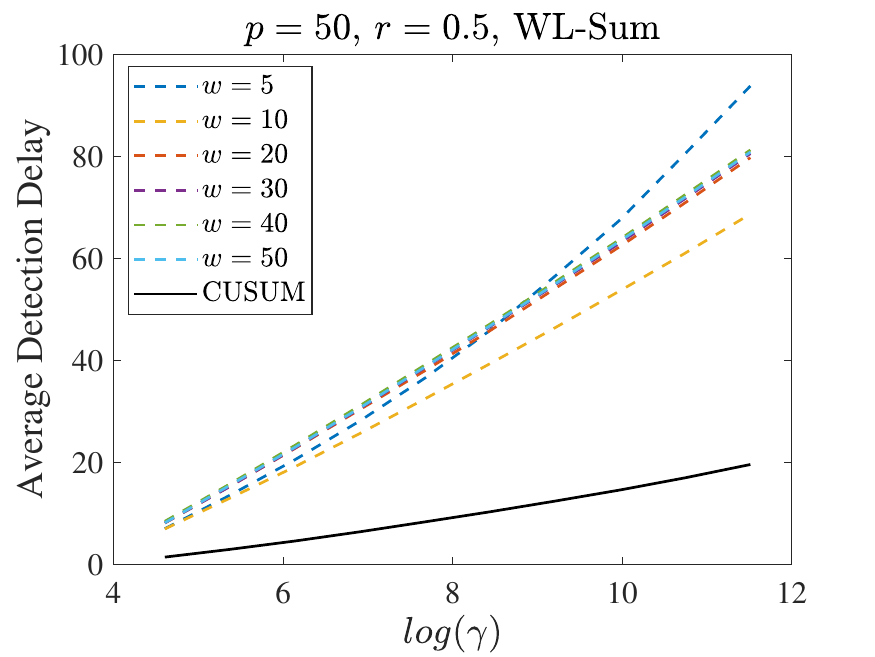}
		\label{fig:w_change_g}
	\end{minipage}
	\begin{minipage}[t]{0.3\linewidth}
		\centering
		\includegraphics[width=2in, height=1.8in,keepaspectratio]{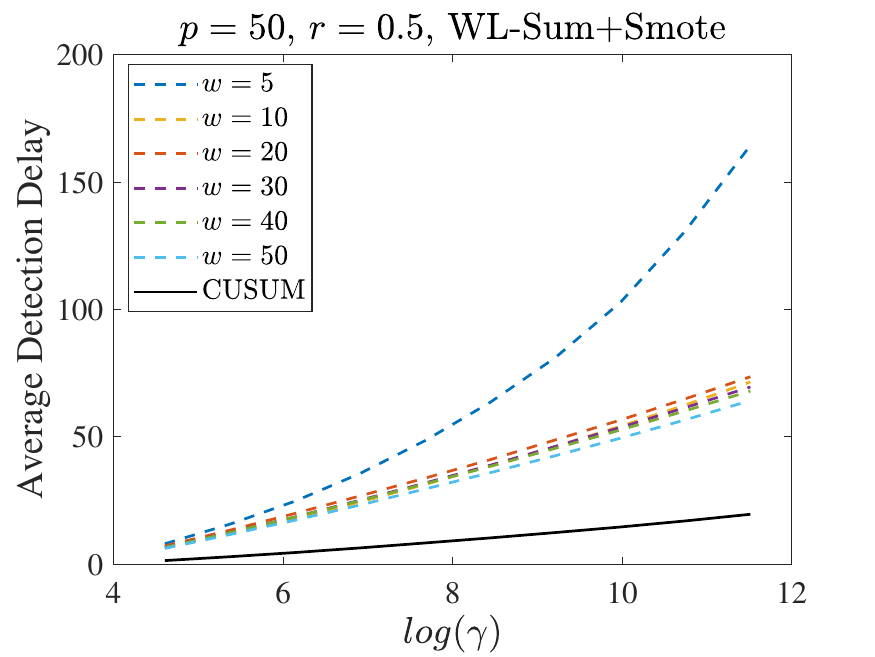}
		\label{fig:w_change_h}
	\end{minipage}
	\begin{minipage}[t]{0.3\linewidth}
		\centering
		\includegraphics[width=2in, height=1.8in,keepaspectratio]{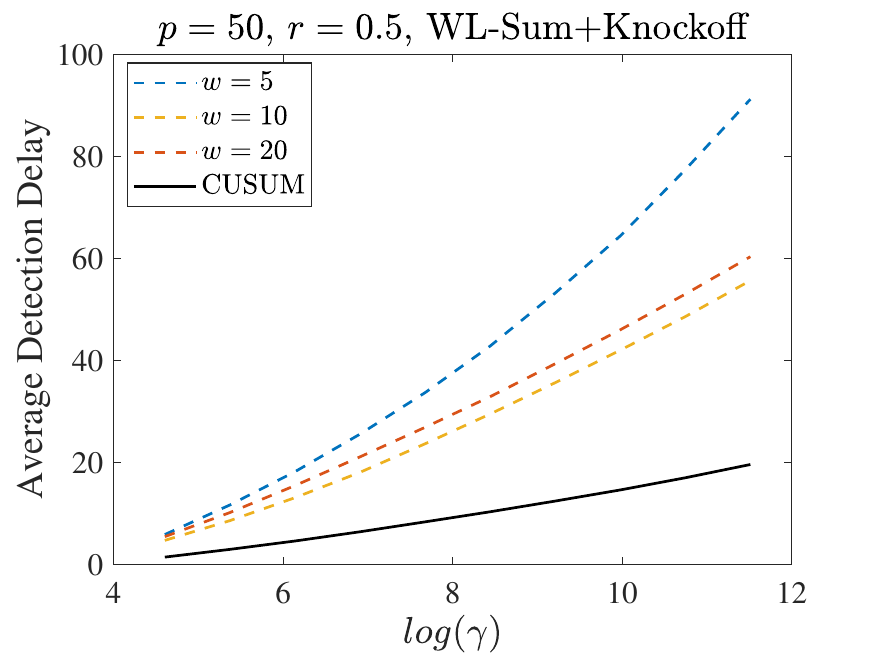}
		\label{fig:w_change_i}
	\end{minipage}
	
	\begin{minipage}[t]{0.3\linewidth}
		\centering
		\includegraphics[width=2in, height=1.8in,keepaspectratio]{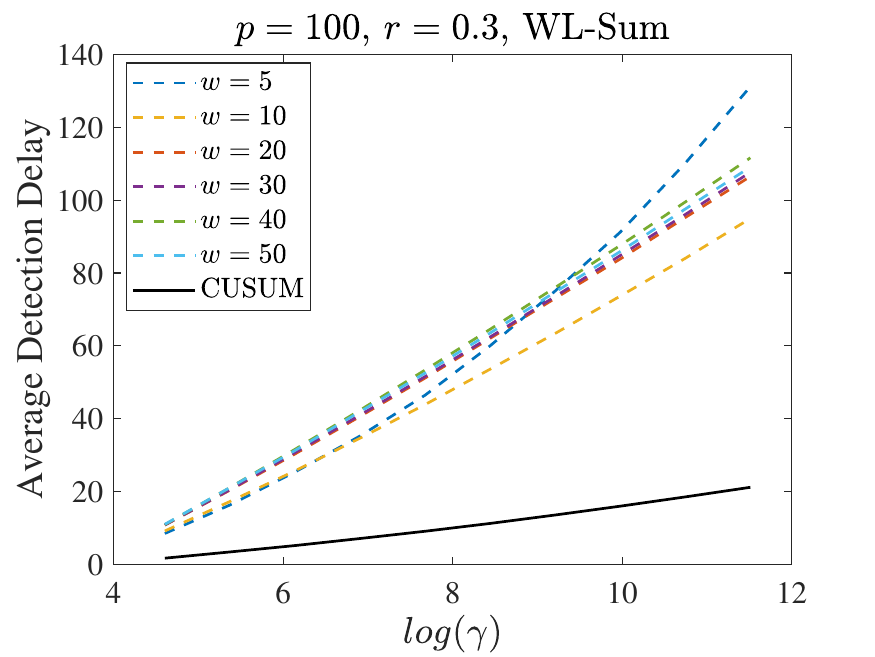}
		\label{fig:w_change_j}
	\end{minipage}
	\begin{minipage}[t]{0.3\linewidth}
		\centering
		\includegraphics[width=2in, height=1.8in,keepaspectratio]{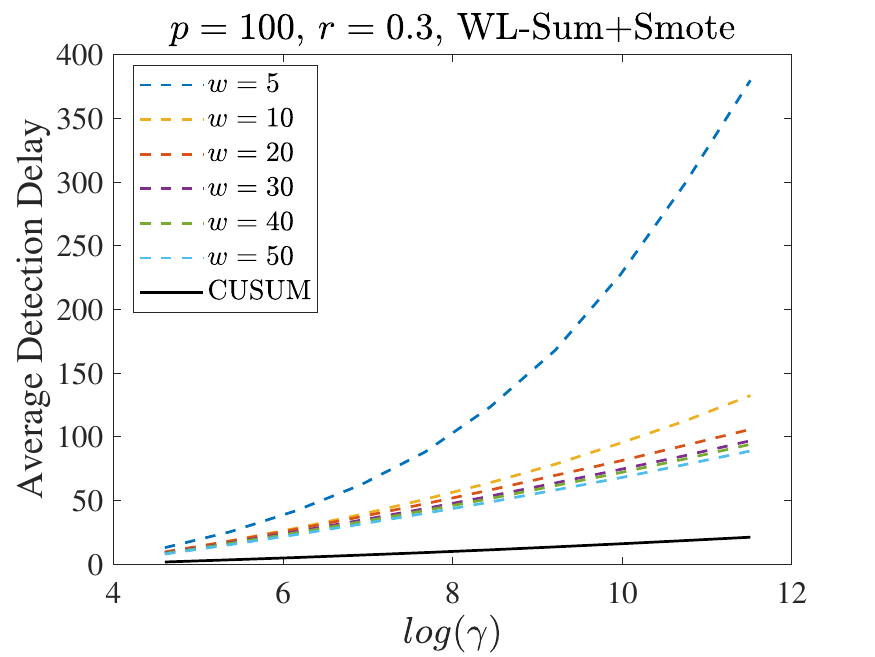}
		\label{fig:w_change_k}
	\end{minipage}
	\begin{minipage}[t]{0.3\linewidth}
		\centering
		\includegraphics[width=2in, height=1.8in,keepaspectratio]{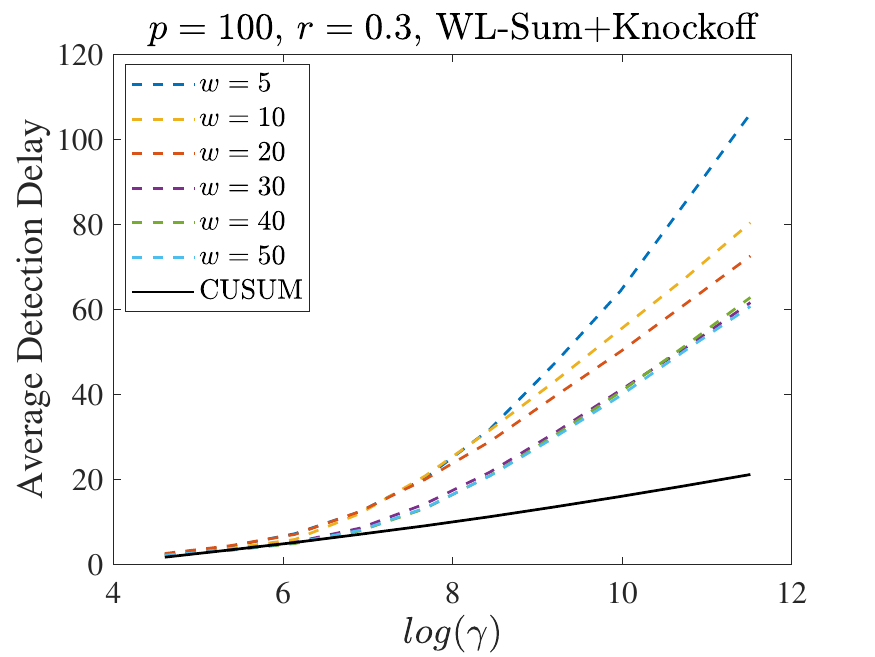}
		\label{fig:w_change_l}
	\end{minipage}
	\caption{Effects of different window sizes $w$. Case 1: first and second rows. Case 2: third and fourth rows.}
	\label{fig:w_change}
\end{figure}

A small simulation is conducted to illustrate the impact of different window sizes $w$ and the results are shown in Figure~\ref{fig:w_change}. We let $w$ varies from 5 to 50, and CUSUM again serves as the theoretical lower bound for detection delay. The first two rows represent data generated from the Case 1 setting defined in Section 5
, while the last two rows correspond to Case 2. For a fixed small ARL, all window sizes exhibit low detection delays, with smaller delays preferred for computational efficiency. However, as ARL increases, the detection delay for smaller window sizes can grow quadratically, necessitating the use of larger window sizes. Comparing the first two rows with the last two, it is evident that a larger window size is required for the more challenging detection scenario in Case 2.

\subsection{Auxiliary simulation 2}
\label{appendix:simu1}
A small simulation study is conducted to show the effects of the SMOTE and knockoff enhancement methods on correlation estimation. Two settings are considered under different ground truth correlation matrices, \begin{itemize}
	\item Setting 1: the true correlation matrix is an identity matrix $\boldsymbol{R}=\mathbf{I}_p$;
	\item Setting 2: the true correlation matrix is $\boldsymbol{R}=[\rho(i,j)]_{i, j=1,\ldots,p}$ with $\rho(i,j) = 1$ for $i=j$, $\rho(i,j)=0.5$ for $i\neq j$, $1\le i,j \le p$.
\end{itemize}  
Table~\ref{table:two tests} compares the means and standard deviations of $\vert \vert \boldsymbol{R} - \boldsymbol{\hat{R}}\vert \vert_F$, where $\boldsymbol{\hat{R}}$ is the sample correlation matrix calculated from the original samples (of size $w$), the SMOTE enhancement samples and the knockoff enhancement samples, respectively. Different combinations of data dimension $p$ and sample size $w$ are considered. Compared to the sample correlation matrix obtained from the original samples, knockoff method can enhance the accuracy of correlation estimation, while the SMOTE method cannot (but the bias is small). This is due to the construction principle of knockoff variables; they keep the same correlation structure with original ones, while SMOTE method only makes sure new SMOTE variables have the same distribution with the original ones.

\begin{table}
	\caption{Mean and standard deviation (Std) of $\vert \vert \boldsymbol{R}_0 - \boldsymbol{\hat{R}}_0\vert \vert_F$.}
	\centering
	\label{table:two tests}
	\begin{tabular}{ccccc|ccc}
		\hline
		&  & \multicolumn{3}{c}{Setting 1} & \multicolumn{3}{c}{Setting 2} \\ \cline{3-8} 
		&  & Original & SMOTE & Knockoff & Original & SMOTE & Knockoff \\ \hline
		\multirow{2}{*}{$p=50,w=20$} & Mean & 11.3541 & 11.9440 & 11.2093 & 8.6080 & 8.9213 & 8.5133 \\
		& Std & 0.2215 & 0.2816 & 0.2224 & 1.6088 & 1.4629 & 1.6260 \\ \hline
		\multirow{2}{*}{$p=100,w=20$} & Mean & 22.8313 & 23.9804 & 22.6955 & 17.1047 & 17.8367 & 17.0168 \\
		& Std & 0.2192 & 0.3524 & 0.2207 & 2.8597 & 2.6303 & 2.8764 \\ \hline
		\multirow{2}{*}{$p=300,w=20$} & Mean & 68.7075 & 72.1942 & 68.5769 & 51.7309 & 53.7253 & 51.6465 \\
		& Std & 0.1981 & 0.7886 & 0.1987 & 9.6532 & 8.7277 & 9.6662 \\ \hline
		\multirow{2}{*}{$p=300,w=50$} & Mean & 42.7811 & 45.7033 & 42.7045 & 32.8976 & 34.6930 & 32.8487 \\
		& Std & 0.1407 & 0.4921 & 0.1409 & 6.2258 & 5.2383 & 6.2366 \\ \hline
		\multirow{2}{*}{$p=300,w=100$} & Mean & 30.0997 & 32.5922 & 30.0469 & 22.5139 & 25.3277 & 22.4771 \\
		& Std & 0.1067 & 0.3467 & 0.1067 & 3.3156 & 4.0502 & 3.3196 \\ \hline
	\end{tabular}
\end{table}

\subsection{Auxiliary simulation 3}
In the simulation results in Section 5
, both the SMOTE and the knockoff enhanced methods have shorter detection delays compared to the original WL-Sum procedure. We use a simple simulation to illustrate the insights behind the effectiveness of SMOTE and knockoff enhanced detection procedures. The pre-change correlation matrix is $\boldsymbol{R}_0= \mathbf{I}_p$, and the post-change correlation matrix has entries $\rho_1(i,j) = 1$ for $i=j$ and $\rho_1(i,j) = 0.5$ for $1\le i \neq j \le p$. We simulate the data of length $M=1000$ under both the pre- and post-change regimes separately, and use the window size $w=20$. 


For $t$ ranging from $2$ to $1000$ (999 time steps in total), the number of times when $t'_{\text{max}}:=\argmax_{t-w\le t'\le t-1} S_t^{(\text{WL-sum})}$ equals to $t-w, t-w+1, t-w+2,$ and $t'_{\text{max}}< t - w/2$, are presented in Table~\ref{table:count_num}. 
Under the pre-change regime, both SMOTE and knockoff-enhanced methods show more than 91.9\% of the times when $t'_{\text{max}}=t-w$, making their estimated results more {\it credible}. In contrast, the original method has a relatively low number of times when $t'_{\text{max}}=t-w$. In addition, the knockoff-enhanced method consistently finds the maximum of test statistics at $t'_{\text{max}} = t-w$. A similar pattern is also observed under the post-change regime. These results align with the detection delay findings presented in Section 5
, where the knockoff-enhancement method exhibits the smallest detection delay, followed by the SMOTE-enhancement method, compared to the detection delay of the original WL-Sum statistics $S_t^{(\text{WL-sum})}$.

\begin{table}
	\centering
	\caption{Number of times of locations of $t'_{\text{max}}$ among 999 times.}
	\begin{tabular}{ccccccc}
		\hline
		&  &  & $t-w$ & $ t-w+1$ & $ t-w+2$ & $ < t-w/2$ \\ \hline
		\multirow{6}{*}{Pre-change} & \multirow{3}{*}{$p=50$} & SMOTE & 919 & 73 & 7 & 999 \\
		&  & Knockoff & 999 & 0 & 0 & 999 \\
		&  & Original & 362 & 165 & 112 & 964 \\ \cline{2-7} 
		& \multirow{3}{*}{$p=300$} & SMOTE & 982 & 16 & 1 & 999 \\
		&  & Knockoff & 999 & 0 & 0 & 999 \\
		&  & Original & 914 & 72 & 13 & 999 \\ \hline
		\multirow{6}{*}{Post-change} & \multirow{3}{*}{$p=50$} & SMOTE & 852 & 110 & 30 & 999 \\
		&  & Knockoff & 998 & 1 & 0 & 999 \\
		&  & Original & 247 & 138 & 94 & 930 \\ \cline{2-7} 
		& \multirow{3}{*}{$p=300$} & SMOTE & 894 & 78 & 16 & 999 \\
		&  & Knockoff & 999 & 0 & 0 & 999 \\
		&  & Original & 311 & 135 & 111 & 953 \\ \hline
	\end{tabular}
	\label{table:count_num}
\end{table}

\subsection{Auxiliary simulation 4}

\begin{figure}[ht!]
	\footnotesize
	\centering
	\includegraphics[width=0.9\textwidth,height=0.5\textwidth]{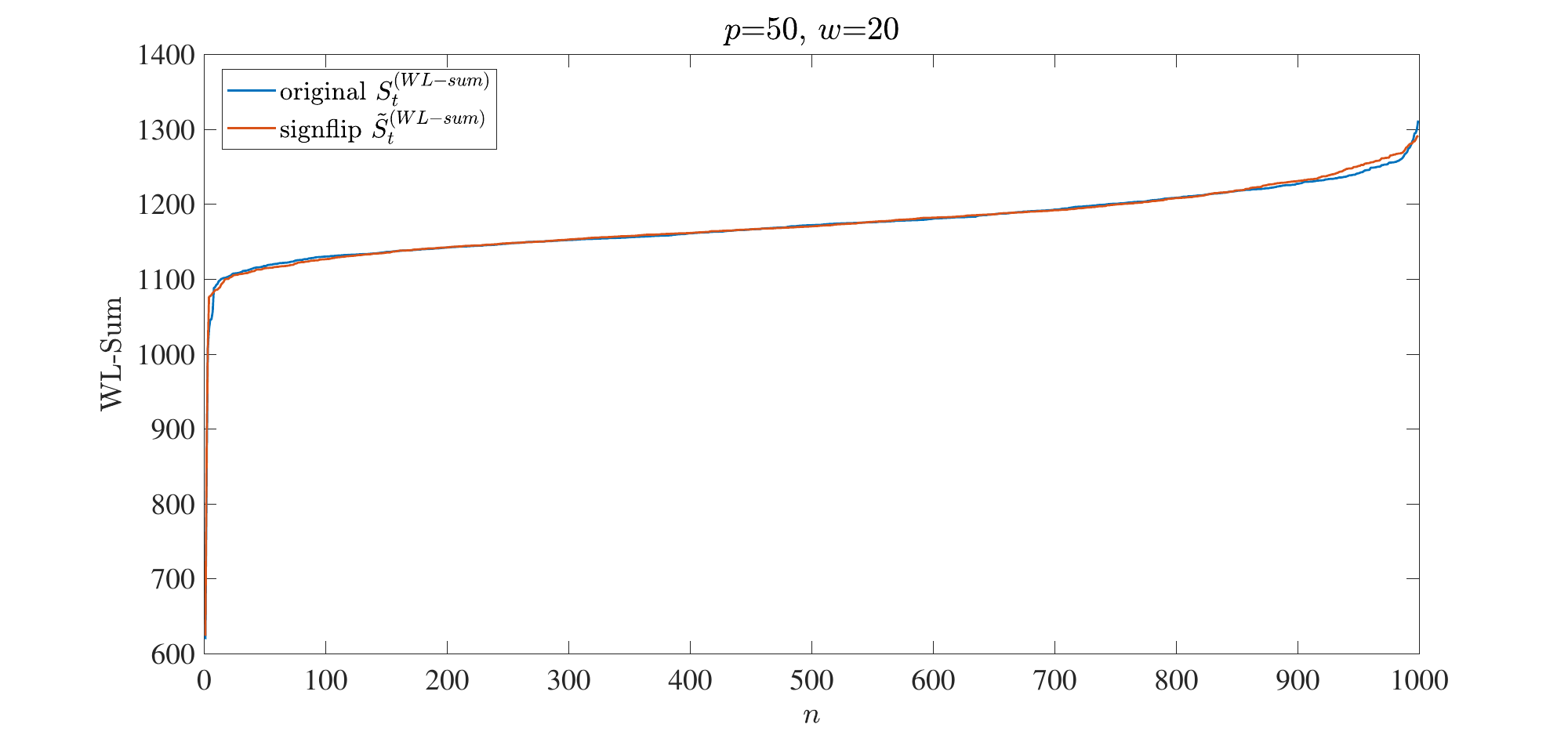}
	\caption{The empirical distributions of increasing-order elements of $S_t$ and $\Tilde{S}_t$.}
	\label{fig:signflip_0}
\end{figure}

We illustrate that the sign-flip permutation in Algorithm 2 does not change the distribution of detection statistic $S_t$ under the pre-change measure through a simulation example. We draw pre-change samples from the standard Gaussian distribution with dimension $p=50$. We set the window size $w=20$, the historical data size $H=100$, and the length of the data sequence $M=1000$. We plot the detection statistic $S_t$ in (3.11)
calculated from the original data and that of $\Tilde{S}_t$ from the sign-flipped data in increasing order in Figure~\ref{fig:signflip_0}, and it is obvious that the two distributions are almost identical.

\subsection{Auxiliary simulation 5}
The results of our method for the data generated from Student's $t$ distribution are presented in Figure~\ref{fig:case1,case2,t5}, which compares the ADDs of WL-Sum and its corresponding enhanced methods in Cases 1 and Case 2. Our methods show strong and robust performance under the Student's $t$ distribution as well.

Figure~\ref{fig:case4} compares the ADDs of WL-Sum and its corresponding enhanced methods in Case 4, where $\boldsymbol{R}_0$ is not an identity matrix. Figure~\ref{fig:case4} (a) 
presents the results when the data is from the normal distribution, while Figure~\ref{fig:case4} (b) illustrates the results when the data are from the Student's $t$ distribution. For the general case of a correlation change, our methods still perform well; however, for the Student's $t$ distribution, the performance of our method degrades due to the increased detection difficulty.

\begin{figure}
	\centering
	\subfigure[]{\includegraphics[width=0.45\textwidth]{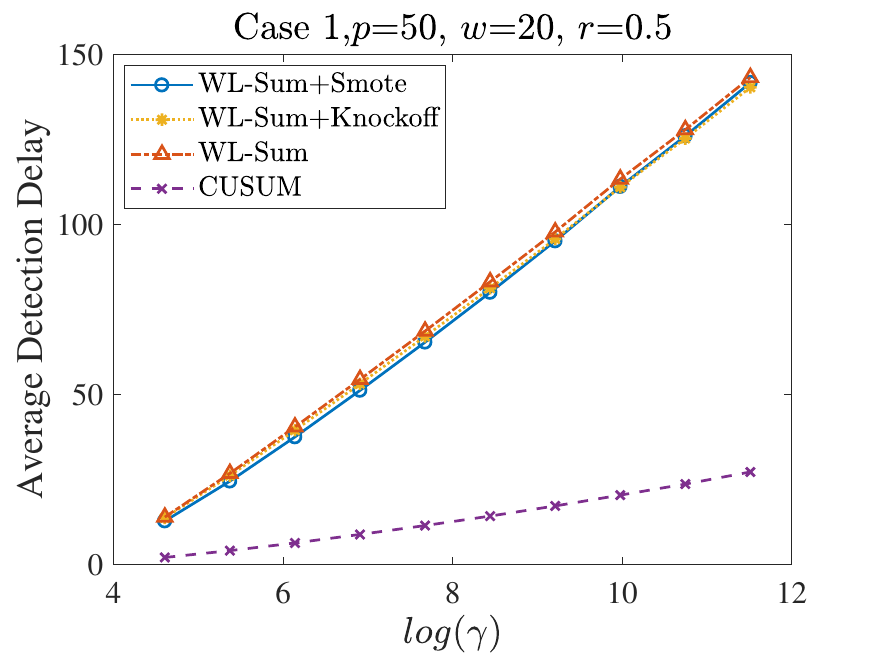}}
	\subfigure[]{\includegraphics[width=0.45\textwidth]{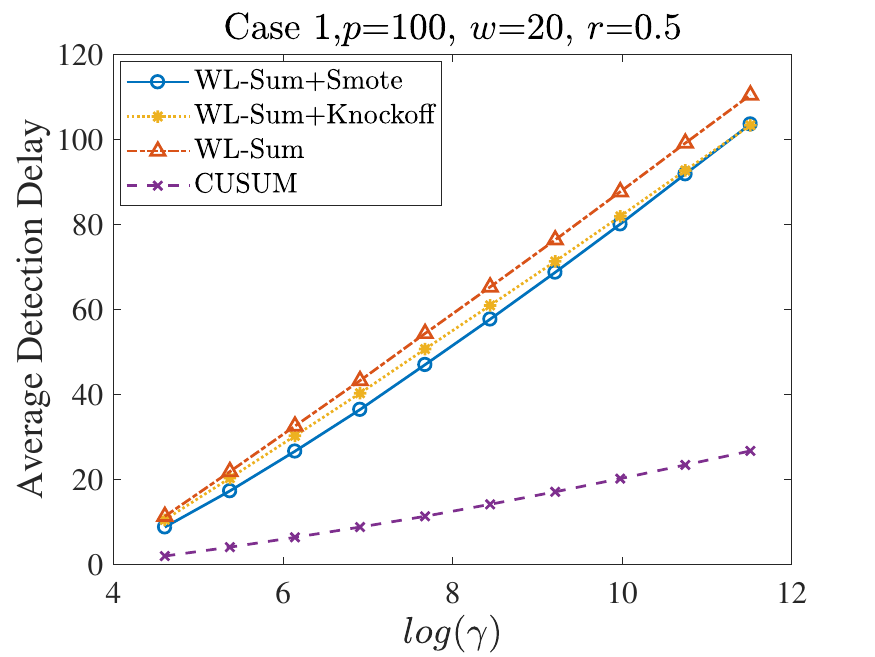}}
	\subfigure[]{\includegraphics[width=0.45\textwidth]{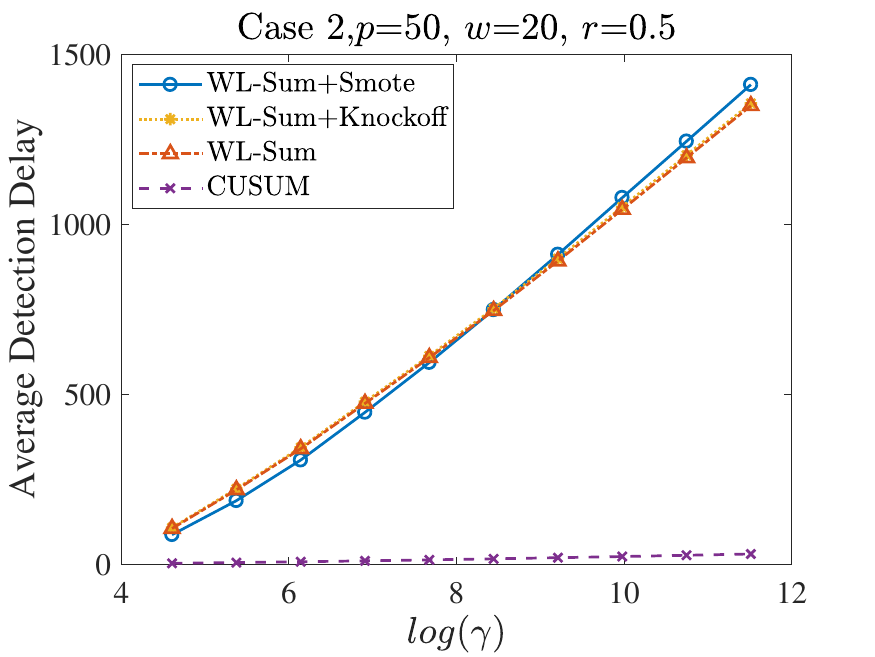}}
	\subfigure[]{\includegraphics[width=0.45\textwidth]{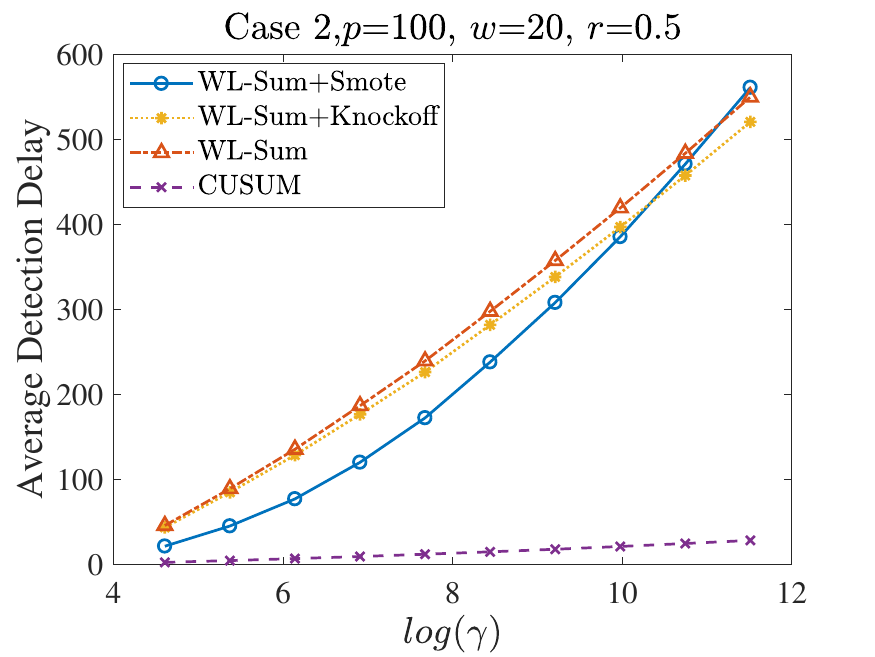}}
	\caption{Comparison of ADDs of WL-Sum, WL-Sum+SMOTE, WL-Sum+Knockoff and CUSUM under Student's $t$ distribution: (a) Case 1, $p=50$; (b) Case 1, $p=100$; (c) Case 2, $p=50$; (d) Case 2, $p=100$. }
	\label{fig:case1,case2,t5}
\end{figure}
\begin{figure}
	\centering
	\subfigure[]{\includegraphics[width=0.45\textwidth]{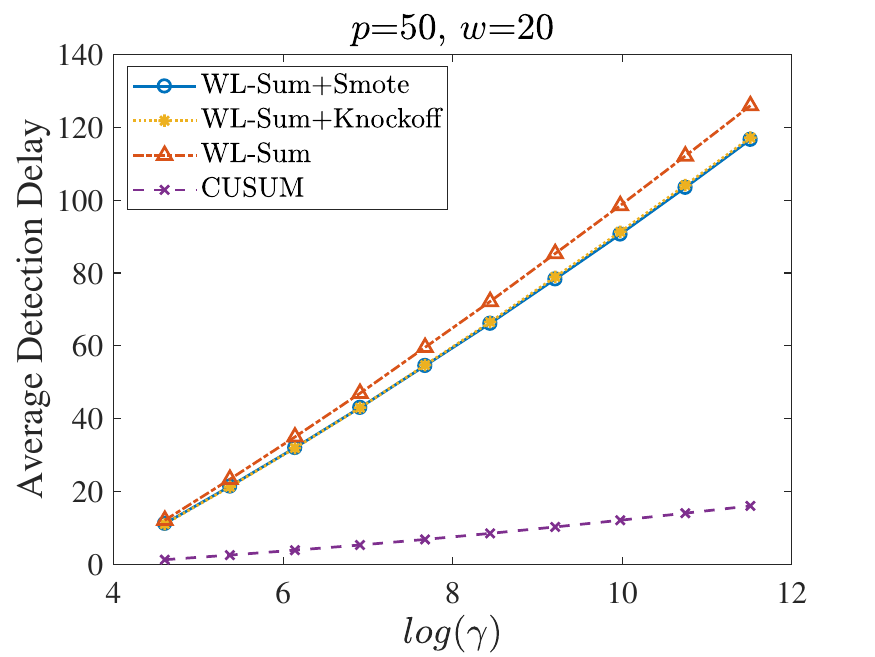}}
	\subfigure[]{\includegraphics[width=0.45\textwidth]{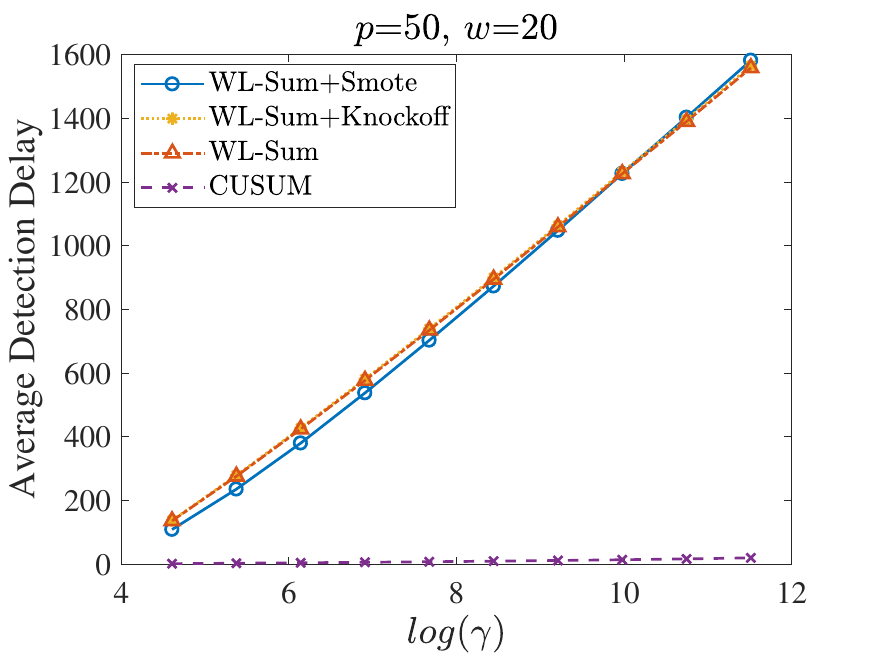}}
	\caption{Comparison of ADDs of WL-Sum, WL-Sum+SMOTE, WL-Sum+Knockoff and CUSUM under Case 4 for: (a) Normal distribution, and (b) Student's $t$ distribution.}
	\label{fig:case4}
\end{figure}

\newpage

\section{Proofs} \label{proofs}
\subsection{Proof of Lemma~\ref{lem:mean_V_zero_mean}}

{\bf Proof.}
For $\boldsymbol{v}_{1,t}(i,j)$, we have
\[
\begin{aligned}
	&(\hat{\rho}_0(i,j) -\hat{\rho}_{1:t}(i,j))^2 =\left(\frac{1}{H+1} \sum_{k=-H}^0 y_{k i} y_{k j}-\frac{1}{t} \sum_{l=1}^t y_{l i} y_{l j}\right)^2 \\
	= & \left(\frac{1}{H+1} \sum_{k=-H}^0 y_{k i} y_{k j}\right)^2+\left(\frac{1}{t} \sum_{l=1}^t y_{li} y_{l j}\right)^2-2 \cdot \frac{1}{(H+1) \cdot t} \sum_{k=-H}^0 \sum_{l=1}^t y_{k i} y_{k j} y_{l i} y_{l j} \\
	= & \frac{1}{(H+1)^2} \sum_{k=-H}^0 y_{k_i}^2 y_{k_j}^2+\frac{1}{(H+1)^2} 
	\sum_{k_1=-H}^0 \sum_{k_2=-H, k_2\neq k_1}^0 y_{k_1 i} y_{k_1 j} y_{k_2 i} y_{k_2 j} +\frac{1}{t^2} \sum_{l=1}^t y_{l i}^2 y_{l j}^2 \\
	& +\frac{1}{t^2} 
	\sum_{l_1=1}^t \sum_{l_2=1, l_2\neq l_1}^t y_{l_1 i} y_{l_1 j} y_{l_2 i} y_{l_2 j}  -\frac{2}{(H+1)t} \sum_{k=-H}^0 \sum_{l=1}^t y_{k i} y_{k j} y_{l i} y_{l j}. 
\end{aligned}
\]
Under the assumption that $\bE(x_{ki})=0$ and $\bE(x_{ki}^2)=1$ for $i=1,\ldots,p, k=1,2,\ldots$, we have $\bE_{\infty}(x_{ki}x_{kj}) = \hat{\rho}_0(i,j) = \rho_0(i,j)$, $\bE_1(x_{ki}x_{kj}) = \bE_1(y_{ki}y_{kj}) = \rho_1(i,j)$, then we take expectation for every term, 
\[
\begin{aligned}
	& \bE\left(\hat{\rho}_0(i,j)\right)^2=\frac{1}{H+1} \beta_{20}+\frac{H}{H+1} \cdot \rho_0^2(i, j), \\
	& \bE\left(\hat{\rho}_1(i,j)\right)^2=\frac{1}{t} \beta_{21}+\frac{t-1}{t} \rho_1^2(i, j), \\
	& \bE\left(\frac{2}{(H+1)t} \sum_{k=-H}^0 \sum_{l=1}^t y_{k_i} y_{k,} y_{l i} y_{l j}\right)= 2\rho_0(i, j) \cdot \rho_1(i, j),\\
\end{aligned}
\]
finally, we have,
\[
\bE\left(\hat{\rho}_0(i,j) - \hat{\rho}_1(i,j)\right)^2 = (\rho_0(i,j) - \rho_1(i,j) )^2 + \frac{1}{H+1}(\beta_{20} - \rho_0^2(i,j)) + \frac{1}{t}(\beta_{21} - \rho_1^2(i,j)).
\\
\]

\subsection{Proof of Remark~\ref{lem:mean_V}}
{\bf Proof.}
Under the assumption that $var(x_{ki})=1$, we have $\boldsymbol{y}_k = \boldsymbol{x}_k - \boldsymbol{\Bar{x}}_{-H:0}$ for $k=-H,\ldots,0$ and $\boldsymbol{y}_k = \boldsymbol{x}_k - \boldsymbol{\Bar{x}}_{1:t}$ for $k=1,\ldots,t$. Due to the unbiasedness of the sample covariance, we have $\bE_\infty[\hat{\rho}_0(i,j) ] = \rho_0(i,j)$ and $\bE_1[   \boldsymbol[{\hat{\rho}}_{1:t}(i,j)] = \rho_1(i,j)$. We first write 
\[
\begin{aligned}
	& (\hat{\rho}_0(i,j) -\hat{\rho}_{1:t}(i,j))^2 = (\hat{\rho}_0(i,j))^2  + (\hat{\rho}_{1:t}(i,j))^2 - 2 \hat{\rho}_0(i,j)\hat{\rho}_{1:t}(i,j).
\end{aligned}
\]
We have $\bE[\hat{\rho}_0(i,j)\hat{\rho}_{1:t}(i,j)] = \rho_0(i,j)\rho_1(i,j)$. For the term $\bE[(\hat{\rho}_0(i,j))^2]$, we can assume $\bE(x_{ki})=0$ for simplicity under both the pre- and post-change regimes, we have (the term $\bE_1[(\hat{\rho}_{1:t}(i,j))^2]$ can be calculated similarly)
\[
\begin{aligned}
	&(\hat{\rho}_0(i,j))^2 =\frac{1}{H^2} \left(\sum_{k=-H}^0 y_{k i} y_{k j}\right)^2 =  \frac{1}{H^2} \sum_{k=-H}^0 y_{ki}^2 y_{kj}^2+\frac{1}{H^2} 
	\sum_{k_1=-H}^0 \sum_{k_2=-H, k_2\neq k_1}^0 y_{k_1 i} y_{k_1 j} y_{k_2 i} y_{k_2 j}.
\end{aligned}
\]
For the first term, we have
\[
\begin{aligned}
	y_{k_i}^2 y_{k j}^2&=\left(x_{k i}-\frac{1}{H+1} \sum_{l=-H}^0 x_{l i}\right)^2\left(x_{k j}-\frac{1}{H+1} \sum_{l=-H}^0 x_{l j}\right)^2 \\
	& =\left[x_{k i}^2-\frac{2}{H+1} x_{k i} \sum_{l=-H}^0 x_{li}+\frac{1}{(H+1)^2}\left(\sum_{l=-H}^0 x_{l i}\right)^2\right] \\
	&\quad \cdot {\left[x_{k j}^2-\frac{2}{H+1} x_{k j} \sum_{l=-H}^0 x_{l j}+\frac{1}{(H+1)^2}\left(\sum_{l=-H}^0 x_{l j}\right)^2\right]} \\
	& =x_{k i}^2 x_{k j}^2-\frac{2}{H+1} x_{k i}^2 x_{k j} \sum_{i=-H}^0 x_{l j}+\frac{1}{(H+1)^2} x_{k i}^2\left(\sum_{l=-H}^0 x_{l j}\right)^2 \\
	&\quad -\frac{2}{H+1} x_{k i} x_{k j}^2 \sum_{l=-H}^{0} x_{li}+\frac{4}{(H+1)^2} x_{k i} x_{k j}\left(\sum_{l=-H}^0 x_{li}\right)\left(\sum_{l=-H}^0 x_{l j}\right) \\
	& \quad-\frac{2}{(H+1)^3} x_{k i}\left(\sum_{l=-H}^0 x_{l i}\right)\left(\sum_{l=-H}^0 x_{l j}\right)^2+\frac{1}{(H+1)^2} x_{k j}^2\left(\sum_{l=-H}^0 X_{l i}\right)^2 \\
	&\quad -\frac{2}{(H+1)^3} x_{k j}\left(\sum_{l=-H}^0 x_{l j}\right)\left(\sum_{l=-H}^0 x_{l i}\right)^2+\frac{1}{(H+1)^4}\left(\sum_{l=-H}^0 x_{l i}\right)^2\left(\sum_{l=-H}^0 X_{l j}\right)^2. 
\end{aligned}
\]
Note that $\bE_\infty[x_{k i}x_{k j}]=\rho_0(i,j)$ and $\bE_\infty[x^2_{k i}]=1$ (after assuming $x_{ki}$ is zero-mean and unit-variance). We define the following notations: 
\[
\begin{aligned}
	&A_1 = x_{k i}^2 x_{k j}^2,\quad A_2=-\frac{2}{H+1} x_{k i}^2 x_{k j} \sum_{i=-H}^0 x_{l j}, \quad A_3 =\frac{1}{(H+1)^2} x_{k i}^2\left(\sum_{l=-H}^0 x_{l j}\right)^2,\\
	&A_4 =-\frac{2}{H+1} x_{k i} x_{k j}^2 \sum_{l=-H}^{0} x_{li}, \quad A_5 =\frac{4}{(H+1)^2} x_{k i} x_{k j}\left(\sum_{l=-H}^0 x_{li}\right)\left(\sum_{l=-H}^0 x_{l j}\right),\\
	&A_6 =-\frac{2}{(H+1)^3} x_{k i}\left(\sum_{l=-H}^0 x_{l i}\right)\left(\sum_{l=-H}^0 x_{l j}\right)^2, \quad A_7 =\frac{1}{(H+1)^2} x_{k j}^2\left(\sum_{l=-H}^0 X_{l i}\right)^2,\\
	&A_8 =-\frac{2}{(H+1)^3} x_{k j}\left(\sum_{l=-H}^0 x_{l j}\right)\left(\sum_{l=-H}^0 x_{l i}\right)^2, \quad A_9 = \frac{1}{(H+1)^4}\left(\sum_{l=-H}^0 x_{l i}\right)^2\left(\sum_{l=-H}^0 X_{l j}\right)^2.
\end{aligned}
\]
It is obvious that $y_{ki}^2 y_{kj}^2$ is decomposed into a summation from $A_1$ to $A_9$, that is, $y_{ki}^2 y_{kj}^2 = A_1 + A_2 + \cdots + A_8 + A_9$, then if we take the expectation of each term, the summation will be $\bE(y_{ki}^2 y_{kj}^2)$. The detailed expectations for each term are as follows.
{\footnotesize 	
	\begin{align*}
		& \bE(A_1)=\beta_{20}, \bE(A_2)=-\frac{2}{H+1} \beta_{20}, \\
		& \bE(A_3)=E\left(\frac{1}{(H+1)^2} x_{k i}^2\left(\sum_{l=-H}^0 x_{l j}^2+\sum_{l_1=-H}^0 \sum_{l_2=-H,l_1 \neq l_2}^0 
		x_{l_1 j} x_{l_2 j}\right)\right)\\
		& \quad=\bE\left(\frac{1}{(H+1)^2}\left[x_{k i}^2 x_{k j}^2+\sum_{l \neq k=-H}^0 x_{k i}^2 x_{l j}^2\right]\right)=\frac{1}{(H+1)^2} \beta_{20}+\frac{H}{(H+1)^2} \cdot 1 ,\\
		& \bE(A_4)=E\left(-\frac{2}{H+1} x_{k i} x_{k j}^2 \sum_{l=-H}^0 x_{l i}\right)=E\left(-\frac{2}{H+1} x_{k i}^2 x_{k j}^2\right)=-\frac{2}{H+1} \beta_{20}, \\
		& \bE(A_5)=E\left(\frac{4}{(H+1)^2} x_{k i} x_{k j}\left(\sum_{l=-H}^0 x_{l i}\right)\left(\sum_{l=-H}^0 X_{l j}\right)\right) \\
		&\quad =E\left(\frac{4}{(H+1)^2} x_{k i}^2 x_{k j}^2+\frac{4}{(H+1)^2} \sum_{l\neq k=-H}^0 x_{k i} x_{k j} x_{l i} x_{l j}\right) \\
		&\quad =\frac{4}{(H+1)^2} \beta_{20}+\frac{4 H}{(H+1)^2} \rho_0^2(i, j), \\
		& \bE(A_6)=E\left(-\frac{2}{(H+1)^3} x_{k i}\left(\sum_{l=-H}^0 x_{l i}\right)\left(\sum_{l=-H}^0 x_{l j}\right)^2\right)\\
		&\quad =E\left(-\frac{2}{(H+1)^3} x_{k i}\left(\sum_{l=-H}^0 x_{l i}\right)\left(\sum_{l=-H}^0 x_{l j}^2+\sum_{l_1=-H}^0 \sum_{l_2=-H,l_1\neq l_2}^0 x_{l_1 j} x_{l_2 j}\right)\right] \\
		&\quad =E\left(-\frac{2}{(H+1)^3}\left[x_{k i}\left(\sum_{l=-H}^0 x_{l i}\right)\left(\sum_{l=-H}^0 x_{l j}^2\right)+x_{k i}\left(\sum_{l=-H}^0 x_{l i}\right) \sum_{l_1=-H}^0 \sum_{l_2=-H,l_1\neq l_2}^0 x_{l_1 j} x_{l_2 j}\right]\right) \\
		&\quad =E\left(-\frac{2}{(H+1)^3}\left[x_{k i}^2 x_{k j}^2+x_{k i}^2 \cdot \sum_{l \neq k=-H}^0 x_{l j}^2+\sum_{l \neq k=-H}^0 x_{k i} x_{k j} x_{l i} x_{l j}\right]\right) \\
		&\quad =-\frac{2}{(H+1)^3} \beta_{20}-\frac{2 H}{(H+1)^3} \cdot 1-\frac{2 H}{(H+1)^3} \rho_0^2(i, j), 
\end{align*}}
{\begin{align*}
		&\bE(A_7)  =E\left(\frac{1}{(H+1)^2} x_{k j}^2\left(\sum_{l=-H}^0 x_{l i}\right)^2\right)\\
		&\quad =E\left[\frac{1}{(H+1)^2} x_{k j}^2\left(\sum_{l=-H}^0 x_{li}^2+\sum_{l_1 = -H}^0 \sum_{l_2=-H,l_1 \neq l_2}^0 x_{l_1 i} x_{l_2 i}\right)\right] \\
		&\quad =E\left[\frac{1}{(H+1)^2}\left(x_{k j}^2 x_{k i}^2+\sum_{l\neq k=-H}^0 x_{k j}^2 X_{l i}^2\right)\right] \\
		&\quad =\frac{1}{(H+1)^2} \beta_{20}+\frac{H}{(H+1)^2} \cdot 1,\\
		&\bE(A_8) =-\frac{2}{(H+1)^3} \beta_{20}-\frac{2 H}{(H+1)^3}-\frac{2 H}{(H+1)^3} \rho_0^2(i, j), \\
		& \bE(A_9)=\frac{1}{(H+1)^3}\beta_{20} + \frac{H}{(H+1)^3} + \frac{H}{(H+1)^3}\rho_0^2(i,j). 
\end{align*}}
Then we calculate the expectation of $y_{ki}^2 y_{kj}^2$ as
\[
\bE\left( y_{ki}^2 y_{kj}^2\right) =\frac{H\left(1-H+H^2\right)}{(H+1)^3} \beta_{20}(i, j)+\frac{2 H^2-H}{(H+1)^3}+\frac{4 H^2+H}{(H+1)^3} \rho_0^2(i, j).
\]

Similarly, for the second term $y_{k_1 i} y_{k_1 j} y_{k_2 i} y_{k_2 j}$, notice that for any $k_1\neq k_2$, we can decompose it into the summation of several separate terms, the details are as follows.
{\footnotesize
	\begin{align*}
		&y_{k_1 i} y_{k_1 j} y_{k_2 i} y_{k_2 j} =\left(x_{k_1 i}-\frac{1}{H+1} \sum_{k=-H}^0 x_{k i}\right)\left(x_{k_1 j}-\frac{1}{H+1} \sum_{k=-H}^0 x_{k j}\right) \\
		&\hspace{90pt} \cdot \left(x_{k_2 i}-\frac{1}{H+1} \sum_{k=-H}^0 x_{k i}\right)\left(x_{k_2 j}-\frac{1}{H+1} \sum_{k=-H}^0 x_{k j}\right) \\
		& =\left[x_{k_1 i} x_{k_{k_1 j}}-\frac{1}{H+1} x_{k_1 i} \sum_{k=-H}^0 x_{k j}-\frac{1}{H+1} x_{k_1 j} \sum_{k=-H}^0 x_{k j}+\frac{1}{(H+1)^2}\left(\sum_{k=-H}^0 x_{k i}\right)\left(\sum_{k=-H}^0 x_{k j}\right)\right] \\
		& \quad {\left[x_{k_2 i} x_{k_2 j}-\frac{1}{H+1} x_{k_2 i} \sum_{k=-H}^0 x_{k j}-\frac{1}{H+1} x_{k_2 j} \sum_{k=-H}^0 x_{k i}+\frac{1}{(H+1)^2}\left(\sum_{k=-H}^0 x_{k i}\right)\left(\sum_{k=-H}^0 x_{k j}\right)\right]} \\
		& =x_{k_1 i} x_{k_1 j} x_{k_2 i} x_{k_2 j}-\frac{1}{H+1} x_{k_1 i} x_{k_1 j} x_{k_2 i}\left(\sum_{k=-H}^0 x_{k j}\right)-\frac{1}{H+1} x_{k_1 i} x_{k_1 j} x_{k-2 j}\left(\sum_{k=-H}^0 x_{k i}\right) \\
		&\quad +\frac{1}{(H+1)^2} x_{k_1 i} x_{k_1 j}\left(\sum_{k=-H}^0 x_{k i}\right)\left(\sum_{k=-H}^0 x_{k j}\right) \\
		&\quad -\frac{1}{(H+1)} x_{k_1 i}\big(\sum_{k=-H}^0 x_{k j}\big) x_{k_2 i} x_{k_2 j}+\frac{1}{(H+1)^2} x_{k_1 i}\big(\sum_{k=-H}^0 x_{k j}\big) x_{k_2 i}\big(\sum_{k=-H}^0 x_{k j}\big) \\
		&\quad +\frac{1}{(H+1)^2} x_{k_1 i}\big(\sum_{k=-H}^0 x_{k j}\big) x_{k_2 j}\big(\sum_{k=-H}^0 x_{k i}\big)-\frac{1}{(H+1)^3} x_{k_1 i}\big(\sum_{k=-H}^0 x_{k j}\big)\big(\sum_{k=-H}^0 x_{k i}\big)\big(\sum_{k=-H}^0 x_{k j}\big) \\
		& \quad -\frac{1}{(H+1)} x_{k_1 j}\big(\sum_{k=-H}^0 x_{k i}\big) x_{k_2 i} x_{k_2 j} + \frac{1}{(H+1)^2} x_{k_1 j}\big(\sum_{k=-H}^0 x_{k i}\big) x_{k_2 i}\big(\sum_{k=-H}^0 x_{k j}\big) \\
		&\quad +\frac{1}{(H+1)^2} x_{k_1 j}\big(\sum_{k=-H}^0 x_{k i}\big) x_{k_2 j}\big(\sum_{k=-H}^0 x_{k i}\big)-\frac{1}{(H+1)^3} x_{k_1 j}\big(\sum_{k=-H}^0 x_{k i}\big)\big(\sum_{k=-H}^0 x_{k i}\big)\big(\sum_{k=-H}^0 x_{k j}\big) \\
		&\quad +\frac{1}{(H+1)^2}\big(\sum_{k=-H}^0 x_{k i}\big)\big(\sum_{k=-H}^0 x_{k j}\big) x_{k_2 i} x_{k_2 j}-\frac{1}{(H+1)^3}\big(\sum_{k=-H}^0 x_{k i}\big)\big(\sum_{k=-H}^0 x_{k j}\big) x_{k_2 i}\big(\sum_{k=-H}^0 x_{k j}\big) \\
		&\quad -\frac{1}{(H+1)^3}\big(\sum_{k=-H}^0 x_{k i}\big)\big(\sum_{k=-H}^0 x_{k j}\big) x_{k_2 j}\big(\sum_{k=-H}^0 x_{k i}\big)+\frac{1}{(H+1)^4}\big(\sum_{k=-H}^0 x_{k i}\big)^2\big(\sum_{k=-H}^0 x_{k j}\big)^2. 
\end{align*}}

The expectation of $y_{k_1 i} y_{k_1 j} y_{k_2 i} y_{k_2 j}$ can be calculated in a similar way, 
\[
\bE\left(y_{k_1 i} y_{k_1 j} y_{k_2 i} y_{k_2 j}\right)=\frac{1-2 H}{(H+1)^3}+\frac{3 H}{(H+1)^3} \beta_{20}+\frac{3+H^2+H^3}{(H+1)^3} \rho_0^2(i, j). \\
\]
Based on the expectations of $y_{ki}^2y_{kj}^2$ and $y_{k_1 i} y_{k_1 j} y_{k_2 i} y_{k_2 j}$, we have,
\begin{align*}
	\bE\left(\left(\hat{\rho}_0(i,j)\right)^2\right)&=\frac{H+1}{H^2} \cdot E\left(y_{k i}^2 y_{k j}^2\right)+\frac{(H+1) \cdot H}{H^2} \cdot E\left(y_{k_1 i} y_{k_1 j} y_{k_2 i} y_{k_2 j}\right) \\
	& =\frac{H+1}{H^2} \cdot\left[\frac{H\left(1-H+H^2\right)}{(H+1)^3} \beta_{20}+\frac{2 H^2-H}{(H+1)^3}+\frac{4 H^2+H}{(H+1)^3} \rho_0^2(i . j)\right] \\
	& \quad +\frac{(H+1)}{H}\left[\frac{1-2 H}{(H+1)^3}+\frac{3 H}{(H+1)^3} \beta_{20}+\frac{3+H^2+H^3}{(H+1)^3} \rho_0^2(i, j)\right] \\
	& =\frac{H^2+2 H+1}{H(H+1)^2} \beta_{20}+0+\frac{H^3+H^2+4 H+4}{H(H+1)^2} \rho_0^2(i, j) \\
	& =\frac{1}{H} \beta_{20}+\frac{H^3+H^2+4 H+4}{H(H+1)^2} \rho_0^2(i, j). 
\end{align*}
Similarly, we can obtain,
\[
\bE \left(\left(\hat{\rho}_{1:t}(i,j)\right)^2 \right) = \frac{1}{t-1}\beta_{21} + \frac{(t^2 -2t+5)}{(t-1)t}\rho_1^2(i,j).
\]
Finally, we have,
\[ 
\begin{aligned}
	&\bE\left(  (\hat{\rho}_0(i,j) - \hat{\rho}_{1:t}(i,j)  )^2 \right)
	\\& = \frac{1}{H}\beta_{20} + \frac{1}{t-1}\beta_{21} + (\rho_0(i,j) - \rho_1(i,j) )^2
	+\frac{5-t}{t(t-1)}\rho_1^2(i,j) + \frac{3H+4-H^2}{H(H+1)^2} \rho_0^2(i,j).
\end{aligned} 
\]

\subsection{Proof of lemma~\ref{lem:signflip_indep}}
{\bf Proof.}
Given a vector $\boldsymbol{x}_t=(x_{t1},\ldots,x_{tp})^\top \in \mathbb{R}^{p\times 1}$ and two independent Rademacher vectors $\boldsymbol{\mathcal{R}}_1:=(r_1^{(1)},\ldots,r_p^{(1)})^\top$ and $\boldsymbol{\mathcal{R}}_2:=(r_1^{(2)},\ldots,r_p^{(2)})^\top$ with i.i.d. Rademacher entries, then $\boldsymbol{x}_t \circ \boldsymbol{\mathcal{R}}_1$ and $\boldsymbol{x}_t \circ \boldsymbol{\mathcal{R}}_2$ are sub-gaussian random vectors. $\mathbb{E}(r_i^{(1)})= \mathbb{E}(r_j^{(2)}) =0$ due to the property of Rademacher entries. Let $\boldsymbol{\Tilde{x}}_t = ((\boldsymbol{x}_t \circ \boldsymbol{\mathcal{R}}_1)^\top , (\boldsymbol{x}_t \circ \boldsymbol{\mathcal{R}}_2)^\top)^\top = (x_{t1}r_1^{(1)}, \ldots, x_{tp} r_p^{(1)},x_{t1}r_1^{(2)},\ldots,x_{tp}r_p^{(2)}  )^\top \in \mathbb{R}^{2p\times 1}$, then for $1\le i,j \le p$, the element in the covariance matrix of $\boldsymbol{\Tilde{x}}_t$ is
\[ \footnotesize
\begin{aligned}
	Cov(x_{ti}r_i^{(1)} , x_{tj}r_j^{(2)} ) = \mathbb{E}(x_{ti}r_i^{(1)} x_{tj}r_j^{(2)} ) - \mathbb{E}(x_{ti}r_i^{(1)} ) \mathbb{E}(x_{tj}r_j^{(2)}) = \mathbb{E}(r_i^{(1)}) \mathbb{E}(r_j^{(2)}) Cov(x_{ti},x_{tj}) =0.
\end{aligned}
\]
Then $\boldsymbol{x}_t \circ \boldsymbol{\mathcal{R}}_1$ and $\boldsymbol{x}_t \circ \boldsymbol{\mathcal{R}}_2$ are asymptotically independent.

\subsection{Proof of Lemma~\ref{lem:corr}}
{\bf Proof.}
In this lemma, we calculate the temporal correlation of $S_{t+s}$ and $S_t$, where $S_t$ can represent both sum-type and max-type test statistics. We may unify both sum-type statistics for non-sparse settings and max-type statistics for sparse settings, since they are, in essence, the combination of element-wise entries, with each entry being $C\times (\hat{\rho}_0(i,j) - \hat{\rho}_{t':t}(i,j))^2$ for a constant $C$. $C$ equals 1 for max-type statistics and smaller than $p(p-1)/2$ for sum-type statistics, depending on the change numbers. Besides, for window-limited variant test statistics, the weight $\frac{(t-t')H}{H+t-t'}$ can also be treated as a constant, which does not influence the correlation value. Then the correlation between $S_{t+s}$ and $S_t$ is in essence the correlation between $\boldsymbol{v}_{t-w,t}(i,j)$ and $\boldsymbol{v}_{t+s-w,t+s}(i,j)$. We give the concrete form of the correlation coefficient as
\[
\operatorname{Corr}\left(S_{t+s}, S_t\right)=\operatorname{Corr} 
\left(\left(\hat{\rho}_0(i,j)-\hat{\rho}_{t-w:t}(i,j)\right)^2,\left(\hat{\rho}_0(i,j)-\hat{\rho}_{t+s-w:t+s}(i,j)\right)^2\right) . 
\]
Then we calculate the covariance and variance terms separately, for the covariance term, we have
$$
\begin{aligned}
	&\operatorname{Cov}\left(\left(\hat{\rho}_0(i,j)-\hat{\rho}_{t-w:t}(i,j)\right)^2,\left(\hat{\rho}_0(i,j)-\hat{\rho}_{t+s-w:t+s}(i,j)\right)^2\right) \\
	=&\operatorname{Cov}\left(\left(\frac{1}{w} \sum_{k=-w}^0 y_{k i} y_{k j}-\frac{1}{w} \sum_{k=t+s-w}^t y_{k i} y_{k j}-\frac{1}{w} \sum_{k=t-w}^{t+s-w-1} y_{k i} y_{k j}\right)^2,\right. \\
	& \hspace{30pt}\left.\left(\frac{1}{w} \sum_{k=-w}^0 y_{k i} y_{k j}-\frac{1}{w} \sum_{k=t+s-w}^t y_{k i} y_{k j}-\frac{1}{w} \sum_{k=t+1}^{t+s} y_{k i} y_{k j}\right)^2\right).
\end{aligned}
$$
For simplicity, we denote $A =\frac{1}{w} \sum_{k=-w}^0 y_{k i} y_{k j}-\frac{1}{w} \sum_{k=t+s-w}^t y_{k i} y_{k j}$, $B =\frac{1}{w} \sum_{k=t-w}^{t+s-w-1} y_{k i} y_{k j}$, $C =\frac{1}{w} \sum_{k=t+1}^{t+s} y_{k i} y_{k j}$, then
$$
\begin{aligned}
	\operatorname{Cov}&\left(\left(\hat{\rho}_0(i,j)-\hat{\rho}_{t-w:t}(i,j)\right)^2,\left(\hat{\rho}_0(i,j)-\hat{\rho}_{t+s-w:t+s}(i,j)\right)^2\right) =\operatorname{Cov}\left((A-B)^2,(A-C)^2\right)\\
	= & \operatorname{Var}\left(A^2\right)-2\bE C\left(\bE A^3-\bE A^2 \bE A\right)-2 \bE B\left(\bE A^3-\bE A^2 \bE A\right)+4 \bE B \bE C \operatorname{Var}(A).
\end{aligned}
$$

Since $  \operatorname{Var}  \left(\left(\hat{\rho}_0(i,j)-\hat{\rho}_{t-w:t}(i,j)\right)^2\right) = \operatorname{Var}  \left(\left(\hat{\rho}_0(i,j)-\hat{\rho}_{t+s-w:t+s}(i,j)\right)^2\right)$ due to the independent assumption, then we only calculate the first one of the variance terms.
$$ 
\begin{aligned}
	\operatorname{Var} & \left(\left(\hat{\rho}_0(i,j)-\hat{\rho}_{t-w:t}(i,j)\right)^2\right) = \operatorname{Var}\left(\left(A-B \right)^2\right) \\
	= & \operatorname{Var}\left(A^2\right)+4 \bE A^2 \bE B^2+\operatorname{Var}\left(B^2\right)-4 \bE B\left(\bE A^3-\bE A^2 \bE A\right) 
	\\& \quad -4 \bE A\left(\bE B^3-E B^2 \bE B\right)-4(\bE A)^2(\bE B)^2  \\
	= & \operatorname{Var}\left(A^2\right)+\operatorname{Var}\left(B^2\right)-4 \bE B\left(\bE A^3- \bE A^2 \bE A\right) 
	-4 \bE A\left(\bE B^3- \bE B^2 \bE B\right) + 4\operatorname{Var}(AB).
\end{aligned}
$$

Let $\bE_{\infty}\left(y_{k_i} y_{k j}\right)=\rho_0(i, j)$, $\bE_{\infty}\left(y_{k i} y_{k j}\right)^2=\beta_{20}$, $\bE_{\infty}\left(y_{ki} y_{k j}\right)^3=\beta_{30}$, $\bE_{\infty}\left(y_{k i} y_{k j}\right)^4=\beta_{40}$, then we take expectation for each term as follows.
{
	\begin{align*}
		& \bE(A)=\rho_0(i, j)-\frac{w-s}{w} \rho_0(i, j)=\rho_0(i, j) \frac{s}{w}, \\
		& \bE(B)=\rho_0\left(i,j\right) \frac{s}{w}, \quad \bE(C)=\rho_0(i, j) \frac{s}{w},\\
		& \bE\left(B^2\right)=\left(\beta_{20}-\rho_0^2(i,j)\right) \frac{s}{w^2}+\rho_0^2(i, j) \frac{s^2}{w^2}, \\
		& \bE\left(B^3\right)=\left(\beta_{30}-3\beta_{20} \rho_0(i, j)+2 \rho_0^3(i, j)\right) \frac{s}{w^3}+3\left(\beta_{20} \rho_0(i, j)- \rho_0^3(i, j)\right) \frac{s^2}{w^3}+\rho_0^3(i, j) \frac{s^3}{w^3} ,\\
		& \bE\left(B^4\right)=\left(\beta_{40}-4\beta_{30} \rho_0\left(i,j\right)-3\beta_{20}^2-6 \rho_0^4\left(i,j\right)+12 \beta_{20} \rho_0^2\left(i,j\right)\right) \frac{s}{w^4} \\
		&\quad +\left(4\beta_{30} \rho_0(i,j)+3\beta_{20}^2-18 \beta_{20} \rho_0^2(i, j)+ 11\rho_0^4(i, j)\right) \frac{s^2}{w^4} \\
		&\quad +\left(6\beta_{20} \rho_0^2(i, j)-6 \rho_0^4(i, j)\right) \frac{s^3}{w^4}+\rho_0^4(i,j) \frac{s^4}{w^4}, \\
		& \bE\left(A^2\right)=\left(2\beta_{20}-2\rho_0^2(i, j)\right)\cdot \frac{1}{w}+\left(\rho_0^2(i,j) - \beta_{20}\right) \frac{s}{w^2}+\rho_0^2\left(i,j\right) \frac{s^2}{w^2} + \left(2\beta_{20} - 2\rho_0^2(i,j) \right) \frac{1}{w^2}, \\
		& \bE\left(A^3\right)=\left(6\beta_{20} \rho_0(i,j)-6 \rho_0^3(i, j)\right) \frac{s}{w^2}+\left(\beta_{30}+3\beta_{20} \rho_0\left(i,j\right)-4 \rho_0^3\left(i,j\right)\right)  \frac{s}{w^3}\\
		&\quad +\left(3 \rho_0^3\left(i,j\right)-3\beta_{20} \rho_0\left(i,j\right)\right) \frac{s^2}{w^3}+\rho_0^3\left(i,j\right) \frac{s^3}{w^3}, \\
		& \bE A^4=\left(-24\beta_{20}\rho_0^2(i,j) + 12\rho_0^4(i,j) + 12\beta_{20}^2 \right)\frac{1}{w^2} + \left(24\beta_{20}\rho_0^2(i,j) -12\rho_0^4(i,j) -12\beta_{20}^2 \right) \frac{s}{w^3}\\
		& \quad + \left(12\beta_{20}\rho_0^2(i,j) -12 \rho_0^4(i,j)  \right)\frac{s^2}{w^3} + \left(2\beta_{40} - 8\beta_{30}\rho_0(i,j) +18\beta_{20}^2 -24\beta_{20}\rho_0^2(i,j) +12\rho_0^4(i,j) \right)\frac{1}{w^3} \\
		& \quad+ \left(-\beta_{40} + 4\beta_{30}\rho_0(i,j) -9\beta_{20}^2 +12\beta_{20}\rho_0^2(i,j) - 6\rho_0^4(i,j) \right)\frac{s}{w^4}\\
		& \quad + \left(4\beta_{30}\rho_0(i,j) + 3\beta_{20}^2 -6\beta_{20}\rho_0^2(i,j) -\rho_0^4(i,j) \right)\frac{s^2}{w^4} \\
		&\quad + \left( -6\beta_{20}\rho_0^2(i,j) + 6\rho_0^4(i,j) \right) \frac{s^3}{w^4} + \rho_0^4(i,j)\frac{s^4}{w^4}   
		+ \left(2\beta_{40} - 8\beta_{30}\rho_0(i,j)+6\beta_{20}^2 \right) \frac{1}{w^4},\\
		& \operatorname{Var}(A)=\left(2\beta_{20}-2\rho_0^2(i,j)\right)  \frac{1}{w}+\left(\rho_0^2\left(i,j\right)-\beta_{20}\right) \frac{s}{w^2} + \left(2\beta_{20} - 2\rho_0^2(i,j) \right)\frac{1}{w^2} ,\\
		& \operatorname{Var}\left(A^2\right)= 8\left(\beta_{20} - \rho_0^2(i,j) \right)^2\frac{1}{w^2} - 8 \left(\beta_{20} - \rho_0^2(i,j) \right)^2 \frac{s}{w^3}  \\
		&\quad + \left(8\beta_{20}\rho_0^2(i,j) - 8\rho_0^4(i,j) \right)\frac{s^2}{w^3} + \left(2\beta_{40} - 8\beta_{30}\rho_0(i,j) +10\beta_{20}^2 -8 \beta_{20}\rho_0^2(i,j) +4 \rho_0^4(i,j)  \right) \frac{1}{w^3} \\
		&\quad + \left(-\beta_{40} + 4\beta_{30}\rho_0(i,j) - 5\beta_{20}^2 + 4 \beta_{20}\rho_0^2(i,j) - 2\rho_0^4(i,j)  \right)\frac{s}{w^4} \\
		&\quad +\left(4\beta_{30}\rho_0(i,j) + 2\beta_{20}^2 -8\beta_{20}\rho_0^2(i,j) + 2\rho_0^4(i,j)  \right)\frac{s^2}{w^4}
		\\& \quad + \left(-4\beta_{20}\rho_0^2(i,j) + 4\rho_0^4(i,j) \right)\frac{s^3}{w^4} + \left(2\beta_{40} + 2\beta_{20}^2 -8\beta_{30}\rho_0(i,j) + 8\beta_{20}\rho_0^2(i,j) - 4\rho_0^4(i,j) \right) \frac{1}{w^4} , \\
		& \operatorname{Var}\left(B^2\right)=\left(\beta_{40}-4\beta_{30} \rho_0\left(i,j\right)-3\beta_{20}^2-6 \rho_0^4\left(i,j\right)+12 \beta_{20} \rho_0^2\left(i,j\right)\right) \frac{s}{w^4} \\
		& \quad+\left(4\beta_{30} \rho_0(i,j)-16\beta_{20} \rho_0^2(i,j)+10 \rho_0^4(i,j) + 2\beta_{20}^2\right) \frac{s^2}{w^4}+\left(4\beta_{20} \rho_0^2(i,j)-4 \rho_0^4(i,j)\right)\frac{s^3}{w^4} ,\\
		& \operatorname{Var}(AB) = 2\left(\beta_{20} - \rho_0^2(i,j) \right)^2\frac{s}{w^3} + 2\left(\beta_{20}\rho_0^2(i,j) - \rho_0^4(i,j)  \right)\frac{s^2}{w^3} \\
		&\quad +\left(-\beta_{20}^2 - 3\rho_0^4(i,j) + 4\beta_{20}\rho_0^2(i,j) \right) \frac{s^2}{w^4} + 2(\beta_{20} -\rho_0^2(i,j))^2\frac{s}{w^4}.
\end{align*}}

Since the correlation is represented as follows,
$$
\begin{aligned}
	&\operatorname{Corr}\left( S_{t+s}, S_t \right) \\
	&=1+\frac{4 \bE B \bE C \operatorname{Var}(A)-\operatorname{Var}\left(B^2\right)+4 \bE A\left(\bE B^3-
		\bE B 
		\bE B^2\right) - 4\operatorname{Var}(AB)}{\operatorname{Var}\left(A^2\right)+\operatorname{Var}\left(B^2\right)-4 \bE B\left(\bE A^3-\bE A^2 E A\right)-4 \bE A\left(\bE B^3-\bE B^2 \bE B\right) + 4\operatorname{Var}(AB)} ,
\end{aligned}
$$
the numerator equals
$$
\begin{aligned}
	& 4 \bE B \bE C \operatorname{Var}(A)-\operatorname{Var}\left(B^2\right)+4 \bE A\left(\bE B^3-\bE B E B^2\right) - 4\operatorname{Var}(AB) \\
	&\quad  = \left(-\beta_{40} +4\beta_{30}\rho_0(i,j) - 5\beta_{20}^2 + 4\beta_{20}\rho_0^2(i,j) - 2\rho_0^4(i,j) \right)\frac{s}{w^4} -8\left(\beta_{20} - \rho_0^2(i,j) \right)^2\frac{s}{w^3}\\
	&\quad + \left(2\beta_{20}^2 - 4\rho_0^4(i,j) + 5\beta_{20}\rho_0^2(i,j) \right) \frac{s^2}{w^4},
\end{aligned}
$$
and the denominator takes value
$$
\begin{aligned}
	&\operatorname{Var}\left(A^2\right)+\operatorname{Var}\left(B^2\right)-4 \bE B\left(\bE A^3-\bE A^2 \bE A\right)-4 \bE A\left(\bE B^3-\bE B^2 \bE B\right) + 4\operatorname{Var}(AB) \\
	& \quad = 8\left(\beta_{20} - \rho_0^2(i,j) \right)^2 \frac{1}{w^2} + \left(2\beta_{40} - 8\beta_{30}\rho_0(i,j) +10\beta_{20}^2 - 8\beta_{20}\rho_0^2(i,j) + 4\rho_0^4(i,j) \right)\frac{1}{w^3}\\
	& \quad + (-12\beta_{20}\rho_0^2(i,j) + 12\rho_0^4(i,j) )\frac{s^2}{w^4} \\
	& \quad + \left(2\beta_{40} + 2\beta_{20}^2 -8\beta_{30}\rho_0(i,j) + 8\beta_{20}\rho_0^2(i,j) -4rho_0^4(i,j) \right)\frac{1}{w^4} \\
	& \quad + \left(-8\beta_{20}^2 -8\rho_0^4(i,j) + 16\beta_{20}\rho_0^2(i,j) \right) \frac{s}{w^4}. \\
\end{aligned}
$$ 
Combining these results yields the correlation result
\[
Corr = 1-\frac{s}{w} + o(\frac{s}{w}),
\]
and complete the proof. 

\subsection{Proof of Theorem~\ref{theo:ARL}}
{\bf Proof.}
The proof is based on a general method for computing first passage probabilities first introduced in \citet{pollak1998new} and further developed in \citet{siegmund2000tail} and \citet{siegmund2010tail}, and commonly used in similar problems \citep{xie2013sequential}, \citet{li2015m}, \citet{cao2018multi}. First of all, it is worth mentioning that the probability measure in the following proof always stands for the nominal case where all samples are from the same distribution $d$. For the test statistic defined in this paper 
$$
S_t^{(\text{WL-sum})}=\max _{t-w \leqslant t^{\prime} \leqslant t-1} \frac{\left(t-t^{\prime}\right) H}{H+t-t^{\prime}}\left\|\boldsymbol{v}_{t^{\prime}, t}\right\|_1,
$$
if we suppose $H=w$, and $\frac{\left(t-t^{\prime}\right) H}{H+t-t^{\prime}}\left\|\boldsymbol{v}_{t^{\prime}, t}\right\|_1$ reaches its maximum value when $t^{\prime}=t-w$ (as shown in Table~\ref{table:count_num}, this is the most common case), then
$$
\begin{aligned}
	& S_t^{\text {(WL-sum) }}=\frac{w}{2} \cdot \frac{p(p-1)}{2} \cdot \left(\hat{\rho}_0(i,j)-\hat{\rho}_{t-w:t}(i,j)\right)^2 \\
	& S_t^{\text {(WL-max) }}=\frac{w}{2} \cdot \left(\hat{\rho}_0(i,j)-\hat{\rho}_{t-w:t}(i,j)\right)^2,
\end{aligned}
$$

let $V_t^w = S_t = c \boldsymbol{v}_{t-w,t}(i,j) =  c \left(\hat{\rho}_0(i,j)-\hat{\rho}_{t-w:t}(i, j)\right)^2
=c \left(\frac{1}{w} \sum_{k=-w}^0 y_{k i} y_{k j}-\frac{1}{w} \sum_{k=t-w}^t y_{k i} y_{k j}\right)^2,$  where $c = \frac{w}{2} \frac{p(p-1)}{2}$ for WL-sum statistic and $c= \frac{w}{2}$ for WL-max statistic.

From the existing calculation, we have
$$
\begin{aligned}
	& \mu:= E\left(V_t^w\right)= \frac{2c}{w}\left(\beta_{20}-\rho_0^2(i, j)\right), \\
	& \sigma_d^2:=\operatorname{Var}\left(V_t^w\right)  =8\left(\beta_{20} - \rho_0^2(i,j) \right)^2 \frac{c^2}{w^2} \\
	& \quad + \left(2\beta_{40} - 8\beta_{30}\rho_0(i,j) +10\beta_{20}^2 - 8\beta_{20}\rho_0^2(i,j) + 4\rho_0^4(i,j) \right)\frac{c^2}{w^3}\\
	& \quad + (-12\beta_{20}\rho_0^2(i,j) + 12\rho_0^4(i,j) )\frac{s^2c^2}{w^4} \\
	& \quad + \left(2\beta_{40} + 2\beta_{20}^2 -8\beta_{30}\rho_0(i,j) + 8\beta_{20}\rho_0^2(i,j) -4rho_0^4(i,j) \right)\frac{c^2}{w^4}     \\
	& \quad + \left(-8\beta_{20}^2 -8\rho_0^4(i,j) + 16\beta_{20}\rho_0^2(i,j) \right) \frac{sc^2}{w^4}.
\end{aligned}
$$

Here we denote the moment generating function as 
$$\psi_w(\theta)=\log \mathbb{E}\left(\exp \left\{\theta V_t^w\right\}\right),$$
and select $\theta=\theta_w$ by solving equation $\dot{\psi}_w(\theta)=b$. Since $V_t^w$ is defined by a function of $2w+2$ independent random samples, $\psi_{w}$ converges to a limit as $w \rightarrow \infty$ and $\theta_w$ converges to a limit, denoted by $\theta$. The transformed distribution for all sequences before position $t$ and window size $w$ is denoted by
$$
d\mathbb{P}_t^w=\exp \left\{\theta V_t^w-\psi_w\left(\theta_w\right)\right\} d\mathbb{P}.$$
Let
$$ \ell(t, w):=\log \left(d \mathbb{P}_t^w / d \mathbb{P}\right)=\theta V_t^w-\psi_w\left(\theta_w\right).
$$
Denote $D =\{(t,w): 0\le t\le m, 1 \le w \le m \}$ be the set of all possible windows in the scan. Let $A = \left\{ \underset{(t,w)\in D}{max} V_t^w \ge b \right\}$ be the event of interests (the event \{$T\le m$\}), i.e., the detection procedure stops before time $m$. By measure transformation, we have
\begin{equation}\label{equa:p(A)} 
	\begin{aligned}
		\mathbb{P}(A) & =\sum_{(t,w) \in D} \mathbb{E}\left[\exp [\ell(t, w)]\left(\sum_{\left(t^{\prime}, w^{\prime}\right) \in D} \exp \left[\ell\left(t^{\prime}, w^{\prime}\right)\right]\right)^{-1} ; A\right] \\
		& =\sum_{(t, w) \in D} \mathbb{E}_t^w\left[\left(\sum_{\left(t^{\prime}, w^{\prime}\right) \in D} \exp \left[\ell\left(t^{\prime}, w^{\prime}\right)\right]\right)^{-1} ; A\right] \\
		& =\sum_{(t,w) \in D} e^{\tilde{\ell}(t,w)-\ell(t,w)} \times \mathbb{E}_t^w\left[\frac{\max_{t^{\prime}, w^{\prime}} e^{\ell\left(t^{\prime}, w^{\prime}\right)-\ell(t, w)}}{\sum_{t^{\prime}, w^{\prime}} e^{\ell\left(t^{\prime}, w^{\prime}\right)-\ell(t, w)}} e^{-\tilde{\ell}(t, w)-\left[\max _{t^{\prime}, w^{\prime}} \ell\left(t^{\prime}, w^{\prime}\right)-\ell(t, w)\right]} ; A\right] \\
		& =e^{-\theta_w \dot{\psi}_w\left(\theta_w\right)+\psi_w\left(\theta_w\right)} \times \sum_{(t, w) \in D} \mathbb{E}_t^w\left[\frac{M(t, w)}{S(t, w)} e^{-\tilde{\ell}(t,w)-\log M(t, w)} ; A\right],
	\end{aligned}
\end{equation}

where
$$
\begin{aligned}
	\tilde{\ell}(t, w) & =\theta_w\left[V_t^w-\dot{\psi}_w\left(\theta_w\right)\right], \\
	M(t, w) & =\max _{t^{\prime}, w^{\prime}} \exp \left\{\theta_w\left(V_{t^{\prime}}^{w^{\prime}}-V_t^w\right)\right\}, \\
	S(t, w) & =\sum_{t^{\prime}, w^{\prime}} \exp \left\{\theta_w\left(V_{t^{\prime}}^{w^{\prime}}-V_t^w\right)\right\} .
\end{aligned}
$$

Consider a sequence of $\sigma-$fields $\hat{\mathcal{F}}$, let $\hat{M}(t,w)$ and $\hat{S}(t,w)$ be approximations of $M(t,w)$ and $S(t,w)$, respectively, which are measurable with respect to $\hat{\mathcal{F}}$. Given $\epsilon>0$, we assume that for all large $b$,
\begin{assumption}\label{assum1}
	$M(t,w)$ and $S(t,w)$ satisfy $0 < M(t,w) \le S(t,w) < \infty$ with probability 1.
\end{assumption}
\begin{assumption}\label{assump2}
	There exist $\hat{M}(t,w)$ and $\hat{S}(t,w)$ measurable with respect to $\hat{\mathcal{F}}$ such that $\Vert \hat{M}(t,w) / M(t,w) -1 \Vert \le \epsilon$ and $\vert \hat{S}(t,w) / S(t,w) -1 \vert \le \epsilon $, with probability at least $1-\epsilon b^{-1/2}$.   
\end{assumption}

\begin{assumption}\label{assump3}
	$\mathbb{E} [ \hat{M}(t,w) / \hat{S}(t,w) ]$ converges to a finite and positive limit denoted by $\mathbb{E}[\mathcal{M} / \mathscr{S}] $.
\end{assumption}

\begin{assumption}\label{assump4}
	There exist $\mu \in \mathbb{R}$ and $\sigma_d \in \mathbb{R}^+$ such that for every $0<\epsilon,\delta$ and for all large enough $b$ the probability of the event 
	$$B = \{ \underset{\vert x\vert \le log(b) }{sup} \left\vert b^{1/2}\mathbb{P}\left( \Tilde{l}(k,w) \in x - log \hat{M}(k,w) + (0,\delta] \vert \hat{\mathcal{F}}  \right) - \frac{\delta}{\sigma_d} \phi(\frac{\mu}{\sigma_d})  \right\vert \le \epsilon \} $$
	
	is bounded from below by $1-\epsilon b^{-1/2}$.
\end{assumption}

Since $t,w$ are fixed in much of the following analysis, we suppress the dependence of the notation on $t,w$ and simply write $\tilde{\ell}, S, M$. Under Assumptions~\ref{assum1}, \ref{assump2}, \ref{assump3} and \ref{assump4}, a localization lemma allows us to simplify the expectation
$$
\mathbb{E}_t^w \left[\frac{M}{S} e^{-\tilde{\ell}-\log M} ; \tilde{\ell}+\log M \geq 0\right]
$$
into a simpler form
$$
\frac{1}{\sqrt{2 \pi \sigma_w^2}} \mathbb{E}\left[\frac{M}{S}\right],
$$
where $\sigma_w^2$ stands for the variance of $\tilde{\ell}$ under measure $\mathbb{P}_t^w$. The reduction relies on the fact that for large $m$, the local processes $M$ and $S$ are approximately independent of the global process $\tilde{\ell}$. Such independence allows the above decomposition into the expectation of $M / S$ times the expectation involving $\tilde{\ell}+\log M$, treating $\log M$ essentially as a constant. 

We first consider the process $M$ and $S$ and derive the expectation $\mathbb{E}[M / S]$ following \citet{siegmund2000tail}.

The covariance between the two terms is given by
$$
\operatorname{Cov}\left(\theta_w V_{t^{\prime}}^{w^{\prime}}, \theta_w V_t^w\right)=\theta_w^2 \mathbb{E}\left[V_{t^{\prime}}^{w^{\prime}}, V_t^w\right]=\theta_w^2 \sigma_d^2\left(1-\frac{|t'-t |}{w} + o\left(\frac{|t'-t |}{w} \right) \right).
$$
When $w$ is large, we have that the correlation depends on the difference $\left|t^{\prime}-t\right|$ in a linear form, which shows that we have the random walk in the change time $t$, and the variance of the increment equals to $\theta_w^2 \sigma_d^2 /w$. Following \citet{siegmund2000tail}, we have
$$
\mathbb{E}[M / S]=\left[\theta_w^2 \sigma_d^2 / w \nu\left(\left[ \theta_w \sigma_d^2 / w\right]^{1 / 2}\right)\right]^2 .
$$
Moreover, the process $\tilde{\ell}$ is zero-mean and has variance $\sigma_w^2=\operatorname{Var}_t^w(\tilde{\ell})=\theta_w^2 \ddot{\psi}\left(\theta_w\right)$ under the measure $\mathbb{P}_t^w$. Substituting the result for the expectations in (\ref{equa:p(A)}) yields
$$ 
\mathbb{P}\left(T \leq m\right)=2 \sum_{w=0}^{m-1}(m-2 w) e^{-\theta_w \dot{\psi}_w\left(\theta_w\right)+\psi_w\left(\theta_w\right)}\left[2 \pi \theta_w^2 \ddot{\psi}_w\left(\theta_w\right)\right]^{-1 / 2}\left[\theta_w^2 \sigma_d^2 / w \nu\left(\left[ \theta_w \sigma_d^2 / w\right]^{1 / 2}\right)\right]^2.
$$
In the limiting case, $V_t^w$ can be well approximated using Gaussian distribution $\mathcal{N}\left(\mu, \sigma_d^2\right)$. The moment generating function then becomes $\psi(\theta)=\mu\theta + \theta^2 \sigma_d^2 / 2$, and the limiting $\theta=(b-\mu )/ \sigma_d^2$, as the solution to $\dot{\psi}(\theta)=b$. Furthermore, the summation term can be approximated by an integral, to obtain
\begin{equation}
	\label{eq:p(T<= m)}
	\footnotesize
	\begin{aligned}
		\mathbb{P}\left(T \leq m\right) & =2 \sum_{w=0}^{m-1}(m-2 w) e^{-(b-\mu)^2 /\left(2 \sigma_d^2\right)}\left[2 \pi (b-\mu)^2 / \sigma_d^2\right]^{-1 / 2}\left[(b-\mu)^2 /\left(w \sigma_d^2\right) \nu\left(\left[(b-\mu) /w \right]^{1 / 2}\right)\right]^2 \\
		& \approx 4 e^{-(b-\mu)^2 /\left(2 \sigma_d^2\right)}\left[2 \pi (b-\mu)^2 / \sigma_d^2\right]^{-1 / 2}\left[(b-\mu)^2 / \sigma_d^2\right]^2 \int_{1}^{m} \nu^2\left(\left[4 b^2 /\left(m t \sigma_d^2\right)\right]^{1 / 2}\right)(1-t) d t / t^2 .
	\end{aligned}
\end{equation}
Here it is assumed that $m$ is large, but small enough that the right-hand side of (\ref{eq:p(T<= m)}) converges to 0 when $b \rightarrow \infty$. Changing variables in the integrand, we can rewrite this approximation as
\begin{equation}
	\label{eq:p(T<= m)2}
	\mathbb{P}\left\{T \leq m\right\} \approx m \times 2 e^{-(b-\mu)^2 /\left(2 \sigma_d^2\right)}\left[2 \pi (b-\mu)^2 / \sigma_d^2\right]^{-1 / 2}\left[(b-\mu)^2 / \sigma_d^2\right] \int_{\left[4 b^2 /\left(w \sigma_d^2\right)\right]^{1 / 2}}^{\left[4 b^2 /\left( \sigma_d^2\right)\right]^{1 / 2}} y \nu^2(y) d y .
\end{equation}
From the arguments in \citet{siegmund1995using}, \citet{siegmund2008detecting}, we know that $T$ is asymptotically exponentially distributed and is uniformly integrable. Hence if $\lambda$ denotes the factor multiplying $m$ on the right-hand side of (\ref{eq:p(T<= m)2}), then for large $m$, in the range where $m \lambda$ is bounded away from 0 and $+\infty$, $\mathbb{P}\left\{T \leq m\right\}-[1-\exp (-\lambda m)] \rightarrow 0$. Consequently, $\mathbb{E}\left[T \right] \approx 1 / \lambda$, thereby we complete the proof. Here we omit some technical details needed to make the derivation rigorous. Those details have been described and proved in \citet{siegmund2010tail}. 

\subsection{Proof of Proposition~\ref{prop:edd}}
{\bf Proof.}
For $\nu=1$ and a given detection threshold $b>0$, when the window size is sufficiently large, at the time of detection (the stopping time $T$) we have 
\[
\E[S_T] \approx \E_T[ \frac{T w}{w+T} \bE[\left\Vert \boldsymbol{v}_{1,T} \right\Vert_1]] \approx \E_T[T\bE[\left\Vert \boldsymbol{v}_{1,T} \right\Vert_1]].
\]
On the other hand, we have $\E[S_T] = b  + \E[S_T-b]$, where $\E[S_T-b]$ is also called the overshoot of the detection procedure and it is of order $o(b)$ as $b\to\infty$ \citep{siegmund1985sequential}. This yields the first-order approximation for the expected stopping time: 
\[
\E[T]\bE[\left\Vert \boldsymbol{v}_{1,T} \right\Vert_1] = b(1+o(1)) \Rightarrow \E[T] = \frac{b(1+o(1))}{\Delta},
\]
where $\Delta=\Delta_1$ when $S_t$ is the sum-type detection statistic, and $\Delta=\Delta_2$ for the max-type detection statistic. 

\end{document}